\documentclass[12pt]{iopart}
\usepackage{natbib}
\usepackage{epsfig}
\usepackage{setstack}
\usepackage{lscape}

\newcommand{\micronn}{\,\hbox{$\mu$m}}

\def\text{\rm}
\def\arcdegg{\hbox{$^\circ$}}
\newcommand{\simge}{\mbox{$\stackrel{>}{_{\sim}}$}}
\newcommand{\simle}{\mbox{$\stackrel{<}{_{\sim}}$}}

\begin{document}

\review[Optical Interferometry in Astronomy]
{Optical Interferometry in Astronomy}

\author{John D. Monnier  
\footnote[1]{Email: monnier@umich.edu}
}

\address{University of Michigan Astronomy Department, 941 Dennison Building,
500 Church Street, Ann Arbor MI 48109} 

\begin{abstract}
  Here I review the current state of the field of optical stellar
  interferometry, concentrating on ground-based work although a brief
  report of space interferometry missions is included.  We pause both
  to reflect on decades of immense progress in the field as well as to
  prepare for a new generation of large interferometers just now being
  commissioned (most notably, the CHARA, Keck and VLT
  Interferometers).  First, this review summarizes the basic
  principles behind stellar interferometry needed by the lay-physicist
  and general astronomer to understand the scientific potential as
  well as technical challenges of interferometry.  Next, the basic
  design principles of practical interferometers are discussed, using
  the experience of past and existing facilities to illustrate
  important points.  Here there is significant discussion of current trends
  in the field, including the new facilities under construction and
  advanced technologies being debuted.  This decade has seen the
  influence of stellar interferometry extend beyond classical regimes
  of stellar diameters and binary orbits to new areas such as mapping
  the accretion disks around young stars, novel calibration of the
  Cepheid Period-Luminosity relation, and imaging of stellar surfaces.
  The third section is devoted to the major scientific results from
  interferometry, grouped into natural categories reflecting these
  current developments.  Lastly, I consider the future of
  interferometry, highlighting the kinds of new science promised by
  the interferometers coming on-line in the next few years.  I also
  discuss the longer-term future of optical interferometry, including
  the prospects for space interferometry and the possibilities of
  large-scale ground-based projects.  Critical technological
  developments are still needed to make these projects attractive and
  affordable.

 \end{abstract}

\maketitle

\section{Introduction}
This review will introduce the theory, technique, and scientific goals
of optical stellar interferometry.  By combining
light collected by widely-separated telescopes, interferometrists can
overcome the diffraction-limit of an individual telescope.  The
angular resolution achieved by current instruments is indeed
astounding, $<5 \times 10^{-9}$ radians (1~milli-arcsecond), as are
the engineering feats of maintaining sub-micron 
optical stability and coherence over hundreds of meters of
pathlength while controlling polarization and dispersion over
broad wavelength bandpasses.
New capabilities are being applied in a wide
variety of astrophysics contexts, including fundamental stellar
parameters, novel ways to measure distances to stars, probing star
formation and evolution, direct detection of extrasolar planets, and
resolving cores of the nearest active galactic nuclei and brightest
quasars.  This current work is pointing the way towards next generation
facilities, and I will close with a discussion of efforts to bring the
angular resolution advantages of interferometers into space.

\subsection{Scope of Review}
The history of stellar interferometry spans more than a century, and a
proper documentation of the rich history is beyond the scope of this
review \citep[see][for a historical overview of the
field]{lawson2000}.  The main purpose of this review will be to
summarize the current state of the field of optical interferometry,
including past scientific and engineering lessons, current
astronomical motivations, and future goals and performance
expectations.  Interested readers may want to consult earlier
reviews which tend to emphasize other topics, especially \citet{sc1992} and \citet{quirrenbach2001}.
In order to restrict the length, I will concentrate on
long-baseline optical interferometry, giving only passing description
to diffraction-limited single-aperture experiments (e.g., Speckle
Interferometry, Aperture Masking, Adaptive Optics).  Further, I
consider ``optical interferometry'' in a restricted sense to mean the
light from the separate telescopes are brought together using optics,
as opposed to heterodyne interferometry whereby the radiation at each
telescope is coherently detected before interference (although I will
discuss briefly the important cases of the Intensity Interferometer
and heterodyne interferometry using CO$_2$ lasers).  In practice then,
``optical interferometry'' is limited to visible and infrared
wavelengths, and I will not discuss recent advances in mm-wave and
sub-mm interferometry.

\subsection{The Organization of Review}
This review is divided into 4 major sections. The first reviews the
basic theory behind optical interferometry and image reconstruction
through a turbulent atmosphere.  The second section explains the basic
designs of interferometers and core modern technologies which make
them work, including descriptions of current facilities.  Major
scientific results are outlined in the third section.  The last
section forecasts near-future science potential as well as the
long-term prospects of optical interferometry on the ground and in
space.

\subsection{Nomenclature}
In this review, ``optical'' does not indicate the visible portion of
the electromagnetic spectrum only, but generally refers to how the
light is manipulated (using optics); in the context of interferometry,
this will limit our discussion to wavelengths from the blue
($\sim0.4\mu$m) to the near-infrared (1-5$\mu$) and mid-infrared
(8-12$\mu$m).  Typical angular units used in this paper are
``milli-arcseconds'', or {\em mas}, where an arcsecond is the standard
$1/3600$ of a degree of angle.  Astronomers often use the magnitude
scale to discuss the wavelength-dependent flux density (power per unit
area per unit bandwidth) from an astronomical source, where the bright
star Vega ($\alpha$~Lyrae) is defined as 0~mag (corresponding to a
10000K blackbody); the magnitude scale is logarithmic such that every
factor of 10 brightness decrease corresponds to a flux ``magnitude''
increase of 2.5 (e.g., a contrast ratio of 10 astronomical magnitudes
is a factor of $10^4$).  In addition, the unit Jansky (Jy) is often
used for measuring the flux density of astronomical objects, 1~Jy
$=10^{-26} W m^{-2} Hz^{-1}$.

A number of other basic astronomical units are used herein.  The
distance between the Earth and Sun is one astronomical unit, 1~AU
$\simeq 1.5 \times 10^{11}$m.  Stellar distances are given in units of
parsecs (1 parsec is the distance to a star exhibiting a parallax
angle of 1~arcsecond): written in terms of other common units of
length, 1 pc $\simeq 3.09\times 10^{16}$m$\sim 3.26$ light years.

Lastly, I want to alert the reader (in advance) 
to Table~\ref{table:all_interferometers}
which will define all the interferometer acronyms used throughout the text.


\section{Basic Principles of Stellar Interferometry}

This section will review the basic principles of stellar
interferometry.  More detailed discussions of optical interferometry
issues can be found in the recent published proceedings of the Michelson
Summer School, ``Principles of Long Baseline Stellar Interferometry''
edited by P. Lawson \citep{mss2000}, and the proceedings of the 2002 Les
Houches Eurowinter school edited by G. Perrin and F. Malbet
\citep{leshouches2002}; earlier such collections also continue to play an
important reference role \citep[e.g.,][]{nrao86, nato1997}.
Other ``classic'' texts on the
subjects of radio interferometry and optics are \citet{tms2001},
\citet{bw65}, and \citet{goodman1985}.

\subsection{Basics of Stellar Interferometry}
The basic principles behind stellar interferometry should be 
familiar to any physicist, founded on the wave properties of light
as first observed by Thomas Young in 1803.  This result is widely known
through Young's ``two-slit experiment,'' although two slits were not used in the
original 1803 work.

\subsubsection{Young's Two-slit Experiment}

\begin{figure}
\begin{center}
\centerline{\epsfxsize=3in{\epsfbox{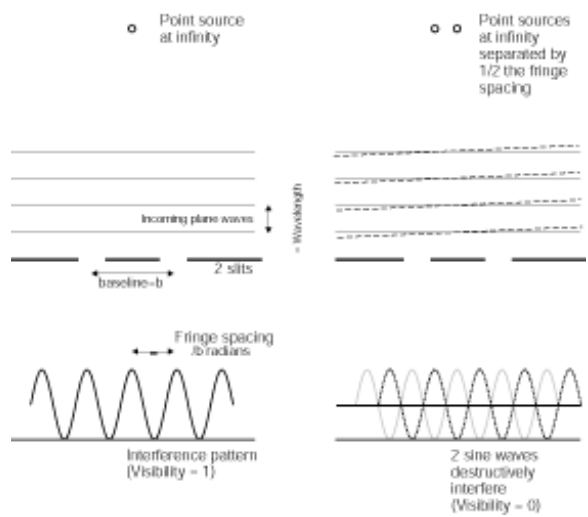}}}
\caption[Simple interferometer]{
  Young's two-slit interference experiment (monochromatic light) is
  presented to illustrate the basic principles behind stellar
  interferometry.  On the left is the case for a single point source,
  while the case on the right is for a double source with the angular
  distance being half the fringe spacing. Note, the interference
  pattern shown represents the {\em intensity} distribution, not the electric
  field.
\label{fig:2slit}}
\end{center}
\end{figure}

In the classical setup, monochromatic light from a distant (``point'')
source impinges upon two slits, schematically shown in the left panel
of Figure~\ref{fig:2slit}.  The subsequent illumination pattern is
projected onto a screen and a pattern of fringes is observed.  This
idealized model is realized in a practical interferometer by receiving
light at two separate telescopes and bringing the light together in a
beam-combination facility for interference (this will be discussed
fully in \S\ref{basicdesigns}; e.g., see
Figure~\ref{fig_realinterferometer}).  The interference is, of course,
due to the wave nature of light \a'a la Huygens; the electric field at
each slit (telescope) propagating to the screen with different
relative path lengths, and hence alternately constructively and
destructively interfering at different points along the screen.  One
can easily write down the condition for constructive interference; the
fringe spatial frequency (fringes per unit angle) of the intensity
distribution on the screen is proportional to the projected slit
separation, or baseline $b$, in units of the observing wavelength
$\lambda$ (see Figure \ref{fig:2slit}).  That is,
\begin{eqnarray}
 {\text{Fringe~Spacing}}  \equiv  \Delta\Theta  =  \frac{\lambda}{b} 
{\text{~radians}} \\
 {\text{Fringe~Spatial~Frequency}}  \equiv  u  =  \frac{b}{\lambda} {\text{~radians}^{-1}}
\end{eqnarray}

Imagine another point source of light (of equal brightness, but
incoherent with the first) located at an angle of $\lambda/(2b)$ from
the first source (see right of panel of Figure~\ref{fig:2slit}).  The
two illumination patterns are out of phase with one another by
180\arcdegg, hence cancelling each other out and presenting a
uniformly illuminated screen.  Clearly such an interfering
device (an ``interferometer'') can be useful in studying the
brightness distribution of a distant ``stellar'' object.  This
application of interferometry was first proposed by Fizeau in 1868
\citep{fizeau68} and successfully applied by Michelson to measure the
angular diameters of Jupiter's moons \citep{michelson90,michelson91}
in 1891 and later (with Pease in 1921) to measure the first angular
size of a star beyond the Sun \citep{michelson21} (see \S\ref{history}
for further details on the early history of optical interferometry).

\subsubsection{Angular Resolution}
The ability to discern the two components of a binary star
system is often used to gauge the spatial {\em resolution} of an
instrument, be it a conventional imaging telescope or a
separated-element interferometer.  Classical diffraction theory has
established the ``Rayleigh Criterion'' for defining the
(diffraction-limited) resolution of a filled circular aperture of
diameter D:
\begin{equation}
 {\text{Resolution~of~Telescope}}  \equiv  \Delta\Theta_{\text{telescope}} 
 =  1.22 \frac{\lambda}{D} 
{\text{ radians}} 
\end{equation}
This criterion corresponds to the angular separation on the sky when
one stellar component is centered on the first null in the diffraction
pattern of the other; the binary is then said to be {\em resolved}.  A
similar criterion can be defined for an interferometer: an
equal brightness binary is resolved by an interferometer 
if the fringe contrast goes to zero at the longest baseline.
As motivated in the last paragraph,
this occurs when the angular separation is $\frac{\lambda}{2b}$, where
b is the baseline.  Hence,
\begin{equation}
 {\text{Resolution~of~Interferometer}}  \equiv  \Delta\Theta_{\text{interferometer}} 
 =  \frac{\lambda}{2b} 
{\text{ radians}} 
\end{equation}

While these two criteria are somewhat arbitrary, they are useful for
estimating the angular resolution of an optical system and are in
widespread use by the astronomical community.

\subsubsection{Complex Visibility}
\label{section:zernike1}
One can be more quantitative in interpreting the fringe
patterns observed with an interferometer.
The fringe contrast is historically called the
{\em visibility} and, for the simple (two-slit) interferometer considered here,
can be written as  
\begin{equation}
V = \frac{I_{\text{max}} - I_{\text{min}}}
{I_{\text{max}} + I_{\text{min}}}
  =  \frac{{\rm Fringe~Amplitude}}{{\rm Average~ Intensity}}
\end{equation}
where $I_{\text{max}}$ and $I_{\text{min}}$ denote the maximum and
minimum intensity of the fringes.  Hence, the left and right fringe
patterns of Figure \ref{fig:2slit} have visibilities of one and zero
respectively. 

The Van Cittert-Zernike Theorem \citep[see][for complete discussion
and proof]{tms2001} relates the contrast of an interferometer's
fringes to a unique Fourier component of the impinging brightness
distribution.  In fact, the visibility is exactly proportional to the
amplitude of the image Fourier component corresponding to
the (spatial) fringe spatial frequency defined above ($u=b/\lambda
\text{ ~rad}^{-1}$).  Also, the phase of the fringe pattern is equal
to the Fourier phase of the same spatial frequency component.

The Van Cittert-Zernike Theorem can be expressed concisely in
mathematical terms. Consider that the astronomical target emits light
at frequency $\nu$ over only a very small portion of the sky with
specific intensity $I_\nu(\theta,\phi)$, so small that the spherical
coordinates $\theta_0+\delta\theta$ and $\phi_0+\delta\phi$ can be
interpreted as Cartesian coordinates $x_\Omega$ and $y_{\,\Omega}$
centered around $\theta_0$ and $\phi_0$ on the plane of the sky.  We
can write the interferometer response (amplitude and phase of the
fringes) as the frequency-dependent complex visibility
$\tilde{\mathcal V}_\nu(u,v)$, defined as the Fourier Transform of the
brightness distribution $I_\nu(\vec{r_{\Omega}})$, normalized so that
$\tilde{\mathcal V}(\frac{\vec{D}}{\lambda}=0) =1$.

\begin{equation}
|{\mathcal V}_\nu\big(\frac{\vec{D}}{\lambda}\big)|e^{-i\phi_{V_\nu}}
        =\frac{
         \int\limits_{\delta\Omega} dx_{\Omega}\,dy_{\Omega}\,
        I_\nu(\vec{r_\Omega})
           e^{-2\pi i\left(\frac{\vec{D}}{\lambda}\cdot\vec{r_\Omega}\right)}}{
        \underbrace{\int\limits_{\delta\Omega}dx_\Omega\,dy_{\Omega}\,
         I_\nu(\vec{r_\Omega})}_{\mbox{\scriptsize Total Specific Flux}}}
\end{equation}

using the following notation:
\begin{eqnarray}
 \vec{r_\Omega}&=&(x_\Omega,y_\Omega)\nonumber\\
 \frac{\vec{D}}{\lambda}&\equiv&
        \mbox{the baseline vector $\vec{D}$ projected onto the plane}\nonumber\\
        &&\mbox{of the sky in units of wavelength $\lambda$}\nonumber\\
        &=&(u,v) ~~~{\rm [Common~Notation]} \label{eq_uv}
\end{eqnarray}

\begin{figure}
\begin{center}

\centerline{\epsfxsize=4in{\epsfbox{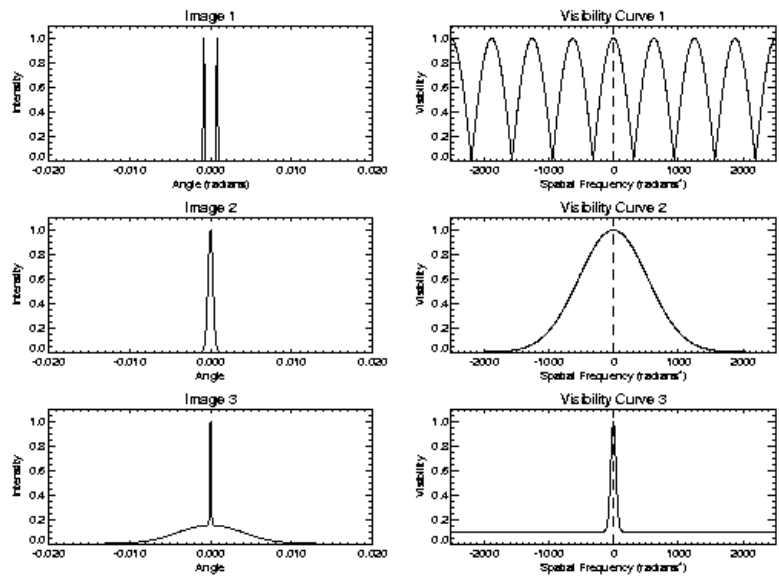}}}
\caption[Example 1-D images and visibility curves]{This figure shows 
simple one-dimensional images and their corresponding visibility curves.
The left panels are the images while the right panels correspond to the
Fourier amplitudes, i.e. the {\em visibility} amplitudes.  
Note that ``large'' structure in image-space result in
``small'' structure in visibility-space.}
\label{fig:visibility}
\end{center}
\end{figure}

Figure\,\ref{fig:visibility} shows some simple examples of
one-dimensional images and the corresponding visibility curves.  The
top panels show the case of an equal binary system, where both
components are unresolved.  The periodicity in the visibility-space
corresponds to the binary separation.  The middle set of panels is
representative of a compact, but resolved, source (such as a star
surrounded by an optically-thick dust shell).  The small image size
means there is more high spatial frequency information, and this is why
the corresponding visibility curve is non-zero even at high
resolution. Lastly, the bottom panels show an image of a unresolved
star (with 10\% of the total flux) surrounded by larger-scale
structure (this is expected when a star is surrounded by an
optically-thin envelope of dust).  The large-scale structure
(containing 90\% of the total flux) can be seen to be ``resolved'' on
short baselines (at low spatial frequency), while the point-source
remains unresolved out to the highest spatial frequency.  Note that
the visibility plateaus at 0.10, corresponding to the fraction of the
total flux which is left unresolved.  This is easy to understand since
the Fourier Transform is linear; that is, the (complex) visibility of
a point-source and extended structure is equal to the visibility of
the point-source plus the visibility of the extended structure
separately.  This property of linearity is very helpful in
interpreting simple visibility curves.

\begin{figure}
\begin{center}
\epsfig{file=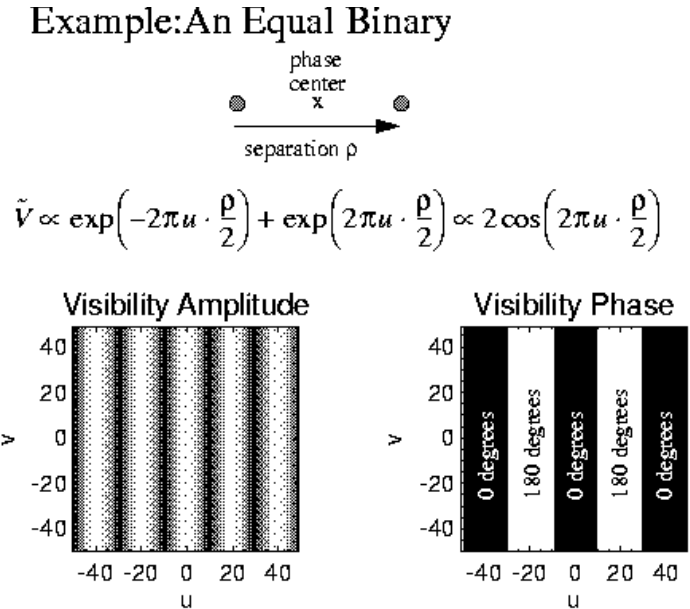,height=8cm,clip}
\caption
{This figure shows the complex visibility for an equal binary system
in the 2-dimension (u,v) plane.
With the above choice for the phase center, the Fourier phases can be
represented simply.  Notice the abrupt phase jumps when the 
visibility amplitude goes through a null. This figure is
reproduced through the courtesy
of the NASA/Jet Propulsion Laboratory, 
California Institute of Technology, Pasadena, California \citep{monnier_mss}.
\label{monnier_eg1}}
\end{center}
\end{figure}

Most astronomical objects are not one-dimensional, and the
two-dimensional space of spatial frequencies is called the
Fourier Plane, or the (u,v) plane, named after the (u,v) coordinates
defined in Eq.\ref{eq_uv}.  Further, in general we must consider both the
visibility amplitude and the visibility phase.
For example,
consider the equal binary system depicted in Figure~\ref{monnier_eg1}.
The complex visibility can be easily written by choosing the origin
midway between the two components.  Note the abrupt phase jump when
the visibility amplitude goes through a null.  These discontinuities
are smoothed out when the two components are not precisely equal.

\subsection{Atmospheric Problems}
\label{section:seeing}
\begin{figure}
\begin{center}
\epsfig{file=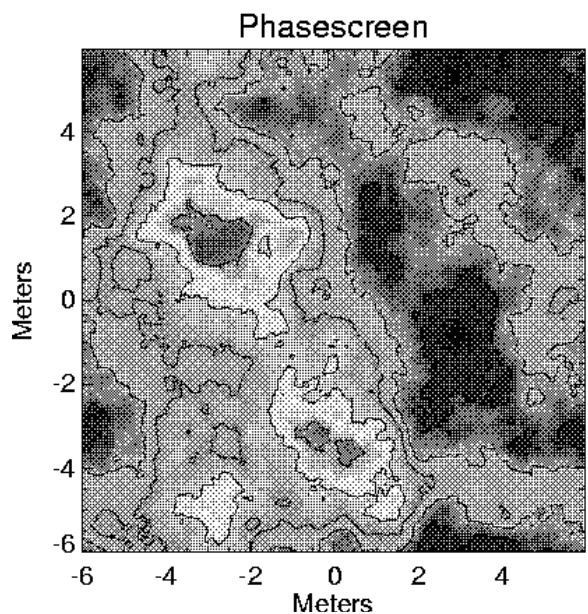,clip,width=3.0in}
\caption[bad atmosphere]{
This figure
shows a typical realization of Kolmogorov turbulence (r$_0 = 50$\,cm at
$\lambda = 2.2$\micronn); each
solid contour line represents $\frac{\lambda}{2}$ of wavefront distortion.
Some areas of the aperture show coherent areas larger
than r$_0$, and some much smaller; r$_0$ is a statistical property of
atmospheric turbulence and wavefront perturbations occur over a wide
range of scales.
\label{fig:phasescreen}}
\end{center}
\end{figure}

An incoming plane wave from a stellar source is corrupted as it
propagates through the turbulent atmosphere.  Variations in the column
density of air along different paths cause the effective pathlength to
vary, introducing wavefront distortion.  If these distortions become a
significant fraction of a wavelength across the aperture of a
telescope, the size of the image formed will not be
diffraction-limited by the primary mirror, but rather by the coherence
scale of the incoming wavefront.  The transverse distance over which
one expects rms pathlength difference to be $\lambda$/2.4 has been
defined as the Fried parameter and is denoted by r$_0(\lambda)$
\citep{fried65}; hence telescope apertures larger than r$_0(\lambda)$
can expect significant degradation of image quality (when observing at
wavelength $\lambda$) due to atmospheric effects.  In fact, for an
r$_0$ diameter circular patch, the rms phase error is $\sim$1.03
radians.  At $\lambda=$500\,nm, r$_0$ is typically 10\,cm (toward
zenith) at average observing sites and hence even small telescopes can
not be used at their diffraction limit in the visible.  In such cases,
the observed angular size of a point source will be determined
entirely by r$_0(\lambda)$ at a given wavelength, and is known as the
seeing disk size, $\Theta_{\rm seeing}(\lambda)$.  The Kolmogorov
theory of turbulence \citep{kolmogorov61} predicts that r$_0(\lambda)
\propto \lambda^{6/5}$, and hence the seeing size, $\Theta_{\rm
  seeing}(\lambda) \propto \frac{\lambda}{r_0(\lambda)} \propto
\lambda^{-1/5}$, is only weakly dependent on the wavelength
\citep{fried65}.  An example of the phase delays associated with a
snapshot of Kolmogorov turbulence can be seen in
Figure\,\ref{fig:phasescreen} for 12-m square, corresponding roughly
to the size of the largest telescopes today (e.g., the Keck
telescopes).

Another consequence of turbulence is that the image distortion varies
across the sky, although stars located close together suffer similar
seeing effects.  The angle over which image distortions are correlated
is called the ``isoplanatic'' angle, and is only a few arcseconds in
the visible and about an arcminute in the near-infrared.  This angle
is determined by the vertical distribution of the turbulence --
obviously low-level turbulence would induce correlated image
distortions over larger sky angles than the same turbulent layer
located higher up in the atmosphere. The isoplanatic angle is a
critical parameter for the field-of-view of adaptive optics systems
which actively sense and correct for atmospheric turbulence in
realtime using a deformable mirror.

Another important atmospheric diagnostic is the coherence time, t$_0$.
Typically, one assumes a ``frozen'' turbulence model in which the
atmospheric density perturbations are assumed constant over the time
it takes wind to blow them across a given aperture (also known as
Taylor's hypothesis of frozen turbulence).  This motivates a convenient
estimate for the coherence time: $t_0(\lambda) \equiv r_0(\lambda) /
v_{\rm wind}$, where $v_{\rm wind}$ is the wind speed.  At most sites,
wind speeds are $\sim$10\,m\,s$^{-1}$ and so $t_0 \sim 10$\,ms at
500\,nm.  Further discussion of atmospheric turbulence and degradation
of astronomical images can be found in \citet{kolmogorov61},
\citet{roddier1981}, and \citet{roddier82}.

These parameters, r$_0(\lambda)$ and t$_0(\lambda)$, are extremely
important for the design of an interferometer, because the value of
r$_0$ limits the useful size of the collecting aperture, while t$_0$
limits the coherent integration time.  Both of these are crucial for
predicting an interferometer sensitivity to faint objects and much
debate surrounds the best estimates for these parameters at various
sites \citep[e.g.,][]{dyck83a, roddier90, theo1994, treuhaft95}.  This
topic is revisited in \S\ref{sensitivity} when I discuss the limiting
magnitude of current interferometers.

It is well-known that r$_0(\lambda)$ and t$_0(\lambda)$ depend greatly
on the observing site, and we now consider the unique seeing
conditions of Mauna Kea, Hawaii, as an example.  The coherence scale
is unusually long due to the highly laminar flow of the Pacific winds
over the peak of the mountain (elevation 4200\,m); r$_0$ usually lies
between 10 and 40\,cm at 500\,nm \citep{wizPC}.  However, the fast
winds of the overhead jet stream result in very short coherence times:
t$_0$ between 1.5 and 10\,ms \citep{wizPC}.  It should be emphasized
that seeing is notoriously difficult to characterize due to large
variations in time (both on short time scales as well as seasonal
ones) as evidenced by the large range of r$_0(\lambda = 500\,nm)$ and
t$_0(\lambda=500\,nm)$ values just given.

\subsubsection{Atmospheric Phase Errors}
\begin{figure}
\begin{center}
\centerline{\epsfig{figure=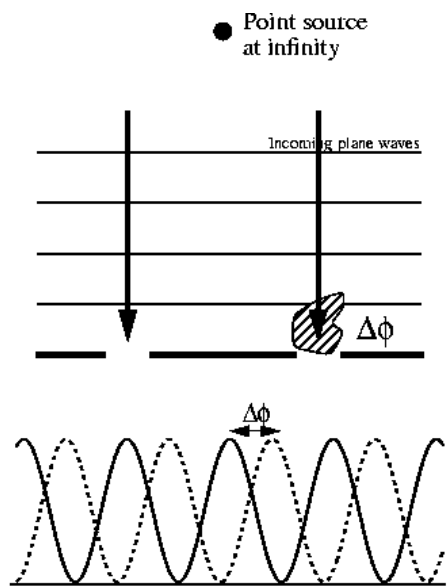,height=3in,clip}}
\caption{Atmospheric time delays or phase errors at telescopes 
cause fringe shifts, as can be seen through analogy with Young's
two-slit experiment. This figure is reproduced through the courtesy
of the NASA/Jet Propulsion Laboratory, 
California Institute of Technology, Pasadena, California \citep{monnier_mss}.
\label{monnier_fig_1}}
\end{center}
\end{figure}

The fluctuating amount of integrated atmospheric pathlength above each
telescope introduce wavefront time delays which show up as phase
shifts in the measured fringes in an interferometer, as illustrated in
Figure~\ref{monnier_fig_1}.  In this figure, an optical interferometer
is represented again by a Young's two-slit experiment, as discussed
earlier in this section.  The spatial frequency of these
fringes is determined by the distance between the slits (in units of
the wavelength of the illuminating radiation).  However if the
pathlength above one slit is changed (due to a pocket of warm air
moving across the aperture, for example), the interference pattern
will be shifted by an amount depending on the difference in pathlength
of the two legs in this simple interferometer.  If the extra
pathlength is half the wavelength, the fringe pattern will shift by
half a fringe, or $\pi$~radians.  The phase shift is completely
independent of the slit (telescope) separation, and only depends on
slit-specific (telescope-specific) phase delays.

The most obvious impact of atmospheric phase delays is that the
assumptions of the van Cittert-Zernike theorem no longer apply, and
that the measure fringe phase can no longer be associated with the
Fourier phase of the sky brightness distribution (the fringe amplitude
retains its original meaning, since phase changes do not change the
measured fringe amplitudes for short exposures).  The corruption of
this phase information has serious consequences, since imaging of
non-centrosymmetric objects rely on the Fourier phase information
encoded in this intrinsic phase of interferometer fringes.  Without
this information, imaging can not be done except for simple  objects
such as disks or round stars.  Fortunately, a number of strategies
have evolved to circumvent these difficulties.

\subsubsection{Phase Referencing}
Possible methods for recovering this phase information using {\em
  phase referencing} techniques are discussed in Chapter~9 (written by
A. Quirrenbach) of ``Principles of Long Baseline Stellar
Interferometry \citep{mss2000}.  Few scientific results have resulted
from phase referencing techniques to date, but this is expected to
change over the coming decade as sophisticated new instruments are
being deployed.  Here, I mention a few of the most promising methods:

\begin{enumerate} 
\item{ {\bf Nearby sources.}  If a bright point reference source (or
    source with well-known structure) lies within an isoplanatic
    patch \citep[see][]{quirrenbach2000}, then its fringes will act as
    a probe of the atmospheric conditions.  By measuring the
    instantaneous phases of fringes from the bright reference source,
    one can correct the corrupted phases on a neaby ``target'' source.
    This has been applied to narrow-angle astrometry where fringe
    phase information is used for determining precise relative
    positions of nearby stars \citep{shao92,colavita99,lane2000a}; see
    Figure~\ref{ptifig} for some preliminary results published by the
    Palomar Testbed Interferometer.  While it would be very valuable
    to use an artificial guide star for phase-referencing a long
    baseline interferometer, current laser beacons are too spatially
    extended (resolved) to produce interferometric fringes.}
\item{{\bf $\Delta\Phi$ Monitoring.}  In the millimeter and
    sub-millimeter, phase shifts caused by fluctuations in atmospheric
    water vapor column density can be monitored by observing its line
    emission.  This information can be used to phase-compensate the
    interferometer, allowing longer coherent integrations and accurate
    fringe phase determination on the target (\citealt{wiedner98}, and
    references therein).  In the mid-infrared, strategies to actively
    monitor ground-level turbulence using temperature sensors are
    being explored by the Infrared Spatial Interferometer group
    \citep{short2002} at Mt. Wilson motivated by recent atmospheric
    studies (e.g., \citealt{bester92}).  \citet{townes2002} recently
    proposed that realtime-monitoring of Rayleigh or Raman
    backscattering might be used to correct for atmospheric column
    density variations in the context of optical interferometers, but
    this method has not yet been validated.  }
\item{{\bf Multi-wavelength.}  Another possibility is to observe a
    target at multiple wavelengths and to use data from one part of
    the spectrum to calibrate another.  For example, one might use
    fringes formed by the continuum emission to phase reference a
    spectral line (e.g., \citealt{gi2t97}).  To use this method, one
    must assume knowledge about the brightness distribution at one of
    the wavelengths being used.}
\end{enumerate}

Currently, phase referencing is not possible with most current beam
combiners in operation, either due to low spectral resolution or
limited field-of-view.  In order to recover phase information, one must
one must make use of the {\em closure phases}.

\subsubsection{Closure Phases}
\label{closurephase}
\begin{figure}
\begin{center}

\centerline{\epsfig{figure=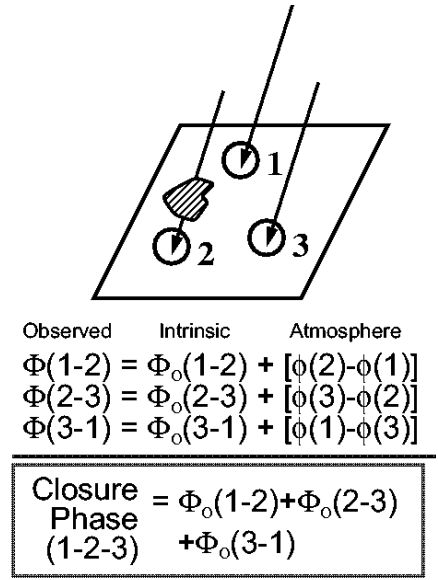,height=3in,clip}}
\caption{This figure explains the principle behind closure phase analysis.
Phase errors introduced at any telescope causes equal but opposite phase
shifts, canceling out in the {\em closure phase} 
(figure after \citealt{readhead88}).  This figure is reproduced 
through the courtesy
of the NASA/Jet Propulsion Laboratory,
California Institute of Technology, Pasadena, California \citep{monnier_mss}.
\label{monnier_fig_3}}
\end{center}
\end{figure}

Consider Figure~\ref{monnier_fig_3} in which a phase delay is
introduced above Telescope~2 of a 3-telescope array.  As discussed
in the last section, this additional
delay causes
a phase shift in the fringe detected between telescopes 1-2.
Note that a phase shift is also
induced for fringes between telescopes 2-3; however, this phase shift
is equal and {\em opposite} to the one for telescopes 1-2.  Hence, the
sum of three fringe phases, between 1-2, 2-3, and 3-1, is insensitive
to the phase delay above telescope 2.  This argument holds for
arbitrary phase delays above any of the three telescopes.  In general,
the sum of three phases around a closed triangle of baselines, the
{\em closure phase}, is a good interferometric observable; that is, it
is independent of telescope-specific phase shifts induced by the
atmosphere or optics.

The idea of closure phase was first introduced to compensate for poor
phase stability in early radio VLBI work (\citealt{jennison58}).
Application at higher frequencies was first mentioned by
\cite{rogstad68}, but only much later carried out in the
visible/infrared through aperture masking experiments
(\citealt{baldwin86}; \citealt{haniff87}; \citealt{readhead88};
\citealt{haniff89}).  Currently three separate-element interferometers
have succeeded in obtaining closure phase measurements in the
visible/infrared, first at COAST (\citealt{coast96}), soon after at
NPOI (\citealt{npoi97}), and most recently at IOTA \citep{traub2002}.

\label{monnier_relations}
How can these {\em closure phases} be used to figure out the {\em
  Fourier~phases} which are needed to allow an image to be
reconstructed?  Each closure triangle phase can be thought of as a
single linear equation relating three different Fourier phases
(assuming none of the baselines are identical), which we desire to
solve for; hence, we must count the number of linear equations
available and compare to the number of unknowns.  For N telescopes,
there are ``N choose 3,"
${{N}\choose{3}}=\frac{(N)(N-1)(N-2)}{(3)(2)}$, possible closing
triangles.  However, there are only
${{N}\choose{2}}=\frac{(N)(N-1)}{2}$ independent Fourier phases;
clearly not all the closure phases can be independent.  The number of
{\em independent} closure phases is only
${{N-1}\choose{2}}=\frac{(N-1)(N-2)}{2}$, equivalent to holding one
telescope fixed and forming all possible triangles with that telescope
\citep[as discussed by][]{readhead88}.  The number of independent
closure phases is always less than the number of phases one would like
to determine, but the {\em percent} of phase information retained by
the closure phases improves as the number of telescopes in the array
increases.  Table~\ref{monnier_table1} lists the number of Fourier
phases, closing triangles, independent closure phases, and recovered
percentage of phase information for telescope arrays of 3 to 50
elements.  For example, approximately 90\% of the phase information is
recovered with a 21 telescope interferometric array (e.g.,
\citealt{readhead88}).  As discussed in the next section, this phase
information can be coupled with other image constraints (e.g., finite
size and positivity) to reconstruct the source brightness
distribution.

In addition to the mathematical (linear algebra) interpretation of
closure phases, there are a few other important properties worth
noting.

\begin{itemize}
\item{For sources with {\em point-symmetry} (otherwise known as
    centro-symmetry), all the closure phases are either 0\arcdegg~ or
    180\arcdegg.  It is easy to prove this by imagining the
    image-center (``phase-center'') at the point of centro-symmetry.}
\item{Closure phases are not sensitive to an overall translation of image.
A translation is indistinguishable from atmospheric phase delays for any
given closing triangle.}
\item{The closure phases are independent of {\em telescope-specific}
    phase errors, however non-zero closure phases from a point source
    can result from having non-closing triangles and phase delays in
    the beam combiner (e.g., for a three-telescope pair-wise beam
    combiner).}
\end{itemize}

\begin{table}
\footnotesize
\caption {Phase information contained in the closure phases alone
\label{monnier_table1}}
\begin{center}
\begin{tabular}{ccccc}
Number of  & Number of & Number of & Number of Independent & Percentage of \\
Telescopes & Fourier Phases & Closing Triangles & Closure Phases & Phase Information \\
\hline
3 & 3 & 1 & 1 & 33\% \\
7 & 21 & 35 & 15 & 71\% \\
21 & 210 & 1330 & 190 & 90\% \\
27 & 351 & 2925 & 325 & 93\% \\
50 & 1225 & 19600 & 1176 & 96\% \\
\end{tabular}
\end{center}
\end{table}

\subsection{Image Reconstruction}
While very few images have been made by today's optical
interferometers, new telescope arrays are now being commissioned which
will make true imaging interferometry straightforward.  Because these
new imaging capabilities are likely to have significant impacts over
the next decade, I wish to review the basic principles of apertures
synthesis imaging.  However, I will restrain myself from excessive
elaboration here,
and instead refer the interested reader to the extensive
radio interferometry literature, especially regarding ``Very Long
Baseline Interferometry (VLBI).''

While modeling visibility and closure phase data with simple models is
useful, one would like to make an image unbiased by theoretical
expectations.  Since any image can be alternatively represented by its
Fourier components, the collection of all ``interesting'' components
can allow the interferometric data to be inverted, thus reconstructing
an estimation of the image brightness distribution.  The collection of a large
number of Fourier components is greatly aided by increasing the number
of telescopes, since independent combinations of telescopes increase
with the number of telescopes to the second power 
($\frac{(N)(N-1)}{2}$; see last section)
  
With a large number of measurements, images of arbitrary complexity
should be attainable using visible/infrared interferometers and
reliable closure phase measurements.  The importance of ``filling up''
the (u,v) plane with measurements when imaging is discussed more fully
in \S\ref{uvplane}.  The next subsections will discuss strategies
currently employed based on the techniques of VLBI in the radio.

\subsubsection{Guiding Principles}
The goals of an image reconstruction procedure can be stated quite
simply: find an image which fits both the visibility amplitudes and
closure phases within experimental uncertainties.  However in
practice, there are an infinite number of candidate images which
satisfy these criteria, because interferometric data is always
incomplete and noisy.  Furthermore, the closure phases can not be used
to unambiguously arrive at Fourier phase estimates as stated above,
even under ideal noise-free conditions.

Additional constraints must be imposed to ``select'' an image as the
best-estimate of the true brightness distribution (to ``regularize''
this ill-posed inverse problem). These constraints introduce
correlations in the Fourier amplitudes and phases, and essentially
remove degrees of freedom from our inversion problem. Some of the most
common (and reasonable) constraints are described below.
\begin{itemize}
\item{Limited Field-of-View.  This constraint is always imposed in
    aperture synthesis imaging, even for a fully-phased array (e.g.,
    VLA).  Limiting the field-of-view introduces correlations in the
    complex visibility in the (u,v) plane.  This is a consequence of
    the Convolution Theorem, a multiplication in image-space
    is equivalent to a convolution in the corresponding
    Fourier-space.}
\item{Positive-Definite.  Since brightness distributions can not be
    negative, this is a sensible constraint (although not appropriate
    in some cases, such as for reconstruction of Stokes/polarization
    components or imaging spectral line absorption).  While clearly
    limiting the range of ``allowed'' complex visibilities, there are
    few obvious, intuitive effects in the Fourier-plane; one is that
    the visibility amplitude is maximum at zero spatial frequency.
    The Maximum Entropy Method (see \S\ref{monnier_mem}) naturally
    incorporates this constraint.}
\item{``Smoothness.''  The Maximum Entropy Method (MEM), for instance,
    selects the ``smoothest'' image consistent with the data. 
See \S\ref{monnier_mem} for more discussion of MEM. }
\item{{\em a priori} Information.  One can incorporate previously
    known information to constrain the possible image reconstructions.
    For instance, a low resolution image may be available from a
    single-dish telescope.  Another commonly encountered example is
    point source embedded in nebulosity; one might want the
    reconstruction algorithm to take into account that the source at
    the center is point-like from theoretical arguments. }
\end{itemize}

For a phased interferometric array (e.g., the Very Large Array) where
the Fourier phases are directly measured (avoiding the need for
closure phases), one can use a number of aperture synthesis techniques
to produce an estimate of an image based on sparsely sampled Fourier
components.  These procedures basically remove artifacts, i.e.
sidelobes, of the interferometer's point-source response arising from
uncomplete sampling of the (u,v) plane.  These procedures do not
incorporate closure phases, but work by inverting the Fourier
amplitudes and phases to make an image.  A brief explanation of the
popular algorithms CLEAN and MEM follow with additional
references for the interested reader.  See \cite{nrao86} for essays on
these topics aimed at radio astronomers.

\subsubsection{CLEAN}
Originally described by \cite{hogbom74}, CLEAN has been traditionally
the most popular algorithm for image reconstruction in the radio
because it is both computationally efficient and intuitively
understandable.  Given a set of visibility amplitudes and phases over
a finite region of the Fourier plane, the ``true'' image can be
estimated by simply setting all other spatial frequencies to zero and
taking the (inverse) Fourier Transform.  As one might expect, this
process leads to a whole host of image artifacts, most damaging being
positive and negative ``sidelobes'' resulting from non-complete
coverage of the Fourier plane; we call this the ``dirty map.''  The
unevenly-filled Fourier plane can be thought of as a product of a
completely-sampled Fourier plane (which we desire to determine) and a
spatial frequency mask which is equal to 1 where we have data and 0
elsewhere.  Since multiplication in Fourier space is identical to
convolution in image space, we can take the Fourier transform of the
spatial frequency mask to find this convolving function; we call this
the ``dirty beam.''  Now the image reconstruction problem can be
recast as a ``deconvolution'' of the dirty map with the dirty beam.

The dirty map is CLEANed by subtracting the dirty beam (scaled to
some fraction of the map peak) from the brightest spot in the dirty
map.  This removes sidelobe structure and artifacts from the dirty
map.  Repeating this process with dirty beams of ever decreasing
amplitudes leads to a series of delta-functions which, when combined,
fit the interferometric data.  For visualization, this map of point
sources is convolved with a Gaussian function whose FWHM values are
the same as the dirty beam; this removes high spatial resolution
information beyond the classic ``Rayleigh'' criterion cutoff.
One major weakness with CLEAN is that this smoothing changes
the visibility amplitudes, hence the CLEANed image no longer
strictly fits the interferometric data, especially the
spatial frequency information near the diffraction limit.
Another weakness is that CLEAN does not
directly use the known uncertainties
in the visibility data, and hence there is no natural method to
weight the high SNR data more than the low SNR data during image
reconstruction.  Further discussion of various
implementations of CLEAN can be found in \cite{clark80}, \cite{schwab84}, 
\cite{cornwell83}, and Chapter~7 of \cite{nrao86} by T. Cornwell.

\subsubsection{MEM}
\label{monnier_mem}
The maximum entropy method (MEM) makes better use of the highest
spatial frequency information by finding the {\em
smoothest} image consistent with the interferometric data.  While
enforcing positivity and conserving the total flux in the frame,
``smoothness'' is estimated here by a global scalar quantity S, the
``entropy.''  If $f_i$ is the fraction of the total flux in pixel i,
then $S=-\sum_{i} f_i \ln \frac{f_i}{I_i}$ after the thermodynamic quantity;
$I_i$ is known as the {\em image prior} and must be specified by the user.
The MEM map $f_i$ will tend toward $I_i$ when there is
little (or noisy) data to constrain the fit.  Often $I_i$ is assumed to
be a uniformly bright background, however one can use other image priors if
additional information is available, such as
the overall size of the source which may be known from previous observations.

Mathematically, MEM solves the multi-dimensional (N=number of
pixels) constrained minimization problem which only
recently has become computationally realizable on desktop computers.
Maintaining an adequate fit to the data ($\Sigma\chi^2 \sim$ number of
degrees of freedom), MEM reconstructs an image with maximum S.  MEM
image reconstructions always contain some spatial frequency
information beyond the diffraction limit in order to keep the image
as ``smooth'' as possible consistent with the data.
Because of this, images typically have maximum spatial resolution a few times
smaller than the typical Rayleigh-type resolution encountered with CLEAN
 (``super-resolution'').
Further discussions of MEM and related Bayesian methods can be found in
\cite{pp92}, \cite{nn86},
\cite{sb84}, \cite{mem83}, and \cite{sivia87}.

Unfortunately, MEM images also suffer from some characteristic artifacts
and biases.  Photometry of MEM-deconvolved images is necessarily biased
because of the positivity constraint; any noise or uncertainty in the
imaging appears in the background of the reconstruction instead of
the source, systematically lowering the estimated fluxes of compact sources.
Also, fields containing a point source embedded in extended emission
often show structure reminiscent of Airy rings, the location of the
rings being influenced by the wavelength of the observation and not inherent
to the astrophysical source.  Fortunately, these imaging artifacts 
are greatly
alleviated for asymmetric structures, when closure phases and not
the visibility amplitudes play a dominant role in shaping the
reconstructed morphology.

\subsubsection{Including Closure Phase Information}

The above algorithms were originally 
designed to use Fourier amplitudes and
phases, not closure phases.  In order to use these algorithms,
one has to come up a way to estimate the Fourier phases, when only
the closure phases are available.
Early image reconstruction algorithms
incorporated closure phase information by using an iterative scheme
(\citealt{thompson86}; \citealt{rw78}).  
The following steps summarize
this process:
\begin{enumerate}
\item{Start with a Fourier ``phase model'' based on either prior
    information or setting all phases to zero.}
\item{Determine candidate phases by using some values from the ``phase
    model'' and enforcing all the (self-consistent) closure phase
    relations (see \S\ref{monnier_relations}).}
\item{Using CLEAN or MEM, perform aperture synthesis mapping on the
    given visibilities and candidate phases.  At this stage, image
    constraints such as positivity and/or finite support are applied.}
\item{Use this image as a basis for a new ``phase model.''}
\item{Go to step 2 and repeat until the process converges to a stable
    image solution.} \end{enumerate}

\begin{figure}
\begin{center}
\epsfig{file=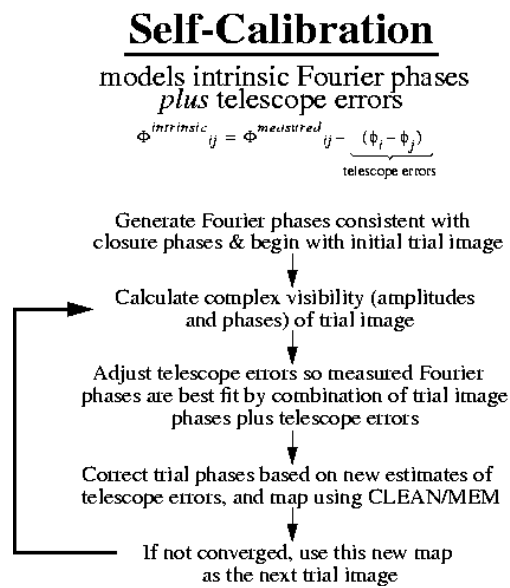,height=3in,clip}
\caption{This is a flow diagram for a incorporating closure phase
information into CLEAN/MEM aperture synthesis imaging algorithms
based on the ``self-calibration'' procedure of \cite{cw81}. 
This figure is reproduced through the courtesy
of the NASA/Jet Propulsion Laboratory,
California Institute of Technology, Pasadena, California \citep{monnier_mss}.
\label{monnier_selfcal}}
\end{center}
\end{figure}

\cite{cw81} introduced a modification of the above scheme by
explicitly solving for the telescope-specific errors as part of the
reconstruction step.  Hence the measured (corrupted) Fourier phases
are fit using a combination of intrinsic phases (which are used for
imaging using CLEAN/MEM) plus telescope phase errors.  In this scheme,
the closure phases are not explicitly fit, but rather are conserved in
the procedure since varying telescope-specific errors can not change
any of the closure phases.  Figure~\ref{monnier_selfcal} shows a flow
diagram for this procedure; thoughtful consideration is required in
order to fully understand the power and elegance of self-calibration,
affectionately known as  ``self-cal.''

Self-calibration works remarkably well for large number of telescopes,
but requires reasonably high signal-to-noise ratio (SNR$\simge$5)
complex visibilities.  Once the SNR decreases below this point, the
method completely fails.  The conceptualization of solving for
telescope-specific errors, while useful for the
radio, is not applicable for visible/infrared interferometry where the good
observables are the closure phases themselves, not corrupted Fourier
phases.  This is because the time-scale for phase variations in the
visible/infrared is much less than a second, as opposed to
minutes/hours in the radio.

Of course, the self-calibration iteration loop can be sidestepped
altogether by fitting directly to all the data, the visibility
amplitudes and closure phases, using MEM or some other regularization
scheme.  This would have the added advantage of allowing all the
measurement errors to be properly addressed, theoretically resulting
in the optimal image reconstruction.  \citet{buscher1994} suggested
this approach, but there has been little demonstrated progress in this
method to date.  I anticipate revived activity as more
interferometers with ``imaging'' capability begin to produce data.

\subsubsection{Speckle Interferometry}
\label{section:speckle}

Another related interferometric technique which permits
diffraction-limited observation through a turbulent atmosphere using a
single filled-aperture telescope is ``speckle interferometry,'' the
promise of which was first realized by \citet{labeyrie70} in 1970.  In
\S\ref{section:seeing}, I claimed that observed angular size of a
point source will be determined entirely by r$_0$ at a given
wavelength, and is known as the seeing disk size, $\Theta_{\rm
  seeing}$.  However, this is only true for a long-exposure image.  A
single short-exposure image of a star actually consists of a network of
small ``speckles'' extending over $\Theta_{\rm seeing}$.

In the original formulation of speckle interferometry, short exposures
of an astrophysical object are made to freeze this ``speckling''
induced by the turbulent atmosphere.  The amount of high-resolution
structure in the speckle pattern, as quantified by its power spectrum,
is a measure of two things: 1) the quality of the atmospheric seeing,
and 2) the high resolution structure in the object of interest.
Observing a nearby point-source star allows the calibration of the
seeing contribution and thus the extraction of interferometric
visibility measurements out to the diffraction limit of the telescope
(i.e., the longest baseline).  In analyzing this situation, one can
think of many virtual subapertures (with size equal to the coherence
length r$_0$) spread across the full telescope, with fringes forming
between all the subaperture pairs.  After the original formulation by
Labeyrie, it was discovered that the Fourier phases could also
be estimated from such data \citep[e.g.,][]{kt74,weigelt77}.

Speckle interferometry data is often reduced using the ``bispectrum,''
which permits a direct inversion from the estimated Fourier amplitudes
and phases.  The bispectrum $\tilde{B}_{ijk}= \tilde{\mathcal V}_{ij}
\tilde{\mathcal V}_{jk}\tilde{\mathcal V}_{ki}$ is formed through
triple products of the complex visibilities around a closed triangle,
where $ijk$ specifies the three aperture locations on the pupil of the
telescope.  One can see the bispectrum is a complex quantity, and that
the bispectrum phase is identical to the closure phase.
Interestingly, the use of the bispectrum for reconstructing
diffraction-limited images was developed independently
(\citealt{weigelt77,hw93}) of the closure phase techniques, and the
connection between the approaches realized only later
\citep{roddier86,cornwell1987}.

\subsection{Other Important Considerations}

\subsubsection{Coherency}
The tolerance for matching pathlengths in an interferometer
depend on the desired spectral bandwidth.  In the limit of monochromatic
light, such as for a laser, interference will occur even when pathlengths of
an interferometer are highly mismatched.  For broadband (``white'') light,
the number of fringes in an interferogram is equal to the inverse of
the fractional bandwidth: $N_{\rm fringes} \sim \frac{\lambda}{\Delta \lambda}$.
Hence, for broad band observations ($\sim$20\% bandwidth) interference is only
efficient if the pathlengths are matched to within a wavelength or so -- 
a stringent requirement.

\subsubsection{Field-of-View}
One consequence of the short coherency envelope for broadband
observations is a limitation on the field-of-view. Bandwidth-smearing,
as it is called, limits the field-of-view to be equal to the
fringe-spacing $\times$ the number of fringes in the coherency
envelope (see last subsection), FOV $ \sim \frac{\lambda}{\rm
  Baseline} \times \frac{\lambda}{\Delta \lambda}$ radians.  This
``field-of-view'' is thus baseline-dependent, leading to confusing
interpretations of data for extended sources. While this effect can be
modeled, it should be avoided by using a spectrometer to limit the
bandwidth of individual observing channels.

Another common limitation of the field-of-view is the primary beam of
an individual telescope.  For most kinds of beam-combiners, flux
outside the diffraction-limited beam is rejected ({\em spatial
  filtering} is described more fully in \S\ref{spatialfiltering}).
For most astronomical objects observed by interferometers, this is not
a serious problem since long integrations by (low-resolution)
individual telescopes can be used to confirm that no significant flux
arises from outside the primary beam.  For an imaging interferometer,
one would like to use narrow enough bandwidths so that
bandwidth-smearing (on the longest baselines) is small enough so that
the entire primary beam can be mapped. The requirement for this is
approximately: $\frac{\lambda}{\Delta \lambda} \sim \frac{\rm
  Longest~Baseline}{\rm Telescope~Diameter}$.

Of course, having a wide field-of-view would be useful for many
studies, such as measuring proper motions of stars at the galactic
center.  As discussed later, a wide field-of-view (beyond the primary
beam) can only be achieved in a so-called Fizeau combiner.  The Large
Binocular Telescope Interferometer is the only interferometer
currently being built which will have this unique and potentially very
powerful faculty.

\subsubsection{Filling the (u,v) Plane}
\label{uvplane}

The ability to make an image depends most strongly on the 
filled fraction of the  (u,v) plane.  Recall
that the visibility amplitude and phase measured by an interferometer
is directly related to a {\em single component} of the Fourier Transform of the
object brightness distribution.  If the object brightness is 
specified on coordinates of Right Ascension (pointing East) and
Declination (pointing North), then the reciprocal Fourier space
has axes referred to as (u,v).  Following astronomical notation, the positive u-axis
typically points to the left on a diagram, just as right-ascension coordinates 
increase towards the left (East).  

\begin{figure}
\begin{center}
\epsfig{file=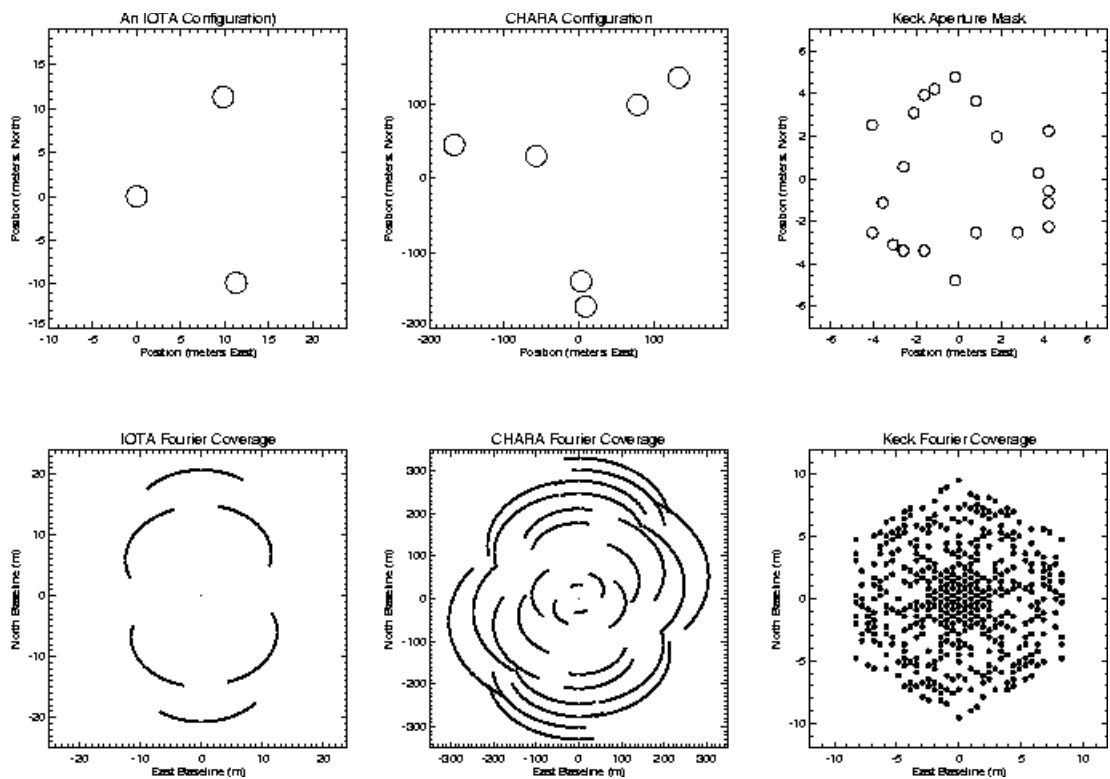,angle=90, height=4in,clip}
\caption{Example of (u,v) plane coverage for different
interferometers.  The top panels 
show the interferometer array configurations, while the bottom panels
show the corresponding (u,v) plane coverage.   For the IOTA and CHARA
interferometers, I have assumed a source at 
45$\arcdegg$ declination observed for three hours both 
before and after transit. The right-most panels show instantaneous
``snapshot'' coverage for an optimized 21-telescope array, a geometry
actually used in the Keck aperture masking experiment 
\citep{tuthill2000}.
Note that the circles in the top plot are not to the same scale as the
individual telescopes diameters but have been enlarged.
\label{figuvplane}}
\end{center}
\end{figure}

For a fixed geometry of telescope locations, the Fourier coverage
varies as the star rises and sets, and as a function of the star's
declination and the interferometer's latitude.
Figure~\ref{figuvplane} shows the Fourier coverage of three actual
interferometers (number of telescopes 3, 6, and 21 for IOTA, CHARA,
and Keck aperture masking respectively) for a declination 45$\arcdegg$
object spanning 3 hours before and after transit (assuming
monochromatic light).  In aperture masking, the pupil plane of a
single telescope is split up into sub-pupils which are allowed to
combine, just like a long-baseline interferometer.  It is obvious that
that the coverage increases rapidly with number of telescopes.  It is
not so obvious that the number of closure phases/triangles also
rapidly increases with array size (equivalent to filling up the
hyper-volume (u$_1$,v$_1$,u$_2$,v$_2$) with closure triangles; see
Table~\ref{monnier_table1}). \citet{tuthillchara2000} studied how
imaging fidelity and dynamic range are affected by differing amounts
of Fourier coverage using real data.  Obviously for imaging it is
absolutely critical to collect as much coverage as possible, and
suitable array design is further discussed in \S\ref{arraydesign}.

\section{Basic Designs of Stellar Interferometers}
\label{basicdesigns}
\subsection{Brief Historical Overview}
\label{history}

Here I give only a brief historical overview of progress in stellar
interferometry drawn partially from the review by \citet{lawson2000}; I refer
the interested reader to the above article for more  information.

Modern interferometry can be traced back to 19th century France.
Hippolyte Fizeau first outlined in 1868 the basic concept of stellar 
interferometry, how interference of light could be used to
measure the sizes of stars.  The first attempts to apply this technique,
akin to modern-day ``aperture masking,'' were carried out
by E. St{\' e}phan soon thereafter, although the telescopes of the time had
insufficient resolution to resolve even the largest stars.

\begin{figure}
\begin{center}
\epsfig{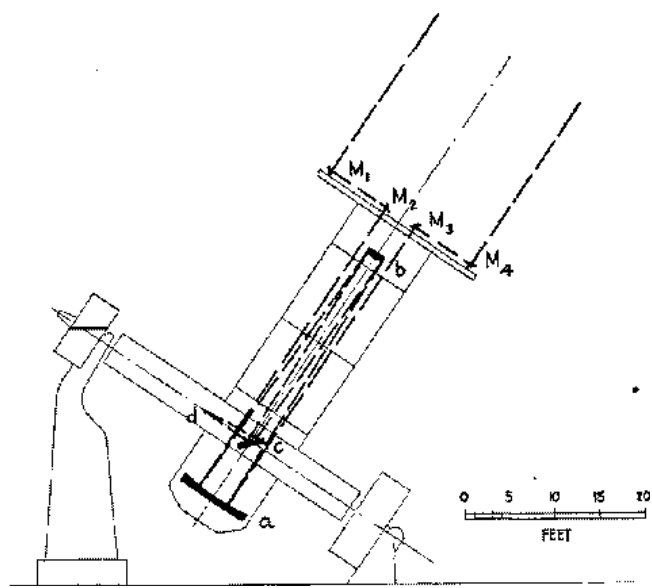}
\caption{This diagram from \citet[][Figure 1]{michelson1921}
illustrates how a 20-foot interferometer beam was installed on
the Mt. Wilson 100'' telescope in order to create, for the
first time, an
interferometer capable of measuring the
diameter of stars beyond the Sun.  Figure reproduced by permission of the AAS.
\label{monnier_michelson100}}
\end{center}
\end{figure}

Albert Michelson developed a more complete mathematical framework for
stellar interferometry in 1890; while apparently Michelson was unaware
of Fizeau's earlier work, more historical investigation is needed to
establish this definitively.  Along with Pease, Michelson
\citep{michelson1921} eventually succeeded in measuring the diameter
of $\alpha$~Orionis (Betelgeuse) in 1920-21 using the Mt. Wilson 100''
telescope \citep[following earlier measurements of Jupiter's
moons;][]{michelson90,michelson91}.  Interestingly, Michelson needed a
baseline longer than 100'' in order to resolve Betelgeuse (uniform
disk diameter $\sim$47~mas), and acquired one by installing a 20-foot
interferometer beam on the Cassegrain cage as illustrated in
Figure~\ref{monnier_michelson100}, reproduced here from their original
paper.  Following the success of the 20-foot interferometer, Pease
(with Hale) constructed a 50-foot interferometer (on Mt. Wilson, but
separate from the 100'' telescope); although some results were
reported, this experiment was not very successful.  Due to its
generally outstanding atmospheric conditions, Mt. Wilson continued to
be a choice site for interferometry projects, subsequently hosting the Mark III,
ISI, and CHARA interferometers.

Following the disappointing results from the 50-foot interferometer,
it would be decades before significant developments inspired new
activity in the optical arena.  Meanwhile, advances in radar during
World War II spurred rapid development of radio interferometry.  We
refer the reader to \citet{tms2001} for a discussion of the
development of radio interferometry beginning with the first radio
interferometer built by Ryle and Vonberg in 1946.

The unexpected success of ``intensity interferometry'' would inspire a host of
new projects.  The basic principle behind the intensity interferometer
was laid out in \citet{hbt1956}, and describes how correlations of
{\em intensities} (not electric fields) can be used to measure stellar
diameters.  First results were reported soon thereafter
\citep{hbt1956b}, leading to the development of the Narrabri Intensity
Interferometer.  With a 188~m longest baseline and
blue-sensitivity, this project had a profound and lasting impact on
the field of optical interferometry, measuring dozens of hot-star
diameters
\citep[e.g.,][]{nii1967a,nii1967b,nii1970a,nii1970b,nii1974}.  The
small bandwidths attainable with Intensity Interferometry limited
the technique to the brightest stars, and pushed the development of
so-called ``direct detection'' schemes, where the light is combined
before detection to allow large observing bandwidths. This group would
go on to develop the SUSI interferometer.

Dr. Charles Townes, inventor of the maser, began a novel
interferometer project during this same time period at University of
California at Berkeley.  He used heterodyne receivers as in radio
interferometry, but the local oscillators were CO$_2$ lasers operating
at frequencies of $\sim$27~THz (or $\sim$10~$\mu$m wavelength), orders
of magnitude higher than radio or microwave oscillators.  First
experiments were performed using the twin McMath auxiliary telescopes
(separation 5.5~m) at Kitt Peak, AZ; first fringes were obtained on
the limb of Mercury in 1974 \citep{jbt1974} and on stars in 1976
\citep{sutton1977,sutton1978,sutton1979,sutton1982} where cool dust
shells were detected around many late-type stars.  heterodyne
detection also suffers from bandwidth limitations (like Intensity
Interferometry) as well as additional noise contribution from laser
shot-noise, which becomes progressively worse at higher frequencies.
The Townes group went on to develop the ISI interferometer on Mt.
Wilson.

Other mid-infrared efforts also are worthy of note.  An independent
project at Arizona using a kind of aperture masking on a single large
telescope (using direct detection) took place almost simultaneously with
the Townes' experiments \citep{mccarthy1975,mccarthy1977,mccarthy1978}.
In France, Jean Gay and collaborators pursued long-baseline
interferometry, both through heterodyne detection
\citep[e.g.,][]{gay1973, assus1979} and later direct detection
efforts \citep[e.g.,][]{rabbia1990}.

Long-baseline interferometry on a star by directly combining the
electric fields before photon detection (``direct detection'') was
first accomplished in 1974 by \citet{labeyrie1975}, using a 12-m
baseline.  This continued the long history of interferometry
innovation in France (starting from Fizeau), and many important
experiments have followed.  I note that ``Speckle Interferometry''
was first described by \citet{labeyrie1970} and these ideas were also
very influential to the field.  However, I will largely limit this
review to separate-element, or long-baseline, interferometry, and will
omit comments on speckle.  Following this 1974 demonstration in Nice,
the project moved to the Plateau de Calern site and become known as
the Interf{\' e}rom{\` e}tre {\` a} 2 T{\' e}lescopes (I2T).  The I2T
made measurements in the visible \citep[e.g.,][]{blazit1977} and in
the near-infrared \citep{benedetto1983,benedetto1985}.  The Grand I2T
\citep[GI2T,][]{gi2t1994} began soon thereafter and was developed in
parallel with the I2T on the same plateau, but with larger telescopes
(1.5~m) and longer maximum baselines (up to 65~m).

At a time when it was a struggle to simply get two telescope
interferometers working, considering the {\em array of telescopes}
needed for imaging was indeed far-fetched. Thus, imaging using optical
interferometry began with aperture masking experiments on large
single-aperture telescopes (in the tradition of Michelson).  In
aperture masking, a pattern of holes (size $\simle$r$_0$ in diameter)
is cut in a plate and placed in the pupil plane of a large telescope.
The interference pattern formed thus simulates one from an array of
telescopes combined like a Young's multi-slit experiment.
\citet{baldwin1986} and \citet{haniff1987} showed how aperture masking
in the visible yield data identical to that expected for an imaging
array, and produced images of binary stars using closure-phase
imaging.  This group, based at the University of Cambridge, England,
would soon begin developing the COAST interferometer, which would
succeed in producing the first image with an aperture synthesis
optical array \citep{baldwin1996}.  Infrared aperture masking at the
Keck Telescope \citep{tuthill2000} also grew out of work from this group 
in collaboration with the U.C. Berkeley ISI team, and excited
unexpected imaging results from this work have led to much enthusiasm
for developing infrared imaging capabilities for long-baseline
interferometers such as CHARA and VLTI.

There is one remaining important interferometer lineage to mention,
one which led to the modern development of ``fringe-tracking''
interferometers such as the NPOI, PTI, and Keck Interferometers, as
well as numerous experimental innovations.  The Massachusetts
Institute of Technology and the Naval Research Laboratory built and
operated a series of prototype interferometers, named the Mark I, Mark
II, and the Mark III.  \citet{shao1980} reported the first successful
active fringe-tracking results, and this group has been most active at
pushing the use of interferometers for precision astrometry.  The
Mark~III was located on Mt. Wilson and was a fully automated
interferometer operating in the visible with baselines up to 20~m
\citep{markiii}.  The high efficiency allowed many astronomical
programs to be carried out until it was shut down in about 1993,
and is widely considered one of the most productive interferometers to date.
Numerous articles were published covering areas of astrometry, angular
diameters, precision binary orbits, and limb-darkening
\citep[e.g.,][]{mozurk1988,hutter1989,mozurk1991,armstrong1992,
hummel1995,quirrenbach1996}.

I should also mention the prototype interferometer IRMA (Infra-Red
Michelson Array), built at University of Wyoming \citep{dyck1993}.
While this instrument did not operate for very long, those involved
were largely responsible for initial success with infrared observing
at IOTA and have had lasting impacts at a number of other currently
operating U.S. facilities, including NPOI, PTI, and Keck
Interferometers.

This section was meant to introduce historical interferometers (ones
no longer in operation) which have had a lasting impact on the field,
and I have left descriptions of currently operating interferometers to
\S\ref{projects}.  As discussed in the opening, this review is not
meant to document all the important results from these first
generation facilities, but rather to give appropriate historical
background for understanding the current state-of-the-field.

\subsection{Overview of Interferometer Design}

\begin{figure}
\begin{center}
\epsfig{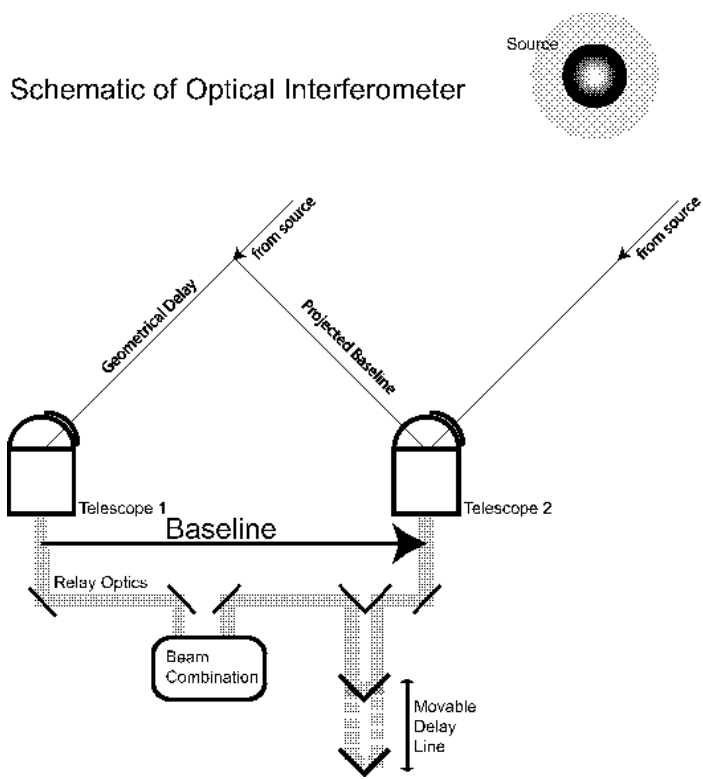}
\caption{This schematic illustrates the major subsystems of a modern
optical interferometer: the Telescopes, the Relay Optics, the Delay Lines,
and the Beam Combination.
\label{fig_realinterferometer}}
\end{center}
\end{figure}

Compare Young's two-slit experiment (Figure~\ref{fig:2slit}) to what
you see in Figure~\ref{fig_realinterferometer}.  We see telescopes
instead of slits and a beam combiner (with relay optics) instead of a
screen for viewing the fringes.  In a real interferometer, we must use
delay lines to compensate for geometrical delay introduced by sidereal
motion of a star across the sky; in this way, we ``point'' the
interferometer at the target.
In order to successfully interfere light together, each interferometer will
have many subsystems, and in this review we will describe the
state-of-the-art developments for the
Telescopes, the Relay Optics, the Delay Lines,
and the Beam Combination.

\label{arraydesign}
Before discussing each of the critical subsystems, the importance of
the physical placement of the telescopes for imaging will be
discussed.  Many of these issues are discussed in more detail by
\citet{mozurk2000}, and here we consider array design from the
perspective of imaging, not for specialized purposes such as nulling
or astrometry.

If there were no practical constraints and telescopes could be placed
optimally, one could consider many possibilities.  Studies have been
published considering distributions based on optimizing uniformity of
(u,v) coverage using three-fold symmetric patterns \citep[used in Keck
aperture masking,][]{golay71}, Reuleaux triangles \citep[used for the
Sub-Millimeter Array,][]{keto1997}, a spiral zoom array (considered
for the Atacama Large Millimeter Array, see ALMA memos \#216, 260,
283, and 291), and a Y-shaped array (adopted by the Very Large Array).
While the first three of these methods offer better Fourier coverage
than the Y-shaped array, the ``imaging'' interferometers of NPOI and
CHARA both use a Y-shaped array -- why?

\begin{figure}
\begin{center}
\includegraphics[clip,width=7in]{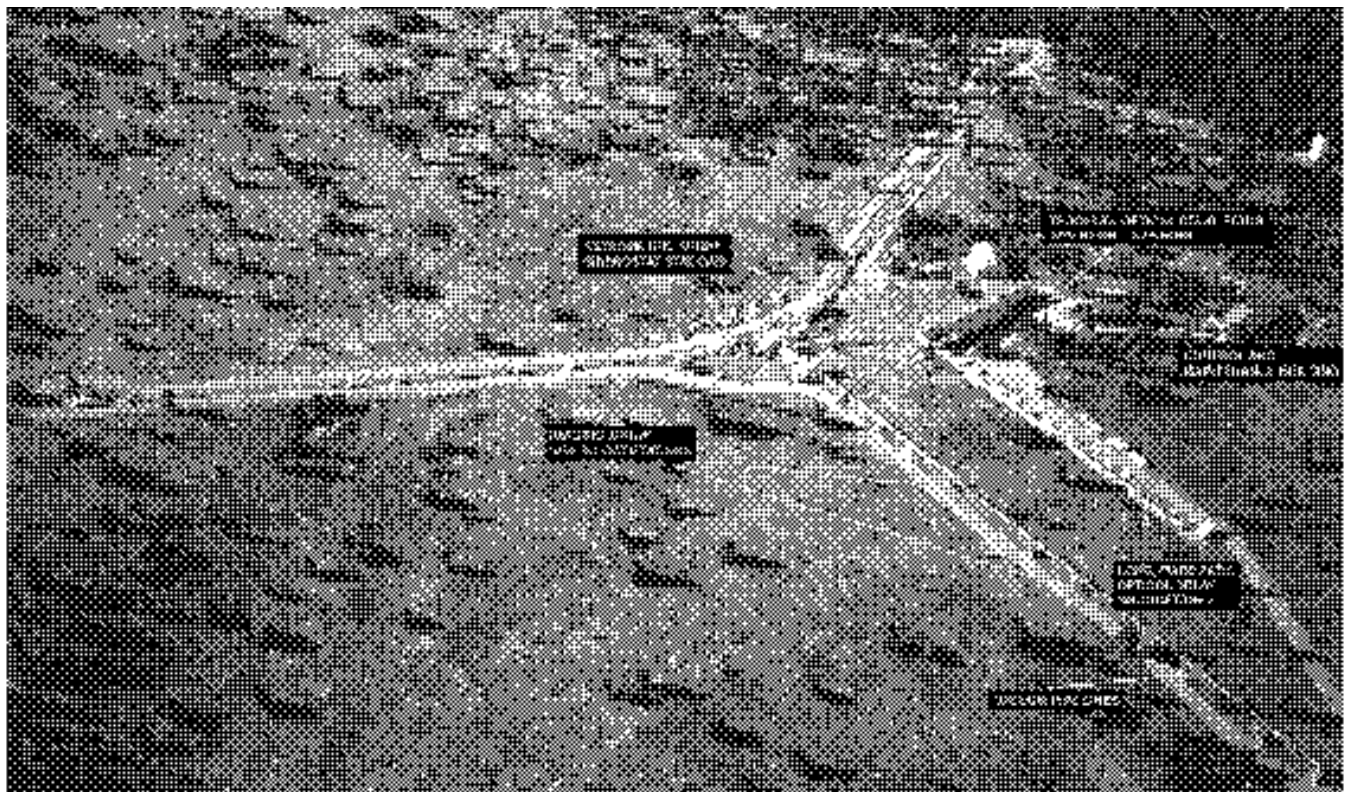}
\caption
{Overhead view of the NPOI interferometer array, to show the
Y-shaped array layout, defined by vacuum pipes extending out to many
possible siderostat ``pads,'' or stations. Photograph reproduced with
permission of the Naval Research Laboratory.
\label{npoi}}
\end{center}
\end{figure}

For the VLA, an important reason for using a a Y-shaped array was a
practical one; it was easy to move the telescopes along railroad
tracks in order to cheaply and easily reconfigure the array geometry.
While optical telescopes in arrays do not generally run on tracks
(except at IOTA), the desire to transport light to a central facility
(see \S\ref{relay}) leads one to a Y-shaped geometry where the
three-arms of the array are defined by vacuum pipes which relay beams
from the telescopes to the delay lines and combiners.  The NPOI
interferometer (shown here in Figure~\ref{npoi}) has many stations
along the three vacuum arms where telescopes can be located, thus
creating a flexible, reconfigurable system capable of pursuing many
astronomical programmes. Another theoretical benefit of this design is
that many telescopes can be arranged along each arm allowing
``baseline bootstrapping'' for imaging highly resolved targets, a
technique where strong fringes measured between close-by telescopes
are used to ``phase-up'' the fringes on the longer baselines.

\subsection{Critical Subsystems Technologies}

\subsubsection{Telescopes}
All interferometers need light collectors of some kind.  In many
cases, simple ``siderostats'' are used, whereby a steerable flat
mirror directs starlight either directly to the interferometer or
first through a beam-compressor (``afocal'' telescope).  A siderostat
has limited sky coverage and makes polarization measurement difficult
(due to the changing, non-normal reflection angles off the flat), but
is thought to offer a more stable structure for minimizing vibrations
and pivot-point drifts for accurate astrometry.  More recent
interferometers, such as CHARA, VLTI, and Keck, have chosen
traditional altitude-azimuth (``alt-az'') telescope designs which give
full-sky coverage and potentially salvaging polarization work.  In
addition, all interferometer telescopes have incorporated fast
``tip-tilt'' guiding which tracks (and corrects) fast jitter of
the stellar image, usually using visible-light ``quad-cell''
detectors.  This corrects the first-order term of the wavefront
perturbations, aligning the wavefronts to allow for stable beam
combination.

Without high-order adaptive optics, there is little use for a
telescope aperture much larger than the atmospheric coherence length
$r_0$ (see \S\ref{section:seeing}). Hence, most telescopes in today's
interferometers are small by ``modern'' (8-m class) telescope
standards.  Dedicated visible-light interferometers (e.g., NPOI, SUSI)
have telescope apertures around 12-14\,cm in diameter; near-infrared
interferometers (e.g., PTI, IOTA, COAST) have apertures diameters
around 45~cm.  The recently-built CHARA interferometer includes 1~m
apertures which can take advantage of excellent seeing conditions in
the infrared, and could benefit from adaptive optics correction; the
Keck and VLT auxiliary telescopes were specified to be 1.8~m for
similar reasons.  However, interferometry is not just for ``small''
telescopes anymore, since the world's largest telescopes, the two Keck
telescopes and also the four VLT telescopes, are now part of the new
generation of optical interferometers.  As of 2002, only the Keck
Interferometer has observed using adaptive optics, although the VLT
Interferometer will soon possess this capability.  See
Table~\ref{table:capabilities} for a summary of telescope apertures of
today's interferometers.

\subsubsection{Relay Optics, Delay Lines, \& Metrology}
\label{relay}

After being collected by the telescopes, the light must be directed to
a central facility for beam combination.  While it may seem trivial to
set up a series of mirrors for this purpose, there are many subtle
issues that must be addressed.  \citet{traub1988} discussed how the
geometry of the relay optics must not corrupt the relative
polarization of the beams, due to differential phase shifts between
the s- and p-wave reflections from the mirror surfaces for non-normal
incidence.  One must pay attention to the issues of mirror and
window coatings as well as geometry.

In addition, due to the long path lengths between the telescope and
central beam combining facility, significant differential chromatic 
dispersion
occurs if the light is propagating in air.  In order to combine broad
bandwidths, one must either transport the light through a vacuum or
construct a dispersion compensator \citep{tango1990,theo1995}, whereby
wedges of glass are inserted into the beam to compensate for air's
index of refraction; a combination of partial vacuum plus dispersion
compensation is also possible.  The size of the mirrors in this optics
chain is also important for limiting the effect of diffraction
\citep{horton2001}, and often also sets the field-of-view of the
interferometer.  Lastly, because of the many reflections, high
reflectivity of the relay optics must be maintained to maximize
throughput and sensitivity; a side-benefit of evacuated relay optics
is that the mirrors stay clean.

Because of the Earth's rotation, the apparent position of an
astronomical object is constantly changing.  In order to track this
sidereal motion, a movable delay line is needed to compensate for
changing geometrical delay between wavefronts reaching any two
telescopes.  The diagram in Figure~\ref{fig_realinterferometer} shows this delay line as a right-angle
retroreflector, although most interferometers do not actually use this
geometry.  The requirements on this system are amazingly stringent:
nanometer-level precision moving at high speeds ($>$ 1 cm/s) and over
long distances ($>$100~m) -- a dynamic range of $>10^{10}$!

\begin{figure}
\begin{center}
\epsfig{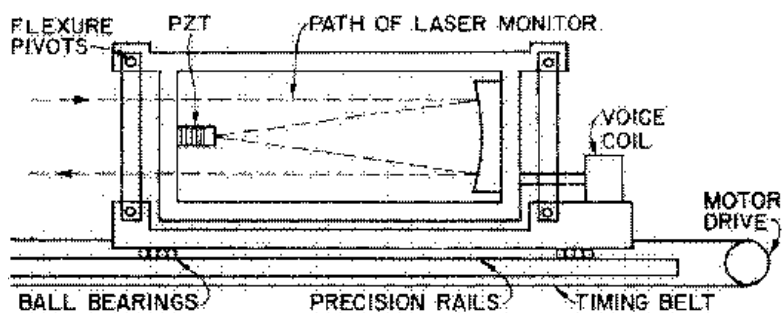}
\caption
{Diagram of the most standard delay line architecture used
in optical interferometry, originally from the Mark~III interferometer.
Figure reproduced from \citet[][Figure 4]{markiii} with permission of 
ESO.
\label{jpl_dl}}
\end{center}
\end{figure}

By far the most popular architecture today for the moving delay line
is based on the solution implemented by the Mark~III interferometer
\citep{markiii,colavita1992}.  The retroreflection is produced by
focusing the incoming beam to a point coincident with a small flat
mirror (attached to a piezo-electric stack), which reflects and 
re-collimated; a practical optical system is illustrated in
Figure~\ref{jpl_dl}.  This mirror system is mounted on a flexible
stage which can be translated using a voice coil.  Lastly, this whole
stage is mounted on a wheeled-cart, which is driven on a rail by
linear motors.  This system has three nested feedback loops, driven by
laser metrology: precise sub-wavelength control is maintained by the
piezo-driven small mirror, when this mirror exceeds its normal
operating range ($\sim$50$\mu$m) then offsets are given to the
voice-coil stage, and so on.  This basic architecture is in use at
most interferometers built in the last 10~years; alternate delay lines
include floating dihedrals mirrors on an air table (IOTA) and moving
the beam combination table itself (GI2T).

\subsubsection{Beam Combination \& Fringe Modulation}
\label{combiners}
Once the beams have been delivered to a central combination facility
and have been properly delayed, there are many ways to actually do the
interference.  Here I discuss image-plane and pupil-plane combination
(also know as ``Fizeau'' and ``Michelson'' combination, respectively),
along with spatial versus temporal modulation of the fringes themselves.

\begin{figure}
\begin{center}
\epsfig{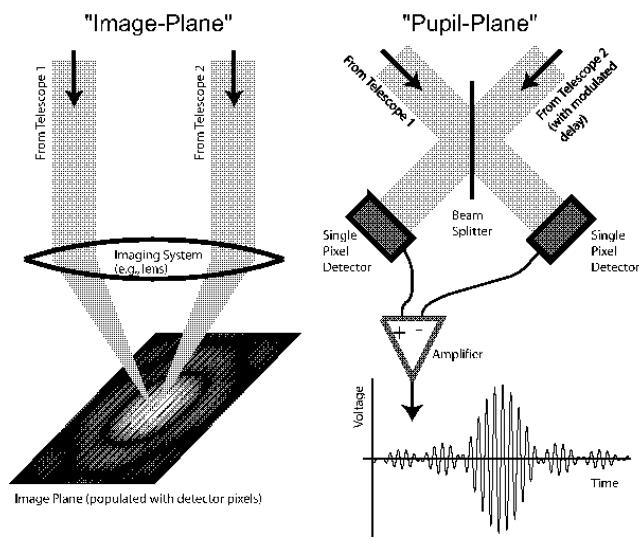}
\caption
{Diagram of Image-plane and Pupil-plane beam combination techniques.
  The left panel shows image-plane, or Fizeau combination, where light
  from the two telescopes are brought together in an image plane to
  interfere, just like Young's two-slit experiment.  The right panel
  shows how pupil-plane (or ``Michelson'') interferometry superimposes
  the two collimated beams at a beam-splitter.  By modulating the time
  delay of one beam with respect to the other (e.g., with the delay
  line), the interference can be modulated and fringes detected using
  single-pixel detectors. 
\label{monnier_combiner}}
\end{center}
\end{figure}

Figure~\ref{monnier_combiner} show these two different ways of
detecting fringes in a two-element interferometer.  In one case, an
imaging system is used to fill the image plane with the equivalent of
Young's fringes.  As one moves along the image plane, there is a
different relative delay between the interfering beams, and hence the
modulation (fringes).  In this scheme, there is no need to actively
modulate the fringes; in fact, atmospheric turbulence will introduce
relative delays and cause the fringe pattern to ``slide'' back and
forth, smearing out the fringes on short time-scales if not
stabilized.  This combination scheme most closely follows the
``two-slit'' interferometer analogy developed in earlier sections.

The second scheme, and currently the most common one, is pupil-plane
combination, or ``Michelson''-style combination.  Interestingly, this
method is named after Michelson, not because of his stellar
interferometry work (which used ``Fizeau'' combination, see
\S\ref{history}), but because of the interferometer used in the
Michelson-Morley experiment.  In this method, the wavefronts from the
two collimated telescope beams are overlapped on a 50/50 beamsplitter.
Depending on the phase relationship of the waves, differing amounts of
energy will be transmitted or reflected at the beamsplitter.
Single-pixel detectors can then be used to measure the energy on both
sides of the beamsplitter (the sum of which is conserved).  The
popular adoption of this method results largely from the
signal-to-noise benefits of using single pixel detectors.  In order to
measure the amplitude of the coherence, a dither mirror (often in the
delay line) sweeps through a linear pathlength difference of many
wavelengths. The white-light fringe, or interferogram, can then be
recorded, as long as the scanning takes place faster than an
atmospheric coherence time.

When dealing with an array of telescopes, there are more options.
\citet{brummelaar1993} outlined some forward-looking beam combiner
designs in the context of the CHARA array and useful articles
by \citet{mozurk2000} and \citet{mariotti1992} also contain extended
discussion on the subject; here I only mention the highlights.  The
image-plane method can be extended to arbitrary number of telescopes,
as long as the spacings between the beams are {\em non-redundant}, so
that each beam-pair will have a unique fringe spatial frequency in the
image-plane.  \citet{labeyrie1996} elaborates on the concept of
{\em pupil densification,} an idea finding increasing application in
modern interferometry.  The pupil-plane method can also be extended,
either by combining the beams ``pair-wise'' or ``all-in-one.''  In a
pairwise-scheme, each telescope beam is split using beamsplitters and
then various combinations are created to measure all the baselines.
In the all-in-one scheme, more than two beams are superimposed and the
fringes from different pairs are distinguished by modulating the
delays such that each baseline pair has a unique fringe temporal
frequency in the readout.  There are methods which combine pair-wise
with all-in-one and are called ``partial-pairwise.''  Each method has
its advantages and disadvantages, depending on the availability of
focal plane arrays, the level of readnoise vs. photon noise, the
required calibration precision, etc.  However, in general,
``pair-wise'' detection is the worst method for large number of
telescope because the light has to be split more times
\citep[see][although beware of some important simplifications made in
this analysis]{buscher_thesis}.

Coherent beam combination can be discussed more generally depending 
whether the interference occurs in the image/pupil plane and whether the
telescope beams are co-axial or multi-axial.  I refer the reader to
the influential internal ESO report by \citet{mariotti1992}, which
explains and defines the useful vocabulary in common use by the European
interferometry community.

We contrast the many imperfect beam combination strategies in the
optical with those adopted in radio interferometry.  At radio and
microwave frequencies, the signals from each telescope can be split
and re-amplified without introducing additional noise after the
initial coherent detection (radio interferometers do not operate
close to the Poisson limit).  Hence, a pair-wise combination
scheme can be employed without any loss in signal-to-noise ratio.  In
addition, the electric field at each telescope can be truly
cross-correlated with that from all others leading to a kind of Fourier
Transform spectroscopy.  Further, this can all be done using digital
electronics after fast digitization of the signals.  For more
information, see the description of the Hat Creek mm-wave
correlator by \citet{hudson85}.  At the end of this process, the
digital correlators can recover all baselines with arbitrary spectral
resolution without lost sensitivity -- a dramatically superior
situation than possible in the photon-starved visible and infrared
regime!

\subsubsection{Fringe Tracking}
An increasingly popular and powerful capability for optical
interferometry is called ``Fringe Tracking.''  To do so, the white-light fringe
has to be actively tracked because atmospheric fluctuations cause the
location of the fringe to vary by up to hundreds of microns on
sub-second time scales.  There are two levels of tracking these
fringes, one is called ``coherencing'' and the other is called
``fringe tracking,'' although these terms are often used rather
loosely.  

In ``coherencing,'' the interferometer control system will track the
interferogram location to a precision of a few wavelengths. In a
scanning interferometer, this will be sufficient to keep the full
interferogram within the scanning range of the delay line.  In an
image-plane combiner, this will ensure you are near the peak of the
white-light fringe (inside the coherence envelope set by the spectral
bandpass).  This can be done on a rather leisurely timescale, since
large optical path distance (OPD) fluctuations tend to occur of slower
time scale: update rates of $\sim$1~Hz are sufficient except for the
worst seeing conditions.

True ``Fringe Tracking'' (also called ``co-phasing'') requires
tracking OPD fluctuations within a small fraction of wavelength in
real-time, and hence requires orders of magnitude faster response (a
time scale which depends on the wavelength and seeing conditions).  In
the most common implementation \citep[the ``ABCD'' method; see][]{shao1977},
two beams are
combined pairwise while a mirror is stepped at quarter-wavelength
intervals.  The broadband white-light fringe is detected at one of the
beamsplitter outputs, and fringe data is recorded synchronous with the
dither mirror, resulting in four measurements (A,B,C,D) representing
four different fringe phases.  A discrete Fourier Transform
(effectively) can be rapidly applied to the data, resulting in a
fringe phase estimate. This offset can be sent to the interferometer
delay line control system to nearly instantaneously correct for
atmospheric turbulence \citep[details in][]{markiii,colavita1999}.
The light from the other beamsplitter output is usually dispersed and
multi-wavelength data is collected.  I also refer the reader to
\citet{lawsonchap2000}, where the ABCD method (and other ``phase''
estimators) or compared to to ``Group Delay'' tracking methods, which
use phase measurements at different wavelengths to measure
interferometer delay offsets.

Historically speaking, active fringe tracking has been important only
for the Mark~III interferometers and its successors (NPOI, PTI, 
Keck Interferometer). One reason fringe tracking has not been more widely
pursued is because the sensitivity limit of a fringe
tracking interferometer is less than fringe-envelope scanning interferometer.
This is because very short integration times are required to stay on the
fringe and hence the source must be fairly bright; in the fringe envelope
scanning method, one has to only keep the interferogram in the scanning range and
thus any given fringe measurement can have a lower signal-to-noise ratio.
In practice, this amounts to sensitivity difference of a 
few magnitudes.

As interferometers become more powerful and seek greater capabilities,
fringe tracking is becoming a standard feature.  High spectral
resolution interferometry data is possible with fringe tracking
systems, because a broadband white-light fringe can be used for fringe
tracking while the remaining output channels can be dispersed.
Normally, this data would have very low signal-to-noise ratio, but if
the fringe tracking essentially ``freezes'' the turbulence, the
dispersed fringes can be detected by integrating on the detector much
longer than the typical atmospheric coherence time.  Hence, fringe
tracking is a kind of ``adaptive optics'' for interferometry.

\subsubsection{Detectors}
\label{detectors}
The most desired properties for detectors used in optical interferometry
are low noise and high readout speed, two qualities usually not found
at the same time.  At the beginning of optical interferometry, the
only visible-light detector was photographic film and infrared detectors
were only just invented.  Detectors have made incredible advances over
the last few decades, and are operating near their fundamental limit
in most wavelength regimes (the near-infrared is a notable exception).

After years of struggling with custom-built photon-counting cameras
for visible-light interferometry work, such as the PAPA camera
\citep{papa1985,lawson1994} and intensified CCDs
\citep[e.g.,][]{blazit1987,foy1988}, commercial devices are being sold
aimed at the adaptive optics market which have high quantum
efficiencies ($>$50\%), kilohertz frame times, and read noise of only
a few electrons (fast readout CCDs).  For beam combinations schemes
where single pixel detectors are suitable, Avalanche Photo-Diodes
(APDs) have as high quantum efficiency as CCDs but can photon-count at
rates up to 10~MHz, although the best commercial devices seem to have
an expensive tendency to stop working.  This covers wavelengths from
the blue to approximately the silicon cutoff ($\sim$1~$\mu$m).

In the near-infrared (1-5$\mu$m), there has been amazing progress this
decade.  After early work with single-element detectors (e.g, using
material InSb), modern interferometers have taken advantage of
technology development at Rockwell in near-IR focal plane arrays made
of HgCdTe, such as the NICMOS3, PICNIC, and HAWAII chips.  These
arrays have high quantum efficiency ($>$70\%) and can be clocked at
$\sim$MHz pixel rates with as low as 15~e- readnoise.  While not
optimal, this represents orders-of-magnitude improvement over
photodiodes and has allowed new kinds of astronomical sources to be
observed (most notably, young stellar objects).  The noise can be
further reduced by reading each pixel many times, a novel mode known as
``non-destructive'' readout.  Hence, by reading a pixel $n$ times
before resetting, one can reduce the effective readnoise by
approximately $\sqrt{n}$, for n$\simle$20.  Interferometry benefits
greatly from this capability, since only a few pixels need be readout,
allowing large number of ``reads'' to be made in a short period of
time \citep[interferometry reference][]{rmg1999b}.

Traditionally, the HgCdTe detectors had a cutoff wavelength of
2.5$\mu$m, but recent Molecular Beam Epitaxy (MBE) processes allow
this cutoff to be tuned to much longer (or shorter) wavelengths
(allegedly even beyond 5$\mu$m).  For a 2.5$\mu$m cutoff, these
detectors must be operated at liquid nitrogen temperatures (77~K) in
order not to be saturated with dark current from thermally-generated
electrons.  An important recent development is that Raytheon has begun
to compete with Rockwell in this market, and we can hope for even
greater advances in HgCdTe arrays in the coming years as well as
possibly even price reductions.

Other materials, such as InSb, can be used for even longer wavelength
performance.  At 5$\mu$m and longer wavelengths, thermal background
levels are sufficiently high that these detectors must be readout very
rapidly, and can usually work in background-limited mode (despite
$>$500~e- readnoise).  This means that Poisson fluctuations in the
thermal background flux dominate over other sources of noise (e.g.,
read noise); the only way to reduce the effect of this background
noise is to increase the quantum efficiency of the detector or to
reduce the thermal background load on the detector.  Various companies
have sold focal plane arrays in the ``mid-infrared''
($\sim$8-25$\mu$m) over the years, and are not all independent efforts
after a complicated series of company sales (e.g., Hughes, Santa
Barbara Research Center, Raytheon, Boeing).  Recently Raytheon
has been offering Si:As Impurity Band Conduction (IBC) 320x280 focal
plane arrays, which also operate at the background-limit.  These
detectors must be cooled below 77~K to avoid high dark currents, and
generally use liquid Helium.  Uniquely, the ISI interferometer uses a
single-element HgCdTe photodiode with high signal level (using CO$_2$
laser local oscillator) which have up to 25\% quantum efficiency and a
5~GHz output bandwidth.

\subsubsection{System Control}

It is not trivial to control all the important subsystems of an
interferometer.  Many current interferometers (e.g., ISI, IOTA, PTI,
Keck, VLTI) use the VME realtime architecture under the vxworks
operating system (Wind Rivers).  This allows different subsystems to
be easily synchronized at the millisecond (or better) level.  VME systems are
fairly expensive, and some groups (in particular, CHARA) have adopted
the RT (RealTime) Linux OS running on networked personal computers.

\subsection{Sensitivity}
\label{sensitivity}

Optical interferometers are orders-of-magnitude less sensitive than
single-dish telescopes.  At visible wavelengths where sensitivity is
the worst, current interferometers have a similar limiting magnitude
as the human eye (e.g., V mag $\sim$6 at NPOI). In this section, we explore
the current and future sensitivities of optical interferometers.

\subsubsection{What sets the limiting magnitude?}
There are three major problems which limit the sensitivity of today's
interferometers: the atmosphere, optical transmission,
detector/background noise.

The sensitivity is most dramatically limited by the atmosphere which
restrict the coherent aperture size and coherent integration time.  We
can use the notion of a {\em coherent volume} of photons which can be
used for interferometry, with dimensions set entirely by the
atmosphere.  The coherent colume has dimensions of $r_0 \times r_0
\times c \tau_0$, and hence is very sensitive to the seeing.  Consider
average seeing conditions in the visible ($r_0 \sim$ 10\,cm, $t_0
\sim$ 10\,ms), we can estimate a limiting magnitude by requiring at
least 10 photons to be in this coherent volume.  Assuming a bandwidth
of 100~nm, 10 photons ($\lambda\sim$ 550\,nm) in the above coherent
volume corresponds to a V magnitude of 12.6, which is more than
10~magnitudes brighter than faint sources observed by today's 8-m
class telescope.  Because the atmospheric coherence length and time
scale approximately like $\lambda^{\frac{6}{5}}$ for Kolmogorov
turbulence, the coherent volume $\propto \lambda^\frac{18}{5}$.

Current interferometers can not achieve this
limiting magnitude because of additional problems.  The most important
in the visible is low optical throughput due to the large
number of reflections between the telescopes and the final detector.
The number of reflections easily exceeds 10 and is often closer to 20.
Even with high quality coatings of 97\% reflectivity, we see that
$\sim$50\% of the light would be lost after 20 bounces
($0.97^{20}=.54$).  In practice, current interferometers have
visible-light transmission between 1\% and 10\%, due the fact that
coatings degrade with time, the need for dichroics and filters with
relatively high losses, and some diffractive losses during beam
transport.  Of course, detectors also do not have 100\% quantum
efficiency.  The COAST interferometer has achieved the faintest
limiting magnitude in the visible of $\sim$9~mag, by optimizing
throughput, detector quantum efficiency, and bandwidth (as another
example, the NPOI interferometer which fringe-tracks and uses narrower
bandwidths has a limiting magnitude around $\sim$6).

Throughput issues can be improved multiple ways.
Lawrence Livermore Laboratory is researching new coatings for mirrors
which will have $\simge$99\% reflectivity at most near- and mid-infrared
wavelengths.  In addition, simplified beam trains with few
reflections are being designed for next generation interferometers.
Lastly, the use of fiber and integrated optics could potentially
lead to high throughput systems in the future; these developments are
discussed more fully in \S\ref{newtechnology}.

The last major limitation is noise associated with the detection.
Some visible light detectors, such as the photon-counting Avalanche
Photo-Diodes, are almost perfect in this regard, boasting very low
``dark counts'' ($<$100 ct/s) and high quantum efficiency. However,
this is not true in the infrared.  Even the best infrared detectors
have $\sim$10~e$^{-}$ noise per read. While normal (incoherent)
astronomers can afford to integrate for minutes or hours to collect
photons, interferometrists must readout pixels within the atmospheric
coherence time and thus are strongly limited by readnoise.  As one
moves further into the infrared (5-10$\mu$m), then thermal background
fluctuations dominate the noise budget.  Again, the relatively short
coherence times of atmospheric turbulence directly result in a poor
limiting magnitude compared to incoherent detection (i.e.,
photometry).  The best published near-infrared performance of a
two-element interferometer was reported by IOTA \citep{rmg1999b} using
a NICMOS3 detector: J mag (1.25$\mu$m) 6.9, H mag (1.65$\mu$m) 6.9,
and K' mag (2.2$\mu$m) 6.2, where J, H are dominated by readnoise and
K' is dominated by fluctuations of the thermal background.  Soon,
these limiting magnitudes will be eclipsed by the
adaptive-optics-corrected Keck and VLT Interferometers which should be
able to observe fainter than 10th magnitude.

There is not much experience yet with mid-infrared observations using
direct detection.  The ISI heterodyne interferometer has observed
stars as faint as $\sim$360~Jy \citep[LkH$\alpha$,101][]{tuthill2002},
corresponding to a N band mag of $\sim$-2.2, limited largely by narrow
bandwidths ($\Delta\lambda\sim0.002\mu$m). The VLTI mid-IR instrument
MIDI will be capable of broadband combination and is forecast to have
a limiting magnitude of $\sim$1~Jy (N band mag $\sim$4) using the 8\,m
VLT telescopes (assuming the thermal background fluctuations can be
well-calibrated for systematic errors).  Shortly before this article
went to press, VLTI reported first fringes with the MIDI instrument.

A number of new technologies are being explored to push down the
limiting magnitude of optical interferometers, and some of these are
described in the next section.

\subsection{New Technologies and Techniques}
\label{newtechnology}

One exciting aspect to the field of optical interferometry
is the aggressive implementation
of new technologies to extend the limits of the sensitivity and 
calibration precision.  In this section, I will discuss 
new developments which are impacting optical interferometry.

\subsubsection{Spatial Filtering and Single-mode Fibers}
\label{spatialfiltering}
The idea to use single-mode fibers in optical interferometry was
originated by \citet{froehly1982}, and work began to implement these
ideas in both France \citep[e.g.,][]{connes1987,reynaud1992}, and in
the United States \citep{shaklan1987,shaklan1989}.  Following initial
fringe detection in 1991 using the Kitt Peak McMath telescopes
\citep{foresto1992}, the FLUOR experiment as implemented on the IOTA
interferometer was a real breakthrough; the amazing improvement in
calibration precision was documented in \citet{foresto1997} and
\citet{perrin1998}.  Currently, the advantages of spatial filtering
are being implemented at virtually all interferometers, and here I
briefly explain why it is so important.

\begin{figure}
\begin{center}
\centerline{\epsfxsize=2in{\epsfbox{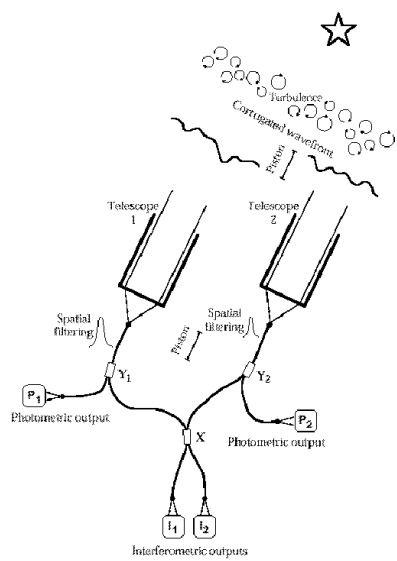}}}
\caption{
This figure shows how the FLUOR beam combiner uses 
spatial filtering and photometric monitoring to allow
precision calibration of fringe visibilities.
Figure reproduced from \citet[][Figure1]{foresto1997} with permission of
ESO.
\label{foresto97fig}}
\end{center}
\end{figure}

Figure~\ref{foresto97fig} shows a schematic of a fiber-based
interferometer, as sketched by \citet{foresto1997}.  When coupling
starlight into a single-mode fiber, the coupling efficiency depends on how
coherent the wavefront is from an individual telescope
\citep{shaklan1988}.  A single mode fiber thus essentially converts
phase errors across the telescope pupil into amplitude fluctuations in
the fiber.  Once coupled into the single-mode fiber, the light can be
partially split in order to monitor the amount of coupled light
(``photometric'' outputs), and also can be interfered with light from
another fiber using a coupler, the fiber equivalent of a
beamsplitter.  The transfer function of the fiber coupler is very
stable and not dependent on the atmosphere; only the input coupling
efficiency at each fiber is dependent on the atmosphere.  Hence, the
visibility can be measured very precisely \citep[$<$0.4\% uncertainty on V$^2$
reported by][]{perrin2003} by measuring the fringe amplitude and
calibrating with the ``photometric'' signals.

\begin{figure}[tbhp]
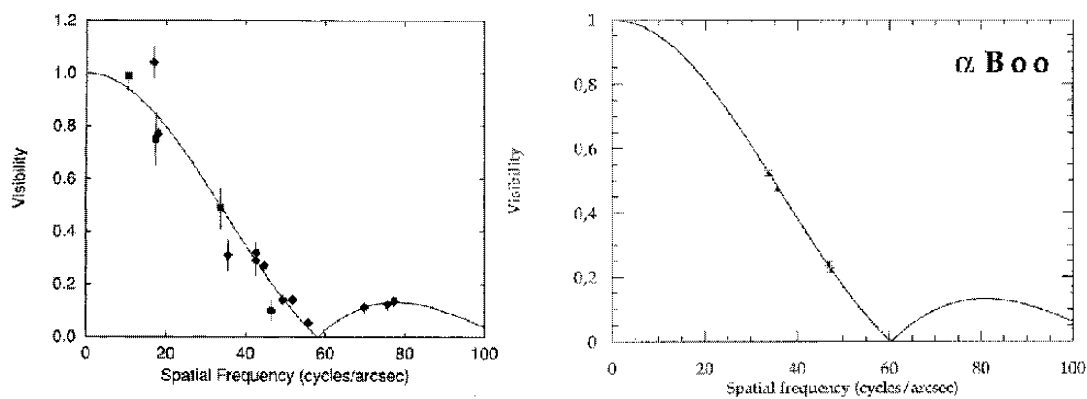

\begin{center}
\mbox{
\includegraphics[clip,angle=0,height=2.in]{Figures/JDM_irma93gif.epsi}
\includegraphics[clip,angle=0.,height=2.in]{Figures/JDM_perrin98gif.epsi}}
\caption{\footnotesize {\em (left panel)}
  a. This figure shows visibility data for $\alpha$~Boo by the CERGA
  interferometer (diamonds) and the IRMA interferometer (squares), and
  originally appeared in the Publications of the Astronomical Society
  of the Pacific \citep[Copyright 1993, Astronomical Society
  of the Pacific;][reproduced with permission of the Editors]{dyck1993}.
  b. The incredible gain in calibration using spatial filtering and
  photometric monitoring is evident in this figure reproduced from
  \citet[][Figure 2a]{perrin1998} with permission from ESO.
\label{comparecal}}
\end{center}
\end{figure}

This method strongly mitigates the dominant source of calibration
error in most optical interferometers, the changing atmosphere.  The
atmospheric turbulence must be monitored in some way when observing
with an interferometer, since the coherence between two wavefronts
from two telescopes strongly depends on seeing. However, this is not
easy to measure with a typical interferometer in realtime, and hence
one must settle for interleaving ``science'' targets with
``calibrator'' sources to calibrate seeing drifts during the night.
With fibers, the changing seeing conditions directly cause variations
in the fiber coupling efficiencies which are monitored in realtime and
corrected for.  Figure~\ref{comparecal} shows near-infrared visibility
data on the calibrator star $\alpha$~Boo using both ``conventional''
interferometry and the FLUOR fiber optics beam combiner.  The
improvement to calibration is indeed dramatic and has had
far-reaching effects on the direction of the whole field of optical
interferometry.

There are other ways to implement these calibration advantages than
the FLUOR method shown in Figure~\ref{foresto97fig}.
\citet{monnier2001} showed how the signal-to-noise can be somewhat
improved by using an asymmetric coupler instead of separate
photometric signals.  Also, \citet{keen2001} compared single-mode
fibers with spatial filtering by small pinholes in order to determine
which method is superior under different conditions. 

I should emphasize that there are many problems and limitations
associated with using single-mode fibers, most notably low coupling
efficiencies, high dispersion and poor polarization stability.  Such
problems have kept fiber optics from playing an important role for
beam transport \citep{simohamed1997}, and currently fibers are used
only for beam combining and spatial filtering at specific wavelengths.
For instance, silica-based (telecom) fibers can generally only be used
at J (1.25$\mu$m) and H (1.65$\mu$m) bands; the FLUOR experiment
utilized Fluoride glass fibers which can transmit at K band
(2.2$\mu$m) and beyond.  Advances in the field of
photonic crystals and photonic bandgap materials could lead to new
fibers with low dispersion and high transmission for new
interferometry applications, and should be aggressively pursued.

\subsubsection{Integrated Optics}
While combining two telescopes together is straightforward using fiber
optics, it becomes very difficult for multiple telescopes. This is
because the light has be split many times, combined together many
times, and the fiber lengths must be precisely matched and maintained
to correct for differential chromatic dispersion and birefringence
effects.

An elegant solution to this problem, while maintaining the advantages
of spatial filtering, is the use of {\em integrated optics}, the
photonics analog to integrated circuits.  P. Kern and an active group
centered at Grenoble Observatory have pioneered this technique
\citep[e.g.,][]{kern1997,malbet1999, berger1999} and it is finding
successful application at a number of observatories, including IOTA
\citep{berger2001} and VLTI. In these combiners, many fibers can be
mated to a small planar element with miniature waveguides etched
in place to manipulate the light (split, combine, etc).  Dozens of
beamsplitting and combinations can all be fit into a few square
centimeters -- and never needs re-aligned!

While integrated optics can solve the problem of how to combine many
beams using guided optics, it has similar difficulties as fibers of
poor transmission, limited wavelength coverage, dispersion, and
birefringence.  While the commercial applications for integrated
optics in telecommunications has driven much of the innovation in this
field, the astronomy community must actively engage with the photonics
engineers to design custom components which can overcome the remaining
problems for next-generation ``astronomical-grade'' devices.

\subsubsection{Adaptive Optics}
One critical advance to improve the sensitivity of infrared
interferometers is the application of adaptive optics on large
aperture telescopes.  Generally, visible light photons are used to
measure the wavefront distortions in realtime, allowing them to be
corrected using a deformable mirror.  Once the aperture is
``phased-up,'' the entire (much larger!) coherent volume can be used
for the infrared fringe detection. This method has already been
applied on the Keck Interferometer, where AO systems on the individual
10\,m telescopes now allow observations approaching K mag 10
(and should allow even fainter objects eventually).  The major
drawback for this is that there must be a ``bright'' visible guide
star in the isoplanatic patch for the AO system to use for wavefront
sensing, not possible for obscured sources such as Young Stellar
Objects and dusty evolved stars where the visible source is often too
faint \citep[a few AO systems do have infrared wavefront sensors to
mitigate this problem, e.g.][]{naos2002}.  The maturation of laser
guide star adaptive optics will allow this gain in coherent volume for
all infrared observing eventually.  Of course, building future
interferometers at the most excellent sites (even in space) will be an
increasing priority.

\subsubsection{Phase Referencing}
Phase Referencing is a kind of adaptive optics for interferometry,
where a bright reference star is used to measure and correct for
atmospheric time delays. This technique is used in radio interferometry
to allow long coherent integrations on targets, via a fast switching scheme.

Unfortunately, the short atmospheric coherence times make a switching
scheme difficult to implement.  A different approach pursued by the Palomar
Testbed Interferometer is to use a ``dual star module'', where light
from two stars are selected and observed simultaneously using different
delay lines. This allows both relative astrometry and phase referencing
to be achieved.

Very few results have been published on this technique so far,
although the technique is being implemented at the Keck Interferometer
and is planned for VLTI. First results from PTI have been published
\citep{lane1999,lane2000a,lane2003}, reporting extending the
atmospheric coherence time to 250~ms and visibility calibration
precision of 3-7\%.  Development of this technique will allow very
faint limiting magnitudes, for a small set of sources located within
an isoplanatic patch ($\sim$30'') of a bright star.

Another method called ``differential phase'' is being applied soon, where
fringes at one wavelength are basically used to stabilize fringes at all the 
others. When a source shows significant wavelength-dependent 
structure, this technique should prove very powerful. 
This is discussed further in the context of extrasolar planet detection in 
\S\ref{precision}.

\subsubsection{Spectroscopy}

Very little has been done in the area of interferometric observations
on spectral lines.  The best science results will be reviewed in the
next section, however here I wanted to mention recent developments.
\citet{bedding1994} discussed methods for combining aperture masking
with spectroscopy, and the design of the MAPPIT instrument offers
lessons for long-baseline interferometrists.  G. Weigelt and
collaborators have developed a spectrometer for use on two element
interferometers, which allows near-infrared molecular bandheads of CO
and H$_2$O to be spatially probed
\citep[e.g.][]{weigelt2000,hofmann2002}.  More interestingly, the
AMBER instrument for the VLTI will boast three different spectral
resolutions (up to R $\sim$10000 across the infrared), making
observations of individual lines possible
\citep[e.g.,][]{petrov2000}.  While the GI2T has had high spectral
resolution for years, a number of other visible-light interferometers,
including NPOI and COAST, have modified their combiners to allow
H$\alpha$ interferometry, following the fascinating results of the
GI2T in the 1990s \citep[e.g.,][]{vakili1998}; see \S\ref{halpha} for
more discussion on this.

\subsubsection{New Detectors}
Single-pixel visible light detectors are nearly ideal in their performance
(e.g., APDs).
However, new detectors exist with many of the same
advantages of APDs, but which can also measure the energy of each
detected photon \citep[Superconducting Tunnel Junction
detectors,][]{peacock1997}.  Although limited to maximum count rates
of $\sim$10~KHz, current STJ devices offer high quantum
efficiency, timing accuracy, and about $\sim$12\% bandwidth energy
resolution in the visible and have been used on the sky
\citep{perryman2001}.  One obstacle for this technology is that most
astronomers want large-format focal plane arrays with millions of
pixels, and present arrays are $\sim$6$\times$6 pixels.  These STJ
arrays are small, but large enough to be be quite interesting for
optical interferometry; this work should be strongly encouraged.

For some type of interferometer combiners (e.g., high-resolution
spectrometers or 6-telescope imaging), many pixels are needed;
unfortunately, APDs and STJs are not economical for this and CCDs
typically still have larger readnoise for fast frame rates.  In this
regard, a new development by Marconi may be interesting
\citep{mackay2001}. They have produced a kind of ``photon-counting''
CCD, which implements on-chip avalanche gain stages in order to
amplify single electrons into large signals.  \citet{tubbs2002} report
the first use of these new detectors in astronomy, and the results are
promising for interferometry (the COAST interferometer is currently
adapting such a device for their work).

Because of the relatively high readnoise for near-infrared detectors,
improvements in the next decade could easily extend the sensitivity of
interferometers by a factor of 10. The AOMUX detector program by
Rockwell has just begun, and has the goal of a few electron readnoise
at high frame rates.  Keeping pace with these developments will remain
a high priority for optical interferometry.

There are also some developments to create photon-counting
near-infrared detectors, equivalent to Avalanche Photo-Diodes.
Sometimes called SSPMs (Solid State Photo-Multipliers),
\citet{eikenberry1996} described one experiment. Currently, the main
drawback with these devices is the low quantum efficiency, a few
percent.  Alternatively, Superconducting Tunnel Junctions can also be
used in the near-infrared for photon-counting.

\subsubsection{Nulling}
\label{nulling}
Another interferometric technique gaining application is nulling.  By
introducing an achromatic 180$\arcdegg$ phase shift in one beam, the
white-light fringe can be turned into a white-light null
\citep{bracewell1978}.  This has obvious applications for extra-solar
planet searches and zodiacal dust disk characterizations, since
removing the bright central star is essential for detecting faint
circumstellar material and companions.  The only astronomical results
from nulling have come from aperture masking style experiments
\citep[e.g.,][]{hinz1998,hinzthesis,hinz2001},  and have encouraged
aggressive follow-up experiments.  In particular, the Keck
Interferometer is pursuing a mid-infrared nulling project
\citep{serabyn2001} and nulling is a central operational mode for the
Large Binocular Telescope Interferometer \citep{hinz2001b}.

\subsection{Current and Future Facilities}

\label{projects}
In Tables~\ref{table:all_interferometers} \& \ref{table:capabilities},
I have summarized all the current and planned facilities
(ground-based).  Further discussion of the current field, including
documentation of the rising trend of publications, can be found in
\citet{ridgway2000} where I have found some of the information for
these summary tables.  We note that links to all these interferometers
can be found on the well-established ``Optical Long-Baseline
Interferometry News'' website, maintained by Peter Lawson at NASA-JPL
(http://olbin.jpl.nasa.gov). 

Each of the currently operating interferometers have unique
capabilities and achievements of note.  The GI2T and ISI
Interferometers are the longest operating interferometers, both
beginning work in the 1980s; notably, the GI2T has uniquely pursued
observing of H$\alpha$ emission (and remains the only direct detection
interferometer with general high spectral resolution capabilities) and
the ISI is the only (published) mid-infrared interferometer. The COAST
and NPOI interferometers are currently best optimized for imaging,
having incorporated 5 and 6 telescopes respectively into their arrays.
IOTA is noted for groundbreaking fiber optics and detector development
in the infrared.  NPOI and PTI have incorporated elaborate internal
metrology to enable ambitious astrometry goals. SUSI has the
capability of 640~m baselines and is one of the only interferometers
in the southern hemisphere.  Strong progress from the MIRA-I array
marks Japan's recent efforts in long baseline interferometry.

Recent developments include new infrared and visible combiners
for the IOTA \citep[first integrated optics success with
stars][]{berger2001} and GI2T interferometers, third telescope upgrade
projects for the ISI and IOTA interferometers, 6-telescope operation
by NPOI and 5-telescopes for COAST, and first fringes from the Keck,
VLTI, MIRA, and CHARA interferometers. SUSI has also commissioned a
new ``red'' table, allowing packet-scanning interferometry using APDs.
The FLUOR combiner, so successfully used on the IOTA interferometer,
has been moved to CHARA, and we can expect excellent results soon to
take advantage of the greater resolution and sensitivity.

Indeed, it has been a busy decade for construction and implementation.
It is apparent in Table~\ref{table:capabilities}  that
the current and next generation interferometers boast
significantly larger and more numerous telescope apertures and
baselines, and promise to deliver significant new results.  In the
area of imaging, CHARA and NPOI will have 6 telescopes spread over
hundreds of meters, to allow imaging capabilities at milliarcsecond
resolution.  The VLTI and Keck Interferometers will have
$\simge$100\,m baselines with adaptive optics corrected primary
mirrors, allowing many new kinds of science to be pursued. In
particular, we can expect the first extragalactic sources, bright AGN
and quasars, to be measured at near-infrared wavelengths very soon
(probably before this article goes to press).  These new developments
are further discussed in \S\ref{future}.

Lastly, I will mention recent progress on the next generation of
interferometers.  The OHANA project has carried out initial
experiments to couple light from Mauna Kea telescopes into single-mode
fibers, the first step in a plan to link the giant telescopes of the
Hawaii into a powerful optical interferometer.  Major construction for
the Large Binocular Telescope (and Interferometer) has been
progressing for many years and is in an advanced stage now.
Importantly, the final design plans for the Magdalena Ridge
Observatory are shaping up and site work for the $\sim$10~telescope
optical array is expected to begin soon.  You can find more
information on these ambitious projects in the interferometer summary
tables as well.

The next section will review the currently exciting results from
optical interferometry, and give some indication of how the new
facilities will impact many areas of astrophysics.

\begin{landscape}
\begin{table}

\footnotesize
\caption {Current and Future Optical Interferometers: Basics
($\ast$ indicates ``in planning'')
\label{table:all_interferometers}}
\begin{center}

\begin{tabular}{|l|l|l|l|l|}
\hline
\footnotesize
Acronym & Full Name & Lead Institution(s) & Location & Start \\
\hline
CHARA & Center for High Angular Resolution Astronomy & Georgia State University &
 Mt. Wilson, CA, USA & 2000\\
COAST & Cambridge Optical Aperture Synthesis Telescope & Cambridge University & 
Cambridge, England & 1992 \\
GI2T & Grand Interf{\' e}rom{\` e}tre {\` a} 2 T{\' e}lescopes & Observatoire Cote D'Azur  &
Plateau de Calern, France & 1985 \\
IOTA & Infrared-Optical Telescope Array & Smithsonian Astrophysical Observatory, &
Mt. Hopkins, AZ, USA & 1993 \\
 & & Univ. of Massachusetts (Amherst) & & \\
ISI & Infrared Spatial Interferometer & Univ. of California at Berkeley & Mt. Wilson, CA, USA &
1988 \\
Keck-I & Keck Interferometer (Keck-I to Keck-II) & NASA-JPL & Mauna Kea, HI, USA & 2001 \\
MIRA-I & Mitake Infrared Array & National Astronomical Observatory, Japan & Mitaka Campus,
Tokyo, Japan & 1998 \\ 
NPOI & Navy Prototype Optical Interferometer & Naval Research Laboratory, &
Flagstaff, AZ, USA & 1994 \\
 & & U.S. Naval Observatory & & \\
PTI & Palomar Testbed Interferometer & NASA-JPL & Mt. Palomar, CA, USA & 1996 \\
SUSI & Sydney University Stellar Interferometer & Sydney University & Narrabri, Australia & 1992 \\
VLTI-UT & VLT Interferometer (Unit Telescopes) & European Southern Observatory & Paranal, Chile & 2001 \\
\hline
Keck$\ast$ & Keck Auxiliary Telescope Array & NASA-JPL & Mauna Kea, HI, USA & $\sim$2004? \\
LBTI$\ast$ & Large Binocular Telescope Interferometer& LBT Consortium & Mt. Graham, AZ, USA & $\sim$2006 \\
MRO$\ast$ & Magdalena Ridge Observatory & Consortium of New Mexico Institutions, & Magdalena Ridge, NM, USA &
$\sim$2007 \\
&& Cambridge University & &\\
OHANA$\ast$ & Optical Hawaiian Array for Nanoradian Astronomy & Consortium (mostly French Institutions, &
Mauna Kea, HI, USA & $\sim$2006 \\
&& Mauna Kea Observatories, others) && \\
VLTI-AT$\ast$ & VLT Interferometer (Auxiliary Telescopes) & 
European Southern Observatory & Paranal, Chile & $\sim$2004 \\
\hline
\end{tabular}
\end{center}

\end{table}
\end{landscape}

\begin{table}
\footnotesize
\caption {Current and Future Optical Interferometers: Capabilities
($\ast$ indicates planned capabilities)
\label{table:capabilities}}
\begin{center}
\begin{tabular}{|l|l|l|l|l|}
\hline
\footnotesize
        &            \multicolumn{2}{|c|}{Telescope} & Maximum & Wavelength  \\
Acronym &  Number & Size (m) & Baseline (m) & Coverage \\ 
\hline
CHARA &  6$\ast$ & 1.0 & 330 & visible$\ast$,
near-IR \\
COAST &  5 & 0.40 & 47 (100$\ast$) & visible \& near-IR \\
GI2T &  2 & 1.52 & 65 & visible, near-IR \\
IOTA &  3 & 0.45 & 38 & visible, near-IR, 4$\mu$m \\
ISI &  3$\ast$ & 1.65 & 85 ($>100 \ast$) & mid-IR \\
Keck-I &  2 & 10.0 & 85 & near-IR, mid-IR$\ast$ \\
MIRA-I &  2 & 0.25 & 30 & visible \\
NPOI &  6 & 0.12 & 64 ($>250\ast$) & visible\\
PTI &  3 & 0.40 & 110 & near-IR \\
SUSI &  2 & 0.14 & 64 (640$\ast$) & visible \\
VLTI-UT &  4 & 8.0 & 130& near-IR, mid-IR \\
\hline
Keck$\ast$ &  4$\ast$ & 1.8 & 140$\ast$ ? & near-IR \\
LBTI$\ast$ &  2$\ast$ & 8.4 & 23$\ast$ & near-IR, mid-IR \\
MRO$\ast$ &  $\sim$10 & $\sim$1.5 & $\sim$1000 & visible, near-IR \\
OHANA$\ast$ & $\sim$6 & 3.5-10 & $\sim$1000 & near-IR \\
VLTI-AT$\ast$ &  3$\ast$ & 1.8 & 202 & near-IR, mid-IR \\
\hline
\end{tabular}
\end{center}
\end{table}

\section{Summary of Major Scientific Results}

This section is divided up into two major areas: astrophysics of stars
and of circumstellar environments.  Optical interferometers have made
substantial contributions in each, and I will 
outline recent progress.

\subsection{Stellar Astrophysics}
Optical interferometry has made the greatest impact in the area
stellar astrophysics, in particular the study of nearby single stars.  
This is not surprising, given the limited nature of
single-baseline interferometers and the limited sensitivity of
first-generation instruments.   In the last decades of work, an impressive
diversity of investigations have been carried out and here we document
the most successful work.   

\subsubsection{Stellar Diameters and Effective Temperatures}

One of the earliest identified applications for optical interferometry
was directly measuring the effective temperature scale of stars.  
The effective temperature is defined such that
\begin{equation}
L = 4 \pi \sigma R^2 T^4_{\rm eff}
\end{equation}
where $\sigma$ is the Stephan-Boltzman constant, $R$ is the radius of
the star, and $L$ is the total bolometric luminosity.  Hence, by
measuring the angular size of a star and the apparent luminosity, the
effective temperature can be directly calculated.   The above equation 
is often rearranged in terms of directly observable quantities (independent
of distance estimate):
\begin{equation}
T_{\rm eff} = 2341 (F_{\rm bol}/\theta^2_R)^{1/4}
\end{equation}
where $F_{\rm bol}$ is the
total bolometric flux (in $10^{-8}$~$ergs$~$cm^{-2}$~$s^{-1}$)
and $\theta_R$ is the angular diameter in milliarcseconds.  Empirical 
calibration of the effective temperature as a function of spectral type
is important
since $T_{\rm eff}$ is considered a fundamental parameter
of a star, appearing throughout stellar astrophysics most notably on the
Hertzsprung-Russell diagram.

The survey of stellar diameters using intensity interferometry by
\citet{nii1974} still serves as the best resource for the
effective temperature scale of hot main sequence stars.  The technique
of lunar occultations has traditionally been the other main method
for high-resolution measurements of stellar sizes, as represented by
the classic paper by \citet{ridgway1980}.

There are now more than a hundred interferometer diameter
measurements, and this progress is marked in Figure~\ref{tempscales}.
Here we see one of the first major results from Michelson
interferometry from the I2T/CERGA interferometer
\citep{benedetto1987}, the effective temperature scale of giants.
Next to it, is a more recent version of the same diagram showing the
increase in the number of diameter measurements, compiled by
\citet{vanbelle1999}.  The effective temperature scale for late-type
stars is now well-established, and available diameter data has been
generated by many interferometers
\citep[e.g.,][]{mozurk1991,dyck1996a,perrin1998,nordgren1999}.  A
recent cross-comparison found the datasets from different groups to be
statistically consistent \citep{nordgren2001a}; but for
the latest spectral types, the visible photosphere is affected by TiO
absorption and it is believed that infrared sizes are more
representative of the ``true'' photospheric extent
\citep[e.g.,][]{dyck2002}.

\begin{figure}[tbhp]
\begin{center}
\includegraphics[clip,angle=0,width=5in]{Figures/JDM_tempscales.epsi}
\caption{\footnotesize {\em (left panel)}
  a. This figure shows one of the first major Michelson interferometer
  results (by the CERGA/I2T interferometer), the effective temperature
  relations for giants by \citet[][see their Figure~1]{benedetto1987},
  reproduced here with permission of ESO.  {\em (right panel)} b. Here
  is reproduced Figure~2a from \citet{vanbelle1999}, with permission
  from the AAS, showing the huge increase in the number of diameter
  measurements that now can be used for empirically determining the
  effective temperature scale of giants.  The X-axis label ``V-K''
  refers to the brightness of the star at V-band ($\lambda_0$=0.55$\mu$m)
  compared to K-band ($\lambda_0$=2.2$\mu$m); redder stars have larger
  V-K color.
  \label{tempscales}}
\end{center}
\end{figure}

While giant stars have made easy targets for
interferometers due to their large angular sizes and high
luminosities, the census of lower-mass dwarf stars and hotter main
sequence stars remain incomplete.  The PTI interferometer recently
made first contributions to the study of K- and M-dwarfs by measuring
the diameters of 5 such stars \citep{lane2001a}, although greater
precision ($\simle$1\% diameter errors) is needed to stringently test
theoretical models.  The first scientific result from the VLTI
interferometer recently contributed to this precious, limited dataset
of M-dwarf diameters \citep{segransan2003}, also finding sizes
consistent with theory but lacking precision.  To date, there are no
published diameters with baselines as long as the Narrabri Intensity
Interferometer (188m), although the CHARA interferometer did record
fringes on a 330m baseline in 2001.  The new long baseline
capabilities of CHARA, NPOI, and SUSI should allow progress in some of
these areas in the near future.
 
\subsubsection{Limb-Darkening, Atmospheric Structure}
Measuring a so-called ``uniform-disk (UD)'' diameter only requires a single
visibility data point for an isolated star, by fitting a one-parameter
model.  However, stellar photospheres are known to limb-darkened due
to optical depth effects, and thus  UD diameters
must be corrected to yield the correct physical photospheric size. The
first attempts to directly measure this was done using the intensity
interferometer on the A1V star Sirius A \citep{hanburybrown1974b}, however the
results suffered from large errors and were not very definitive.
Other attempts have been made based on looking for wavelength-dependent
angular diameters, a sign of limb-darkening \citep[e.g.,][]{ridgway1982,mozurk1991}.

\begin{figure}[tbhp]
\begin{center}
\includegraphics[clip,angle=0,width=2.5in]{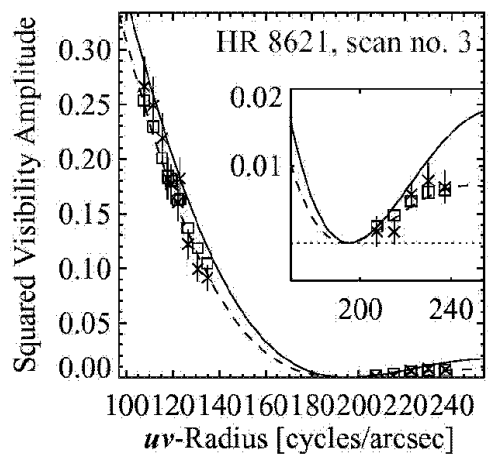}
\caption{\footnotesize This figure shows recent results from the
  NPOI interferometer, which has been optimized to investigate effects
  of limb-darkening on stellar profiles.  A portion of Figure~3 from
  \citet{wittkowski2001} has been reproduced here, with permission
  from EDP Sciences, showing that models accurately predict the
  amount of limb-darkening observed in this K5 star.
\label{limbdarkening}}
\end{center}
\end{figure}

The visibility curve of a uniform-disk star is related to the first
Bessel function, and contains an ever decreasing series of lobes,
separated by nulls, as one observes with increasing angular resolution
(see Figure~\ref{comparecal} for a plot of the first two lobes of this
curve).  The main difficulty for limb-darkening studies is that the
first ``lobe'' of the visibility pattern for a star is insensitive to
limb-darkening effects \citep[mathematically, it probes only the 2nd
moment of the brightness distribution; see][]{lachaume2003};
measurements beyond for the first null must be made to unambiguously
detect limb-darkening effects (see Figure~\ref{limbdarkening} for
representative visibility curves).  However, the fringe contrasts at
high spatial resolution are necessarily low, and thus it has been
difficult to measure these effects.  Precise
measurements of Arcturus ($\alpha$~Boo, K1III) were made by
\citet{quirrenbach1996} using the Mark~III employing a novel
phase-referencing method to increase the signal-to-noise near the
visibility null; they found reasonable agreement with model
expectations.

The NPOI interferometer succeeded the Mark~III, and expanded the
wavelength phase-referencing techniques (using strong fringes at one
wavelength to allow coherent integrations on weak fringes).
\citet{hajian1998} used these advantages first for limb-darkening
studies, followed more recently by \citet{wittkowski2001}. The latter
paper presented data with spatial frequencies sampled well before and
after the null and these exemplary results are reprinted here in
Figure~\ref{limbdarkening}.  Again, the atmospheric models were found
to be in reasonable agreement with the interferometry results. A new
generation of precision tests of stellar atmospheres are now being pursued
using detailed radiative transfer modeling coupled with thoughtful
interferometer measurements at specific wavelengths and with specific
baseline coverage \citep[e.g.,][]{aufdenberg2002}.

\begin{figure}[tbhp]
\begin{center}
\includegraphics[clip,angle=0,width=2.0in]{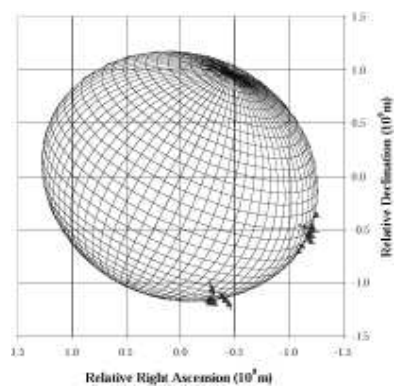}
\caption{\footnotesize Oblate Spheroid Model for the photosphere
of Altair, based on PTI data  
\citep[from Figure~6 of][]{vanbelle2001}, reproduced here with
permission of the AAS.
\label{altair}}
\end{center}
\end{figure}

Until recently, most interferometers had only 2-elements and had
difficulty to directly measure the two-dimensional shape of stellar
photospheres. In most cases, one assumes the star is circularly
symmetric in order to interpret visibility data taken at differently
projected baselines.  The Palomar Testbed Interferometer made the
serendipitous, although not unexpected in retrospect
\citep{hanburybrown1967}, discovery that that the rapidly rotating
star Altair is {\em not} circular.  Figure~\ref{altair} shows the
oblate spheroid model of this source developed by
\citet{vanbelle2001}. The oblateness is caused by centrifugal
``bulging'' along the equator and these measurements offer an
independent measure of the projected stellar rotational velocity
$v~sin~i$.  While red giants and supergiants had been known to deviate
from circular symmetric (see \S\ref{hotspots}), this is the first main
sequence star found to be non-circular; future ``imaging'' work should
allow new probes of other rotational effects, such as
gravity-darkening.

\begin{figure}[tbhp]
\begin{center}
\includegraphics[clip,angle=0,width=5.0in]{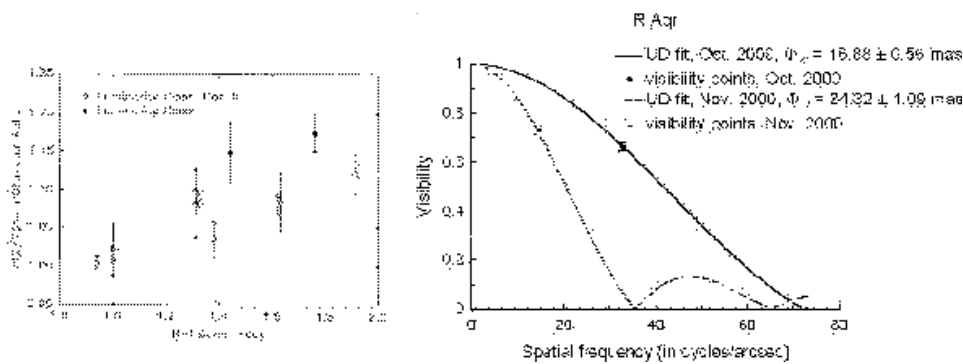}
\caption{\footnotesize {\em (left panel)} a. 
  This figure shows the Mark~III compendium of results measuring
  photospheric diameters in and out of strong TiO bands
  \citep[reproduced from Figure~2 of][with permission of the
  AAS]{quirrenbach1993a}. Redder stars (later spectral types) show
  greater atmospheric extensions in TiO.
{\em (right panel)} b. This figure shows the most recent
data illustrating that Mira stars have strongly wavelength-dependent
diameters
\citep[reproduced from Figure~2 of][with permission of the AAS]{mennesson2002}.
The two curves were fit to visibility data taken at 2.2$\mu$m and at
$\sim$4$\mu$m, and show a greater than factor of two increase in
angular size between these two wavelengths.
\label{atmospheres}}
\end{center}
\end{figure}

In addition to standard limb-darkening profiles, one has to be concerned
with the effects of molecular lines formed in the photospheres of cool
giants (mostly M spectral types), especially for the Mira variables.
It has been known for quite some time (first from speckle
interferometry) that evolved stars appear much larger when observed in
narrow spectral channels coincident with deep TiO bands in the visible
regime \citep[e.g.,][]{labeyrie1977}.  The origin of this extension is
obvious: at the observing wavelength, the optical depth unity surface
is at greater distance from the star, and hence the apparent size is
noticeably larger.  \citet{quirrenbach1993a} made the first systematic
study of this effect for late-type stars covering a range of spectral
types,and these results are shown here in Figure~\ref{atmospheres}a.

A more recent, and unexpected, discovery in this vein is that
molecules with transitions in the {\em near-infrared} are causing
large increases in the apparent sizes also.  The effect of (most
likely) unappreciated H$_2$O lines in the coolest M-stars was
uncovered by many groups using different approaches at about the same
time.  \citet{perrin1999} detected hints of these effects, finding
puzzling deviations from a uniform disk for the O-rich Mira R~Leo.
\citet{tuthill2000b} found that R~Aqr, another O-rich Mira, was
dramatically larger at 3.1~$\mu$m than at shorter wavelengths
\citep[an effect seen also in other O-rich Miras,][]{tuthill1999}.
First results with an L' band (3.75$\mu$m) combiner at IOTA also found
a large diameter increase compared to shorter wavelengths
\citep{mennesson1999}.  

A possible explanation for this effect was separately noted by
researchers analyzing data from the Infrared Space Observatory (ISO),
finding new water features in this part of the spectrum
\citep[e.g.,][]{tsuji1997,matsuura2002}.  Another recent analysis
\citep{jacob2002} coupled a dynamical model to a simple radiative
transfer model and found complex (time-variable) visibility curves due
to molecular effects in the near-IR.  \citet{mennesson2002} have
collected data from multiple infrared bands
\citep[e.g.,][]{chagnon2002}, arguing the presence of ``extended
gaseous layers'' around O-rich miras; see the dramatic difference in
near-infrared sizes observed for R~Aqr in the right panel of
Figure~\ref{atmospheres}.

\citet{thompson2002a} have expanded these studies by measuring the
sizes of O-rich and C-rich miras using narrow spectral channels
($\Delta\lambda\sim 0.1\mu$m) from 2.0-2.4$\mu$m.  While they report
O-rich miras are larger near the edges of the band, a different
behavior is observed for C-rich stars; this tentatively confirms the
role of O-bearing molecules (e.g., H$_2$O).  Moving further into the
infrared, \citet{weiner2000} \citep[following earlier work
by][]{bester1996} actually find stars to be {\em larger} in true
11.15$\mu$m continuum channels (using the ISI interferometer) than at
2.2$\mu$m, a somewhat confusing result since the near-IR wavelengths
are expected to be significantly contaminated by molecular effects.  I
am aware of even more results which have not made it to press yet
(e.g., adaptive optics at the Subaru Telescope, narrow-band IOTA
interferometry), and anticipate rapid progress in this area over the
coming years.

\subsubsection{Pulsating Stellar Atmospheres}

As discussed earlier, the ``continuum'' photospheric size is important
for calculating a meaningful effective temperature scale.  Another
important consequence of angular diameter work is specific to variable
stars: the average physical size should reveal whether a star is
pulsating in the fundamental or first-overtone mode.  Distance
estimates have been combined with interferometry data to
estimate physical diameters; these studies typically found
``large'' sizes consistent with first-overtone pulsation
in most cases \citep[e.g.,][]{hst1995,leeuwen1997,whitelock2000},
although some sources were found to be fundamental pulsators.  This is
at odds with both non-linear pulsation models of Miras
\citep{bessell1996} as well as the persuasive study of variables in
the Large Magellanic Cloud by \citet{wood1996}.  If visible and
near-infrared diameters are indeed contaminated by molecular
absorption as indicated by recent interferometric results discussed in
the last section, it is possible that the true continuum diameters are
small enough to be consistent with fundamental mode pulsation.  The
pulsation mode question of Miras has been debated and ``settled'' many
times, and still more work is needed for a definitive answer.

Pulsating stars, especially the Mira variables (period $\sim$ 1~year),
are also expected to have large changes in the photospheric diameters as
the luminosity varies \citep[e.g.,][]{bessell1989,yaari1996}.
However, because of the difficulty in obtaining a uniform dataset over
many years, it is only recently that good pulsation curves of
diameters have become available.

\begin{figure}[tbhp]
\begin{center}
\includegraphics[clip,angle=0,width=5.0in]{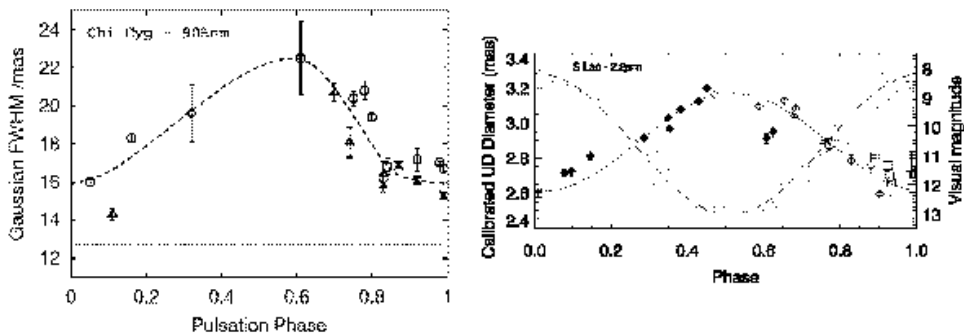}
\caption{\footnotesize Pulsation curves for two different
  Mira Variables (at different wavelengths).  {\em (left panel)} This
  panel shows the large amplitude pulsation of the S-Mira $\chi$~Cyg
  seen at 905nm by COAST, although no obvious pulsations were seen at
  1290\,nm \citep[see Figure~2a in][reproduced here with permission of
  Blackwell Publishing]{young2000a}.  {\em (right panel)} Here we see
  a well-sampled 2.2$\mu$m pulsation curve (both open and filled
  plot symbols) for the O-rich Mira S~Lac, from a recent PTI campaign
  \citep[see Figure~2 in][reproduced with permission of the
  AAS]{thompson2002a}.
\label{pulsations}}
\end{center}
\end{figure}

While first hints of phase-dependent diameter changes were reported
based on a statistical analysis of IOTA data \citep{vanbelle1996}, the
first definitive detection of diameter pulsation came from the COAST
group for the O-rich mira R Leo \citep{burns1998}, where a 35\% change
in diameter was reported.  In Figure~\ref{pulsations}a, we show more
recent results for $\chi$~Cyg (also from COAST) with better temporal
sampling of the pulsation curve \citep{young2000a}.  Most recently,
\citet{weiner2003} report the first detection of pulsation at
mid-infrared wavelengths, in this case for Mira variable $o$~Ceti.

The most significant recent developments are coming from the Ph.D.
dissertation of Thompson at PTI.  The high level of automation of the
PTI has allowed systematic observations of a large number of Miras at
all pulsational phases, and these data are presented in
\citet{thompsonthesis}.  A first look at the data has been published,
and one result from \citet{thompson2002a} appears in
Figure~\ref{pulsations}b; the diameter of this star has been measured
at many phases of the pulsation, providing an unprecedented
opportunity to test non-linear pulsation models.  The full analysis of
the PTI dataset will shed light on pulsation characteristics as a
function of spectral type (O-rich, C-rich Miras) and wavelength, and
we look forward to more of this work in the near future.

\subsubsection{Cepheid Pulsations Calibrate Period-Luminosity Relation}
Although phenomenologically related to measurements of pulsating AGB
stars, observations of Cepheids are quite distinct in their scientific
goals.  As has been discussed in \citet{sasselov1994} and earlier
\citep[e.g.,][]{davis1976}, optical interferometry will play an
important role in independently calibrating the Cepheid distance
scale, a crucial element of the cosmic distance ladder.  By measuring
the changing diameter of a nearby Cepheid and the coeval radial
velocity curve through a pulsation cycle, the distance can be directly
inferred via the Baade-Wesselink method.  A flurry of initial results
have appeared from GI2T \citep{mourard1997}, NPOI
\citep{armstrong2001}, IOTA \citep{kervella2001}, and PTI \citep[the
first definitive detection of Cepheid pulsation;][]{lane2000}.
However, most current published reports only weakly detect the
pulsation, and definitive results will require longer baselines and/or
much higher signal-to-noise ratio fringe measurements.

\begin{figure}[tbhp]
\begin{center}
\includegraphics[clip,angle=0,width=3.0in]{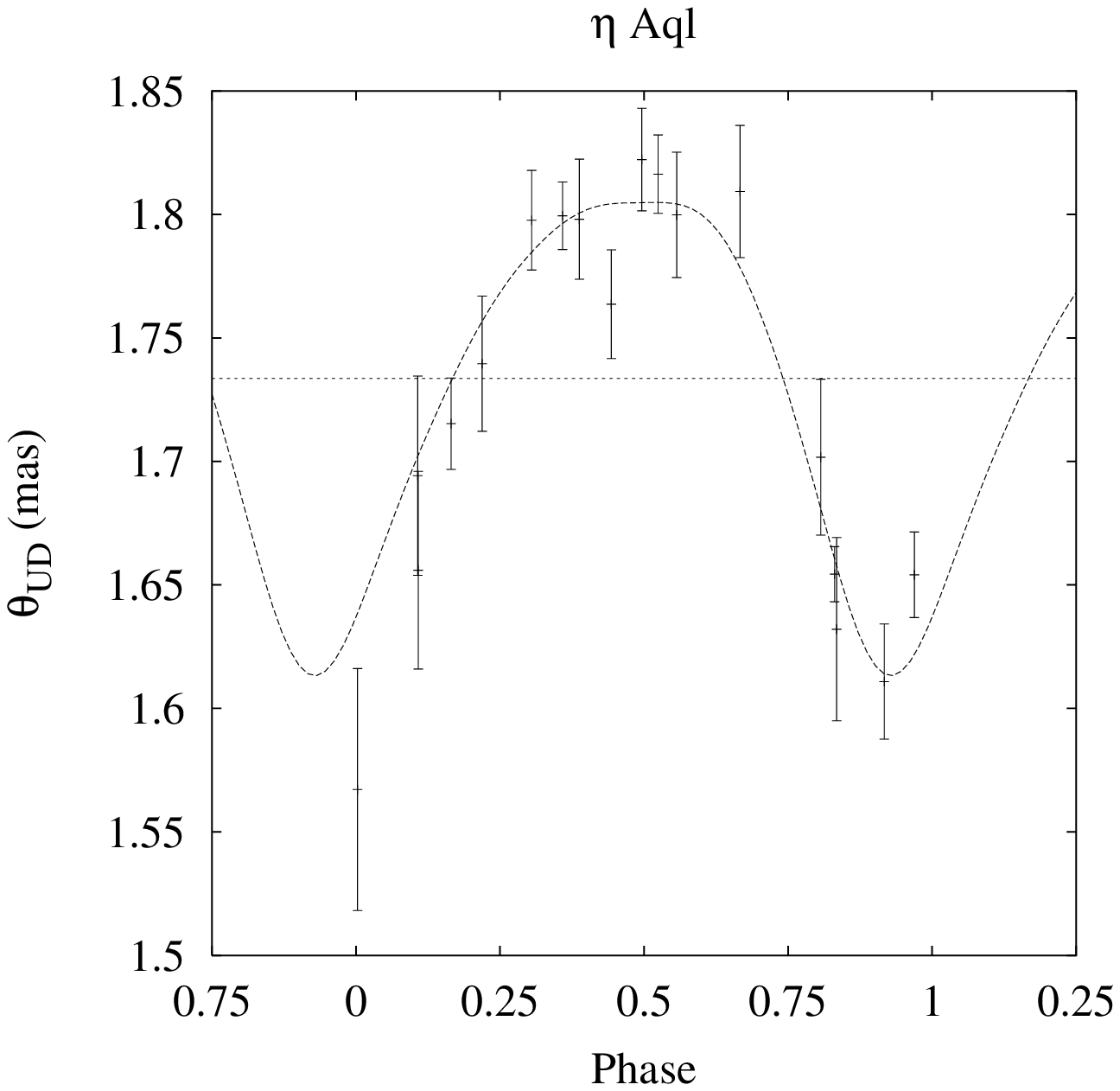}
\includegraphics[clip,angle=0,width=3.0in]{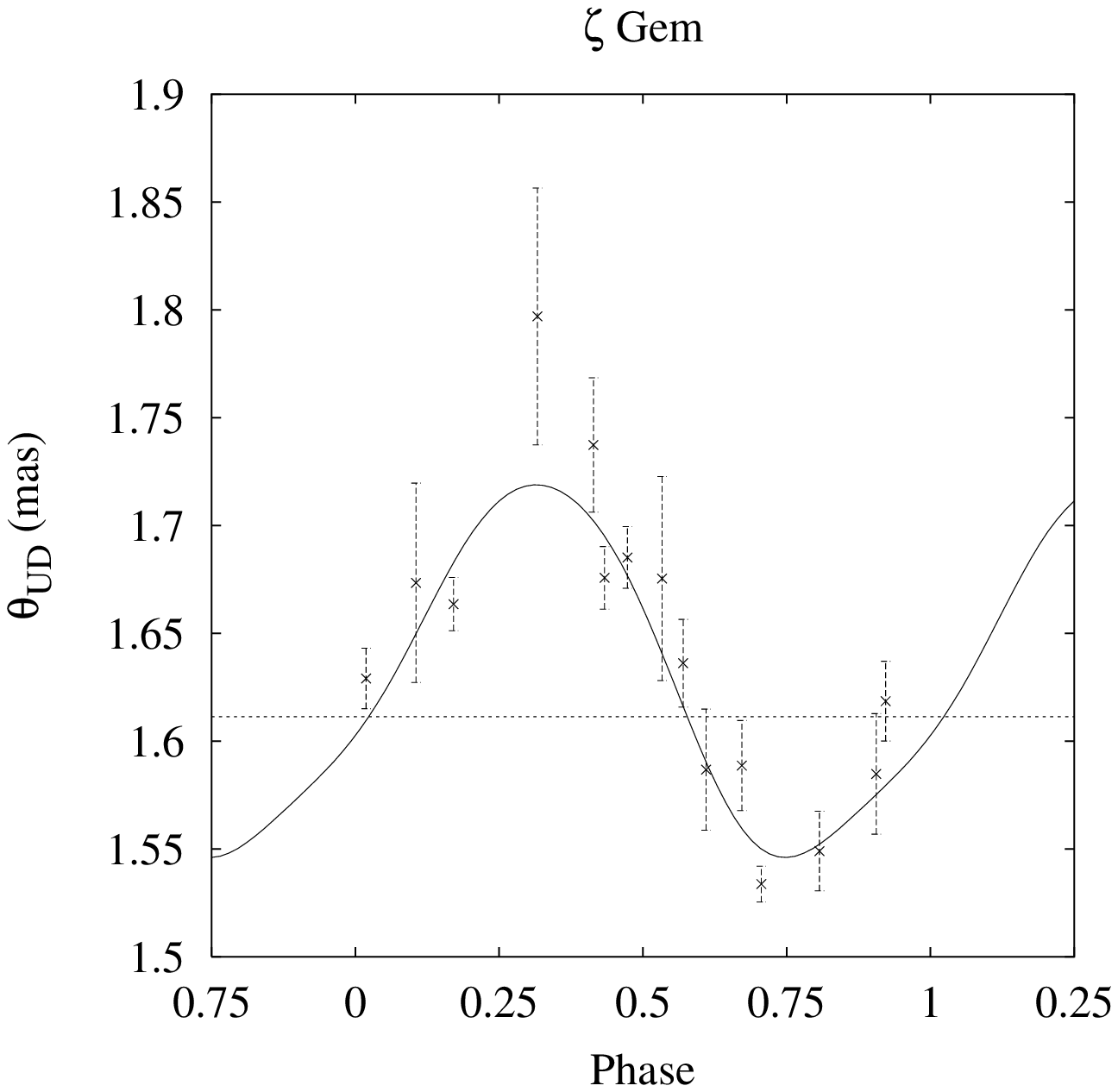}
\caption{\footnotesize 
These data show the most recent pulsation curves for
two Cepheid variables, $\eta$~Aql and $\zeta$~Gem.  
The data was taken by the PTI interferometer at 1.65$\mu$m
and, when  combined with radial velocity data, result in the most
accurate distances to these important primary distance indicators.
This figure reproduced from \citet[][see Figure 1]{lane2002a}, with
permission from the AAS.
\label{cepheids}}
\end{center}
\end{figure}

This field is rapidly developing, both observationally and
theoretically.  PTI workers recently published a second Cepheid
results paper \citep{lane2002a}, and these remarkable data are
reprinted here in Figure~\ref{cepheids}.  The derived Cepheid
distances indeed agree with those measured from the Hipparcos parallax
satellite, but the interferometer results are more precise.  The
distance uncertainty is now as (or more) affected by our uncertainty
in physics of the stellar atmospheres; fortunately,
\citet{marengo2002} report timely theoretical studies of wavelength-
and pulsational-phase-dependent variations in Cepheid limb-darkening
and emphasize the need for careful calibration in order to interpret
interferometry data accurately.  With new observing campaigns underway
at most interferometers, we can expect a rapid development here to
really pin down the pulsational properties of Cepheids in the visible
and near-infrared. In the near future, we can expect the best
calibrated Cepheid distance scale to be interferometric.

\subsubsection{Imaging Stellar Surfaces}
\label{hotspots}
As discussed in \S\ref{history}, the Cambridge group began
interferometry research by using (visible-light) aperture masking on
the William Herschel Telescope (WHT) in the Canary Islands
\citep{baldwin1986,haniff1987} while developing the COAST
interferometer.  Interferometric imaging was performed and early
results showed bright features (strong departure from circular
symmetry) on the surface of Betelgeuse \citep{buscher1990}, confirming
some previous reports \citep[e.g.,][]{roddier1983}.  No long-baseline
(separate-element) interferometer would be able to investigate the
nature of these features for years, and the Cambridge masking group
has spent more than a decade since thoroughly investigating
``hotspots'' on red supergiants and giants.

Over the last decade, it was shown that asymmetries are common
(although not omnipresent) around red supergiants and giants at visible
wavelengths \citep{wilson1992,tuthill1997,tuthill1999a}, that these
hotspots vary on a timescale of months \citep{wilson1997}, and that the
asymmetries become less-pronounced (even disappearing) into the
infrared \citep{young2000b}.  The first image of a stellar photosphere
using the COAST interferometer showed a featureless Betelgeuse
\citep{burns1997}.

\begin{figure}[tbhp]
\begin{center}
\includegraphics[clip,angle=0,width=3.5in]{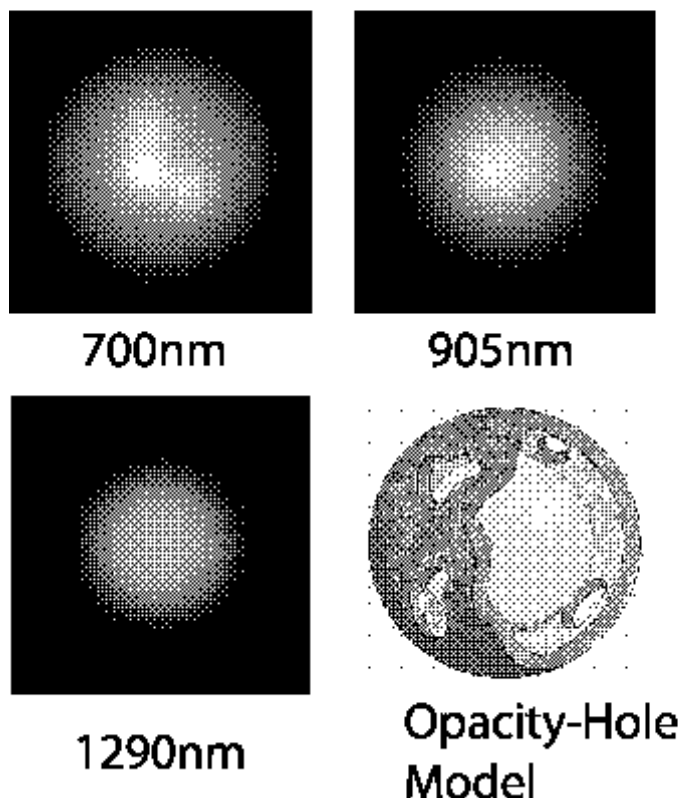}
\caption{\footnotesize This figure shows a summary of the results from
  \citet{young2000b}, the most comprehensive investigation of the
  origin of surface hotspots on evolved stars.  This synthesis
  shows nearly coeval images at different wavelengths of the surface
  of Betelgeuse using a combination of WHT aperture masking and the
  COAST interferometer.  Surface structures (``hotspots'') apparent at
  visible wavelengths disappear when imaged in the infrared.  The
  bottom-right panel shows the schematic model offered by these
  authors, where the hotspots are caused not by literally ``hot''
  patches on the photospheric surface, but rather are caused by
  ``opacity holes'' in the molecular envelope (e.g., TiO) which allow
  visible light to escape in a patchy pattern.  Figures appear here with
  permission of J. Young.
\label{betelgeuse}}
\end{center}
\end{figure}

The hotspots were originally interpreted quite literally, as ``hot''
patches on the photosphere from upwellings of large convective
elements \citep{schwarzschild1975}.  However, \citet{young2000b}
introduced a new paradigm which is illustrated in
Figure~\ref{betelgeuse}. If the bright features were indeed caused by
literal hotspots, one would not expect the features to complete
disappear in the near-infrared.  Here, we see illustrated a model where
the photosphere is surrounded by a molecular blanket (e.g., TiO) which
is optically thick in the visible, but not in the infrared
($\simge$1$\mu$m).  Inhomogeneities (possibly caused by large-scale
convection) allow visible light to escape out of opacity holes.  The
results of \cite{dyck2002} indirectly support this model, by showing
that even ``continuum'' visible diameters appear contaminated by TiO
for late M-stars.

\subsubsection{Binary Stars \& Stellar Evolution}

Binary stars have been an indispensable tool for astronomers for
centuries. Visual and spectroscopic observations yield reliable mass
estimates and form the bedrock of stellar evolution theory
\citep[e.g.,][]{eggen1967}.  The advent of speckle interferometry and
optical long-baseline interferometry has led to a remarkable increase
in the data quality and volume of binaries, including many
short-period binaries for the first time
\citep[e.g.,][]{mcalister1985, hartkopf2001}.  This work allows
unprecedented tests of stellar evolution models on a case-by-case
basis, through sub-1\% precision of stellar parameters.

\begin{figure}[tbhp]
\begin{center}
\includegraphics[clip,angle=0,width=4.0in]{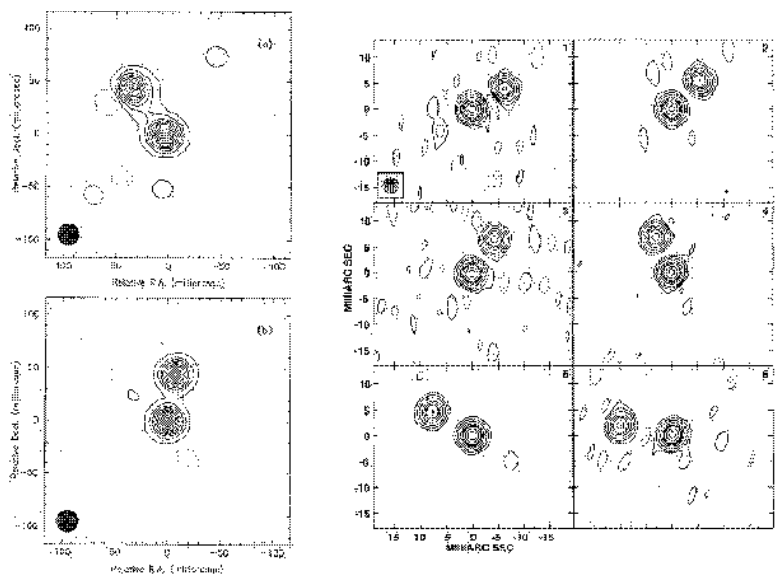}
\caption{\footnotesize Baby Pictures: First true aperture synthesis images using 
long-baseline optical interferometry.
{\em (left panel)} a. 
The binary star Capella reconstructed at two epochs using
the COAST interferometer \citep[see Figure~2 of][reproduced with permission of
ESO]{baldwin1996}.
{\em (right panel)} Six epochs of Mizar A seen by the NPOI interferometer
\citep[see Figure~4 of ][reproduced with permission of the AAS]{benson1997}.
\label{babypictures}}
\end{center}
\end{figure}

In addition to the compelling science, the simple nature of their
brightness distribution have made binary stars prime targets for most
optical interferometers.  Because of the high spatial resolution,
these binaries tend to have short periods and thus full orbital
elements can be determined by tracking the orbit.  Not surprisingly,
the first aperture synthesis images by the COAST and NPOI
interferometers were of binary systems, and these first results are
reproduced here in Figure~\ref{babypictures}.  With sufficient angular
resolution, interferometry yields the angular diameters of the components in addition
to the binary separation vector and flux ratio.

While earlier papers had concentrated on individual systems,
\citet{hummel1995} presented orbits of 8 systems with separations
between 3 and 10 milliarcseconds using the Mark~III interferometer.
For some of the systems, precise mass and luminosity determinations
allowed testing of stellar evolution models.  

\begin{figure}[tbhp]
\begin{center}
\includegraphics[clip,angle=0,width=5.0in]{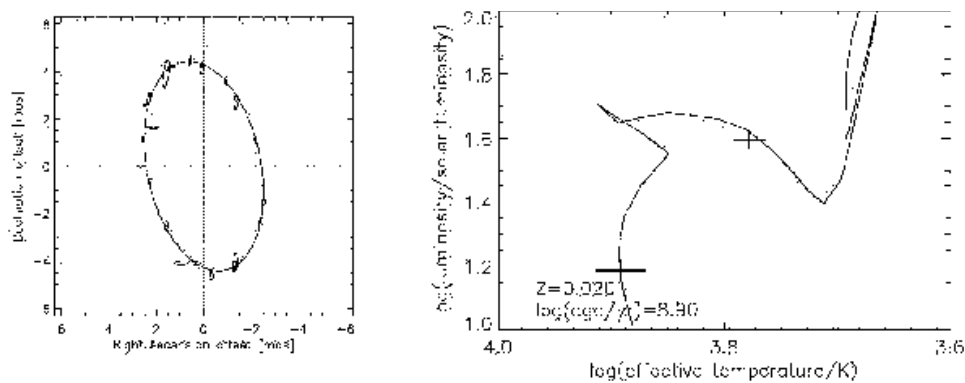}
\caption{\footnotesize Precision binary parameters can be
  derived by combining interferometry and spectroscopy, as shown here
  for $o$~Leo by \citet{hummel2001}.  The left panel shows the derived
  orbit and data from multiple interferometers and the right panel
  shows a matching stellar isochrone along with the effective
  temperatures and luminosities of the two components \citep[see
  Figures~10 \&~11 in][reproduced with permission of the
  AAS]{hummel2001}.
\label{omicronleo}}
\end{center}
\end{figure}

In order to rigorously test stellar evolutionary models, the highest
precision in parameters is needed. This requires combining data from
multiple instruments and techniques in a global fit, and concomitant
attention must be paid to systematic errors.  Ideally, the orbital elements
are fit directly to the visibility data and velocities, as implemented
by \citet{hummel1998} and \citet{boden1999}.  In
Figure~\ref{omicronleo}, the results from a study of $o$~Leo by
\citet{hummel2001} is presented.  This study stands out because it
combines interferometry data from the Mark~III, NPOI, and PTI
interferometers, as well as radial velocity data, resulting in mass
uncertainties of only $\sim$0.5\%.  Stellar evolution isochrones can
be put to a serious test for this system.

While long-baseline interferometers allow very close binaries to be
partially resolved \citep[closest is probably 0.002'' binary
TZ~Tri;][]{koresko1998} and wide systems to be characterized with
incredible precision, this is not always very important.  The most
interesting science lies often in measuring unusual binary systems for
the first time, such as the metal-poor double-lined binary system HD
195987 \citep{torres2002}.  Attractive targets for next generation of
interferometer observations include systems with short-lived
components such as Wolf-Rayet stars or Young Stellar Objects, since
much less is already known about the masses of these objects.

Lastly, the push for high dynamic range imaging of binary stars
has obvious implications for detecting low-mass companions, even 
extrasolar planets, around nearby stars.  This topic will be discussed further
in \S\ref{precision}

\subsection{Circumstellar Environments}

Interferometers can also be used to probe the environments around
stars, both at visible light and infrared wavelengths.  As the
sensitivity of facilities increase, lower surface brightness features
can be measured, opening up new avenues of research.  Advances in
interferometric imaging are particularly relevant here, since gas and
dust around stars might not be distributed uniformly and may be
changing in time.  While there has not been true imaging accomplished
by long baseline interferometers in this area yet, we have included
some of the unexpected recent results from Keck aperture masking
\citep{tuthill2000}.  While equally impressive results have also
appeared using adaptive optics and speckle interferometry (in
particular by the Weigelt group), we highlight the masking results
because the observing methods and data reduction closely parallel that
of optical interferometry and more truly reflect future capabilities;
they directly motivate excitement in the potential of interferometric
imaging with milli-arcsecond resolution and point in new scientific
directions.

\subsubsection{H$\alpha$ Envelopes around Hot Stars}
\label{halpha}
\begin{figure}[tbhp]
\begin{center}
\includegraphics[clip,angle=0,width=4.0in]{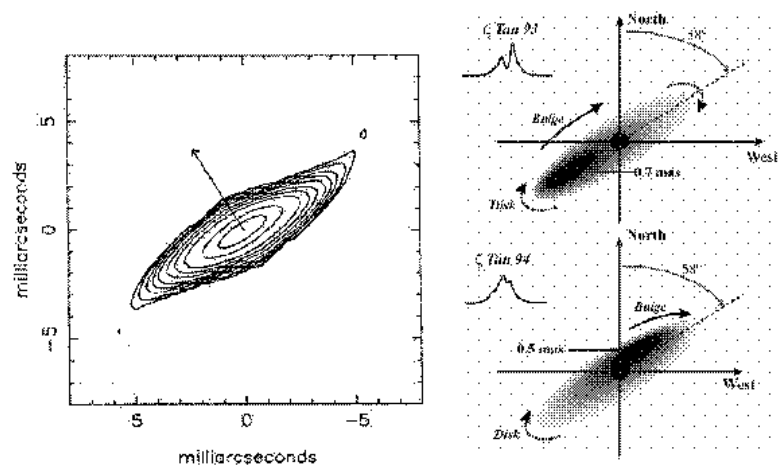}
\caption{\footnotesize 
  Spectral line observations in H$\alpha$ around Be stars have found
  extended, asymmetric envelopes; the full potential of these kind of
  investigations remain untapped. {\em (left panel)} a. Here we reproduce
  the ``image'' of the H$\alpha$ envelope of $\zeta$~Tau from Mark~III
  data \citep[Figure 3 from][]{quirrenbach1994}.  {\em (right panel)} b.
  These schematic illustrations of the same envelope at two later
  dates were based on data from the GI2T interferometer \citep[see
  Figure~4 from][]{vakili1998}.  Both figures are reproduced with
  permission of ESO.
\label{bestars}}
\end{center}
\end{figure}

While almost all the early visible interferometers focused on angular
diameters and binary stars, an interesting exception was observations
of the bright H$\alpha$ line around Be stars.  This emission was
expected to be more extended, and thus more easily resolvable, than
the tiny photosphere itself.  The envelope of $\gamma$~Cas was first
resolved by \citet{thom1986} using the I2T, and \citet{mourard1989}
saw evidence for an envelope in rotation by inspecting multiple
spectral channels across the line itself using the GI2T.  With a good
range of baselines, the Mark~III was able to detect definite
asymmetries in $\gamma$~Cas and $\zeta$~Tau
\citep{quirrenbach1993b,quirrenbach1994}; for the latter, the maximum
entropy method was used to visualize the data as a ``phase-less''
image and this result is shown in Figure~\ref{bestars}a.  This data
lacked Fourier phase information, and thus can not qualify as a true
aperture synthesis image; nonetheless, the maximum entropy procedure
provided an innovative and useful tool for visualizing this asymmetric
envelope.

The high spectral resolution of GI2T later also uncovered asymmetric
emission in these Be star envelopes in H$\alpha$ \citep{stee1995}, and
observed in other lines too \citep{stee1998}.  Figure~\ref{bestars}b
shows results from \citet{vakili1998} \citep[see also][]{berio1999}
which proposed that the emission line region is very one-sided and
time-variable.  The asymmetry in this figure was derived from the
behavior of the Fourier phases and amplitudes across the line profile;
this amounts to a kind of phase-referencing where the continuum
emission surrounding the H$\alpha$ emission is used as a reference
signal allowing the Fourier phases to be recovered.  The origin of this
``one-armed oscillation'' could result from radiative effects, the presence of
a companion, or dynamical behavior in a non-spherical potential; these
interesting results should be confirmed and explored by the
current generation of imaging interferometers.

\subsubsection{Accretion Disks and Young Stellar Objects}

There has been surprising and rapid progress in studies of how dense
accretion disks evolve around pre-main sequence stars.  Just five
years ago, simple accretion scenarios incorporating passively-heated
flared disks \citep[e.g.,][]{hillenbrand1992,hartmann1993,chiang1997}
were widely accepted, 
adequate to explain the spectral energy distributions (SEDs) of
most (low-mass) T~Tauri stars and the higher-mass Herbig Ae/Be
systems.  However, recent observations with higher spatial resolution
suggest a richer set of phenomena, and has produced much excitement.
Direct observational links are now even being made connecting the
fields of star formation and planet formation, focusing on how
accretion disks evolve into protoplanetary disks and finally to debris
disks and planets.

\begin{figure}[tbhp]
\begin{center}
\includegraphics[clip,angle=0,height=2.0in]{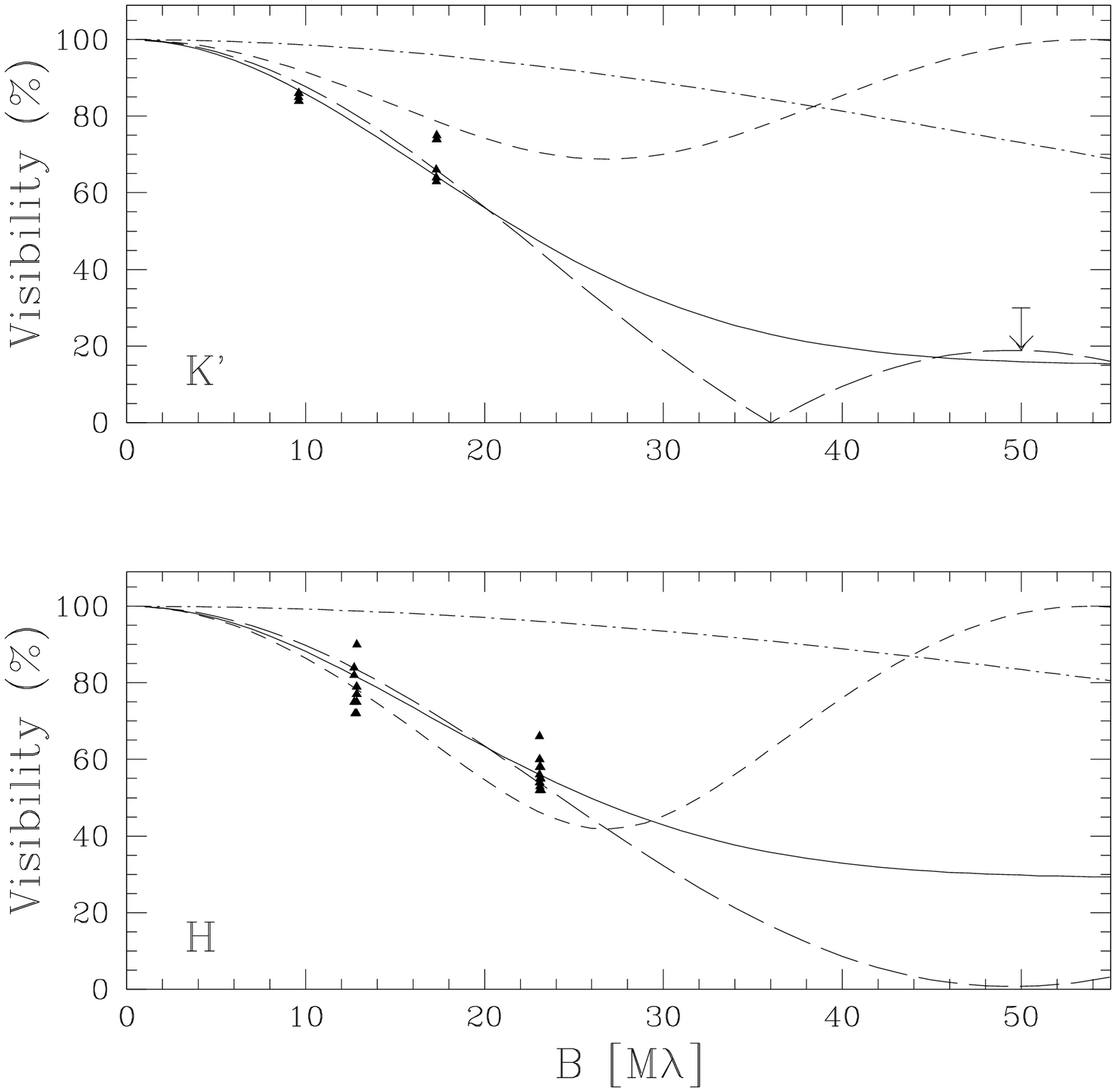}
\includegraphics[clip,angle=0,height=2.0in]{Figures/JDM_akeson2000.epsi}
\caption{\footnotesize {\em (left panel)} This figure shows the
  first Herbig Ae/Be star visibility curve ever measured \citep[see
  Figure~1 by][]{rmg1999a}. The IOTA data of AB~Aur at 1.6$\mu$m (H-band) and
  2.2$\mu$m (K'-band) indicated a much larger dust shell or disk than
  expected from accretion disk models (dash-dotted curves).  {\em
    (right panel)} The first classical T~Tauri stars were observed by
  the PTI interferometer and also showed larger than expected infrared
  sizes.  Here is the original data for T~Tau itself, an important
  source but whose analysis is complicated by uncertain calibration
  due to the presence of a nearby companion in the interferometer
  field-of-view \citep[see Figure~1a by][]{akeson2000}.  Both figures are
  reproduced with permission of the AAS.
\label{ysos}}
\end{center}
\end{figure}

Infrared interferometry is playing an important role in elucidating
the earliest stages of planetary formation by probing the density and
temperature structure of the disks presumably before planets form.
\citet{malbet1998} reported the first resolved Young Stellar Object
(YSO), using near-infrared data from the PTI interferometer; the
source was FU~Ori, a rare type of T Tauri whose emission is dominated
by accretion luminosity, and the disk size was found to be roughly
consistent with expectations.

The first ``normal'' YSO to be resolved with an optical interferometer
was the Herbig Ae/Be star AB~Aur, and \citet{rmg1999a} found the
near-IR emission to be much larger than expected using the IOTA
interferometer.  These young massive stars have high luminosities, and
thus were the brightest/easiest type of young star to study initially.
These workers published a survey of 15 total Herbig Ae/Be stars which
strongly reinforced the initial finding of large near-IR sizes
\citep{millangabetthesis,millangabet2001}. Also, they found no
evidence for disk structures -- intriguingly all the initial data were
consistent with spherical distributions of dust.  The much-awaited
measurements of classical T~Tauri stars would come soon thereafter
from the Palomar Testbed Interferometer \citep{akeson2000,akeson2002}.
The few T~Tauris measured also showed near-IR emission a few times
larger than expected from accretion disk models. Figure~\ref{ysos}
shows the visibility data from these important papers for AB~Aur and
T~Tau.  

\begin{figure}[tbhp]
\begin{center}
\mbox{
\includegraphics[clip,angle=0.,width=3.05in]{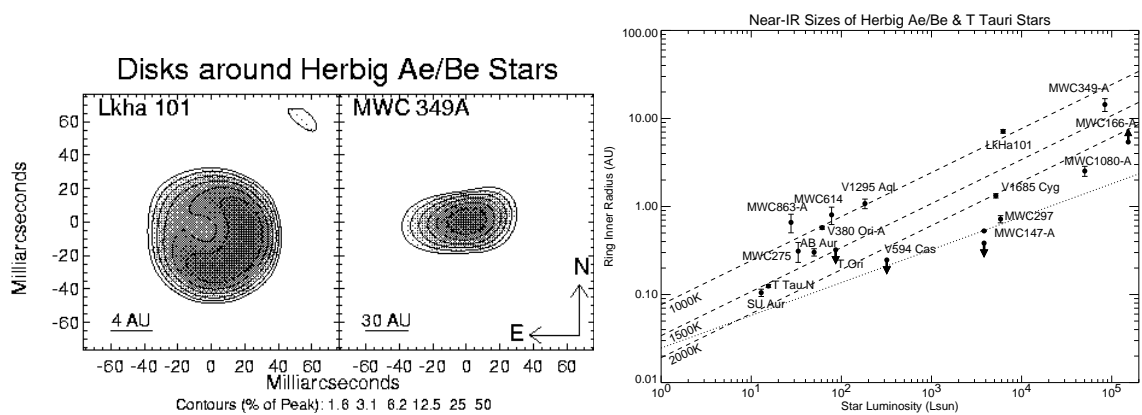}}
\includegraphics[clip,angle=90,width=2.8in]{Figures/JDM_fig_sizes.epsi}
\caption{\footnotesize {\em (left panel)}
  a.  Aperture masking interferometry was used at the Keck telescope
  to create these 2.2$\mu$m images of disks around YSOs.  The left
  panel shows the LkH$\alpha$~101 dust disk with evidence of a central
  hole \citep{tuthill2001a}, while the right panel reveals the MWC
  349A disk viewed nearly edge-on \citep{danchi2001}.  {\em (right
    panel)} b. Compendium of measured sizes of Herbig Ae/Be and T
  Tauri systems with comparison to theoretical dust sublimation radii
  \citep[from Figure~1 of][reproduced with permission of the
  AAS]{monnier2002a}.  Observed sizes are typically many times larger
  than expected from ``classical'' accretion disk models (dotted
  line).  The number of YSO measurements will rapidly grow with the
  current size surveys at the Keck and VLT Interferometers.
\label{herbigfig}}
\end{center}
\end{figure}

While infrared imaging of YSOs has only recently become possible with
new 3+ telescope arrays (and nothing published yet), aperture masking
on the Keck telescope was able to resolve two of the brightest
examples.  \citet{tuthill2001a,tuthill2002} imaged the bright source
LkH$\alpha$~101 and found a bright ring of emission, interpreting it
as the hot dust at the inner edge of an accretion disk.  While the
size was larger than expected for a geometrically-thin,
optically-thick disk model \citep[e.g.,][]{hillenbrand1992}, the
position of the inner edge was consistent with dust evaporating at
temperatures above $\sim$1500~K when illuminated by direct stellar
radiation \citep[the standard paradigm for dust shells around evolved
stars; see][]{rrh1982,dyck1984}.  Further this relation could explain
the ``large'' sizes seen by IOTA and PTI.
Figure~\ref{herbigfig}a
shows images of LkH$\alpha$~101 and also of emission-line star MWC~349
by the Keck aperture maskers.  We note that a subsequent
LkH$\alpha$~101 paper \citep{tuthill2002} included the first mid-IR
measurements of a YSO disk size (using the ISI interferometer), and
also reported unexpected changes in the LkH$\alpha$~101 disk emission.

The first results of nulling interferometry observations of Herbig
Ae/Be stars (see \S\ref{nulling}) have been recently published.  By
masking a single large aperture, \citet{hinz2001} found the spatial
extent of three sources to be unresolved using a $\sim$4\,m nulling
baseline, an unexpected result.  Interestingly, the same simple disk
models which under-predicted the near-infrared sizes were shown to
over-predict the mid-IR sizes.  Clearly, more data is needed to
determine what models are appropriate for disks around young stellar
objects -- none of the ``standard'' ones seem to work at near- or
mid-infrared wavelengths when it comes to spatial observations (the
spectral energy distributions can be fit by a number of models).

Recently, \citet{monnier2002a} summarized the current literature of
near-infrared disk sizes, and one of the figures appear here as
Figure~\ref{herbigfig}b. As indicated above, the most interesting fact
is that nearly all the disks measured are many times larger than
expected from ``standard'' disk models (geometrically-thin and
optically-thick disks), indicating a large dust-free central cavity.
A few theorists are incorporating these results into their models, and
I recommend the papers of 
\citet{natta2001} and \citet{dullemond2001} for further discussion.
Understanding how planetary systems eventually form out of
these disks will not be possible until we are more certain of the
initial physical conditions of material within a few AU of young
stars.  There is a survey already underway at the Keck Interferometer
(and planned soon for the VLTI) to include many more T~Tauri and
Herbig Ae/Be stars, and we can look forward to more developments in
this exciting area.

\subsubsection{Dust Shells \& Molecules in Evolved Stars}
Long before the dusty disks around young stars could be observed,
interferometry techniques were used to characterize dust shells
around evolved stars.  I briefly mentioned the early history of this
work in the mid-infrared in \S\ref{history}, and the capabilities
of 3~m telescopes were exploited in a series of near-infrared
speckle measurements in the 1980s \citep[see especially][]{dyck1984}.

\begin{figure}[tbhp]
\begin{center}
\includegraphics[clip,angle=0,width=5.0in]{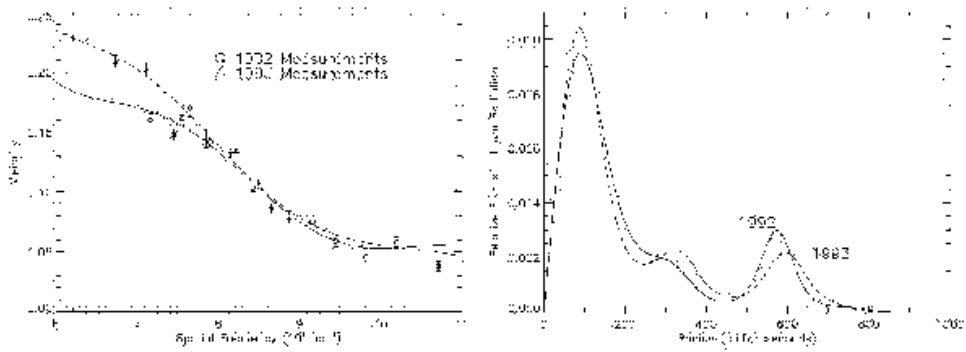}
\caption{\footnotesize {\em (left panel)} This figure shows temporal 
  changes in the IK~Tau visibility curve probing the dust shell at
  11.15$\mu$m as shown by \citet[][Figure~3]{hale1997}.  {\em (right
    panel)} These changes can be visualized through a maximum entropy
  reconstruction of the dust shell, reproduced here from Figure~4 of
  \citet{hale1997}. Both figures are reproduced with permission of the
  AAS.
\label{hale1997}}
\end{center}
\end{figure}

The most important paper on dust shells was published by
\citet{danchi1994}, reporting mid-infrared observations of 13~evolved
stars by the ISI interferometer.  The data were fitted by radiative
transfer models of dust shells, which strongly indicated that
mass-loss appeared continuous around some stars and episodic around
others.  At the time, mass-loss on the AGB was still considered to be
spherically-symmetric and continuous, and these results began to
dislodge this simplified picture.

Subsequent papers concentrated on detailed modeling of individual
sources, as the evidence for non-uniform outflows \citep{monnier1997}
and deviations from spherical-symmetry \citep{lopez1997} accumulated.
\citet{hale1997} found time-variable features in the visibility curve
of the O-rich mira IK~Tau, and attributed it to moving dust shells
expanding in the outflow.  Figure~\ref{hale1997} shows the visibility
data and a maximum entropy reconstruction to help visualize the change
in the dust shell morphology; as for the ``image'' of $\zeta$~Tau by
\citet{quirrenbach1994}, these radial profiles were reconstructed from
phase-less data and so do not necessarily represent a true ``image.''
The changes were consistent with the expected outflow speeds (derived
from maser observations), and allowed an independent
distance estimate to this source.

\begin{figure}[tbhp]
\begin{center}
  \includegraphics[clip,angle=0,width=3.0in]{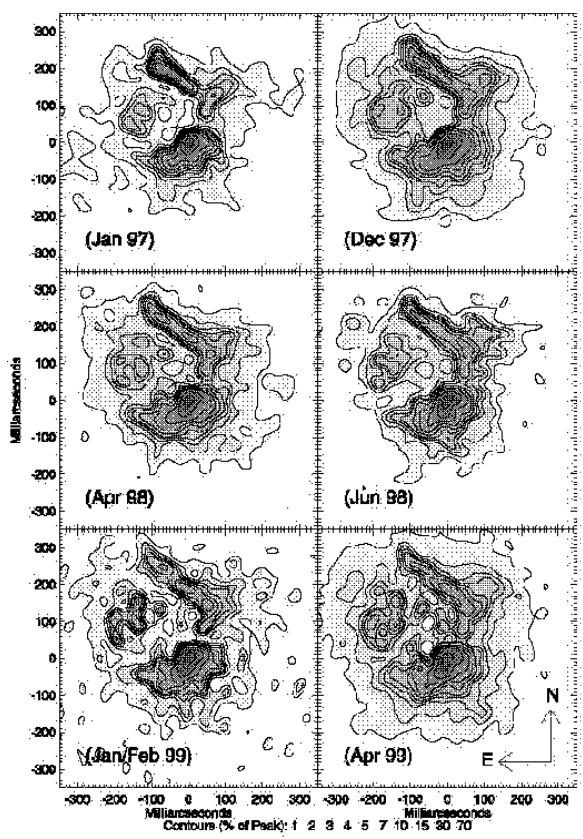}
\caption{\footnotesize Shows temporal evolution of the
  inhomogeneous and clumpy dust shell around carbon star IRC~+10216 as
  seen at 2.2$\mu$m using Keck aperture masking \citep[Figure 1
  from][reproduced with permission of the AAS]{tuthill2000a}.
\label{irc10216}}
\end{center}
\end{figure}

The idea that mass-loss is not uniform in time nor necessarily
spherically symmetric found confirmation using other related
techniques.  At the same time, near-infrared speckle and aperture
masking were being pursued on the new 8-m class telescopes.  Imaging
of the carbon star IRC~+10216 revealed an incredibly inhomogeneous and
asymmetric dust shell \citep{haniff1998,weigelt1998,tuthill2000}, and
enough data has been collected to see the dust structures evolve with
time; I reproduce  a figure from \citet{tuthill2000} which
exploits the diffraction-limit of the world's largest telescope to
image the details of this nebula (see Figure~\ref{irc10216}).  For
readers interested in near-infrared imaging of dust shells using
speckle interferometry and aperture masking, I refer the reader to
many recent papers by the Weigelt group and the Keck masking team,
which will not be reviewed here.

Another recent area of progress is the successful combination of high
spectral resolution with high spatial resolution in the mid-infrared.
A filterbank spectrometer was constructed and installed on the ISI
interferometer allowing interferometry data to be collected on
mid-infrared absorption lines
\citep{monnier2000a,monnier2000b,monnier2000c}.  Polyatomic molecules
such as ammonia and silane form in dense outflows of evolved stars,
but at orders of magnitude greater abundance than expected
\citep[e.g.,][]{keady1993}.  This mystery was partially solved when
the ISI found that the molecules form much further out in the flow
than expected, possibly related (for the case of Silane) to the
depletion of SiS onto grains \citep{bieging1993}.  This work
highlights the potential for studying cosmochemistry using
interferometry.

\subsubsection{Colliding Winds}
\begin{figure}
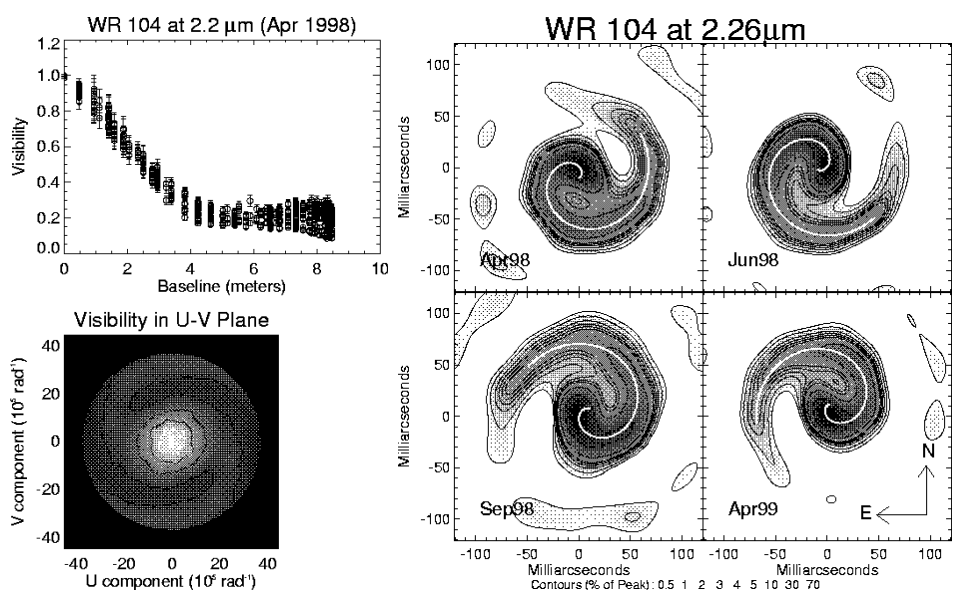

\begin{center}
 \includegraphics[clip,angle=0,height=3.0in]{Figures/JDM_showcase_wr104.epsi}
  \includegraphics[clip,angle=0,height=3.0in]{Figures/JDM_wr104_4ep.epsi}
\caption{
  \footnotesize {\em (left panels)} a. These figures show the
  2.2$\mu$m visibility curves, 1-D and 2-D, of WR~104 observed by Keck
  aperture masking.  {\em (right panel)} b. This figure shows four
  epochs of imaging of WR~104 showing a morphology consistent with an
  Archimedean spiral rotating at a uniform rate \citep[from][Figure
  13.6]{monnier99}
\label{wr104fig}}
\end{center}
\end{figure}

While no Wolf-Rayet dust shells have yet to be observed using a
long-baseline interferometer, the discovery of pinwheel nebulae around
these stars by \citet{tuthill1999b} and \citet{monnier1999} using Keck
aperture masking represents a new direction for the current generation
of infrared imaging interferometers.  An example of the spiral pattern
seen in the visibility data of WR~104 is shown in
Figure~\ref{wr104fig}; from this data (and the closure phases) images
can be reconstructed showing a beautiful spiral structure which
appears to rotate on the sky with a period $\sim$243 days.  These
spinning, spiral dust shells result from dust formation at the
interface of two colliding winds in a Wolf-Rayet and O-star binary.

These sources are usually  obscured by local dust and thus are
faint visible sources (V$\simge$15). The faint V magnitudes make these
targets difficult not only for adaptive optics systems, but also for
all current interferometers (except the ISI) which use visible-light
detectors for star tracking and tip-tilt correction.  This limitation
hopefully will be eliminated at some facilities by using an infrared
star-tracker (for example, IOTA has plans along these lines).

\section{The Future}
\label{future}
\subsection{Exciting Trends and Near-Future Science Potential}
As has been said before, now is a practical time for reviewing the
major achievements of optical interferometry, for we are entering a
new era boasting facilities with significantly greater
sensitivity, angular resolution, spectral resolution, and
wavelength-coverage.  In this section, I will give my views of some of
the new capabilities and the expected science returns.

One important trend that must be bolstered is the inclusion of
theorists and modellers in the observations and interpretations of
interferometry data.  In many areas, the interferometry observations
are outstripping the ready tools for analysis.  For example, the
wavelength-dependent and time-dependent diameters of AGB stars require
a combination of time-dependent hydrodynamical atmospheres and
sophisticated radiative transfer codes, a problem very challenging
even with today's supercomputers.  Understanding the hotspots seen on
the surfaces of stars will required 3-dimensional simulations of
stellar convection.  Accretion disk physics around young stars should
include magnetic fields and demand thoughtful considerations of gas and dust
physics in a 2-D or 3-D context.  Dust production in colliding winds
is very poorly understood, and poses a formidable numerical simulation
problem.  While tackling these difficult physical problems will
require the new high-resolution data from optical interferometers, it
is also true that input from the modellers and theorists is needed to
guide and suggest experiments and observing strategies.

Another general comment is that increasing the angular resolution
usually means probing ever decreasing physical scales.  Since
interferometers often probe scales smaller than an AU, significant
changes in time are expected for even small characteristic velocities
($\sim$km/s).  This poses both a risk and an opportunity: a risk since
data must be taken rapidly and efficiently to accurately capture
snapshots of ever-evolving and changing environs, and an opportunity
to include dynamics and time-evolution into our models and
understanding.  Observing the dynamics of circumstellar and/or stellar
environments allow new physics to be understood, physics that usually
can not be unambiguously reconstructed from typical datasets. Thus, I
hope that new dynamical information will break theoretical stalemates
which paralyze a number of fields.  Interferometers have the
opportunity to revolutionize the way we think of the universe: from
distant ``frozen'' images of the past, to a dynamic and engaging
unfolding of the present.

\subsubsection{New Long Baselines}
New long baselines will allow unprecedented high resolution
measurements on select sources.  With sub-milliarcsecond resolution,
one can measure the diameters of ``small'' sources which have largely
eluded current surveys, such as hot stars and nearby low-mass stars.
Distortions in the photospheric shapes of rapidly rotating stars or
binary stars in nearly Roche-lobe filling systems can be directly
detected.  Limb-darkening studies of important objects, such as
Cepheids, can be accomplished to put the Cepheid distance scale of
firm direct footing.  Further, long baselines make a variety of
exoplanet studies possible, such as directly detecting 51 Peg B-like
planets (``Hot'' jupiters) or resolving planetary transits across the
stellar disk.  The NPOI, CHARA, and SUSI interferometers will possess
the longest baselines in the near-term, while future projects such as
the MRO or 'OHANA (on Mauna Kea, HI) might
someday extend the resolution below even 0.10 milliarcseconds with
$>$1~km baselines.

\subsubsection{Imaging}
Imaging with optical interferometry is currently tedious at best, and
can only investigate simple objects such as resolved photospheres or
binary stars.  The 6-telescope systems of NPOI and CHARA will soon
possess the capability of (comparably) excellent ``snapshot''
coverage, allowing more complicated and higher dynamic range imaging
of select targets.  The CHARA array can not be reconfigured and hence
will only image well targets with the appropriately-sized structures
-- for a maximum baseline of $\sim$330~m at 1.65$\mu$m, the optimum
size scale is a few milliarcseconds.  The NPOI interferometer can be
reconfigured to ``fill-in'' the (u,v)-plane completely over time, and
be adjusted for individual sources to optimally measure the needed
visibility and closure phase information.  In the longer term, the
auxiliary telescope array at the VLTI and the proposed outrigger
telescopes at Keck will allow even fainter infrared targets to be
observed.  For imaging, the MRO is currently the most ambitious
project in the works, hoping to include $>$10 telescopes, which would
make it the premiere imaging interferometer in the world.

Good imaging capabilities would open up new avenues of research,
especially in studies of the circumstellar environments at infrared
wavelengths.  The ability to study disks around young stars and the
time evolution of gaps, rings, or other structures would revolutionize
our understanding of planet formation.  At visible wavelengths,
imaging spots on the surfaces of other stars is a major goal, and
would allow solar physics to be applied in detail to other stars for
the first time.

The unexpected discoveries of Keck aperture masking
justify our optimism that imaging will uncover many new phenomena that
currently are hidden unnoticed in spectral energy distributions.
For example, the Wolf-Rayet dust spirals (see
Figure\,\ref{wr104fig}) have only been observed in a few systems, and
represent a new area of study when imaging interferometric arrays are
fully commissioned.

However, current imaging work using COAST, NPOI, and IOTA
interferometers suffer from the lack of dedicated software resources.
Unfortunately, the decades of software development in radio
interferometry can not be fully leveraged for optical interferometry,
since radio work now relies largely on phase-referencing techniques
not generally available in the optical.  New imaging software is
needed which can take into account the unique nature of optical
interferometry data as well as the different nature of our target
sources.  The recent adoption of a common data exchange format,
defined by the COAST and NPOI interferometer teams, represent an
important first step towards these goals (see
http://www.mrao.cam.ac.uk/~jsy1001/exchange/).

\subsubsection{Precision Interferometry}
\label{precision}

This is a rapidly developing area since the advent of single-mode
fibers for spatial filtering and ``dual-star'' phase referencing.
When a model of the astronomical source is well-known, then incredibly
precise measurements are possible.  The most potential for this is in
the general area of binary stars, where the stars
either are point sources or partially-resolved uniform-disks.  The
case of detecting an exosolar planet is included in this category,
since it can be considered as very high-dynamic-range imaging of a
faint companion.

While there are open questions in binary evolution and stellar
astrophysics which demand such high precision, a more popular reason
to pursue ``Precision Interferometry'' is towards detection of
extrasolar planets around nearby stars.  There are many ways this can
be manifested, and I will outline a few of these.

\begin{figure}
\begin{center}
\includegraphics[clip,width=4in]{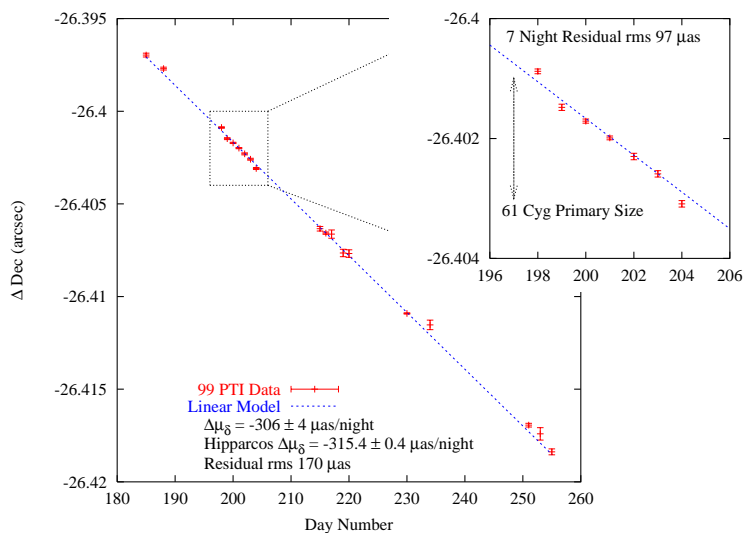}
\caption
{State-of-the-art narrow-angle astrometry of the binary
61~Cyg by the PTI. For a period of one week,
the residual astrometric error in declination was 
$\sim$100~{\em micro-}arcseconds. Figure printed with permission
of SPIE, originally appearing in \citet{lane2000a}.
\label{ptifig}}
\end{center}
\end{figure}

Narrow-angle astrometry is a comparatively ``classical'' way to detect
an exosolar planet. Akin to the doppler shift-radial velocity method,
precision astrometry attempts to detect the minute wobble of the
parent star as a planet proceeds in its orbit.  This can be done by
monitoring the angular distance between a star and a background
reference star.  In this case, the target star is normally quite
bright and used for phase-referencing to a faint star projected within
an isoplanatic patch from the target ($\simle$30'').
\citet{lane2000a} reported the first measurements of this kind using
the PTI (see
Figure~\ref{ptifig}).  For reference, the motion of Saturn and Jupiter
perturb the Sun $\sim$1 milliarcsecond as viewed from 10~pc.  This
technique will be applied by the Keck Interferometer and the VLTI
interferometer for a planet survey, and there is talk of pursuing this
in Antarctica where the isoplanatic patch is larger and the coherence
times longer \citep[e.g.,][]{lloyd2002,swain2002}.

Another method also being aggressively pursued by the Keck and VLTI
interferometers is a multi-wavelength approach to find massive
exoplanets by detecting a very slight photocenter shift between
different infrared bands due to hypothesized absorption bands in the
planet's atmosphere \citep[i.e., the {\em differential phase} method;
e.g.,][]{as1999, lopez2000}.  This method has the advantage of using
the bright target star as its own phase reference.  However, recent
studies of line-of-sight variability of atmospheric water vapor
\citep{akeson2000b} indicate that differential chromatic dispersion
might be more difficult to calibrate for {\em differential phase}
methods than originally expected.

Precision measurements of {\em closure phases} can also be used to
detect faint companions, a method which has not received as much
attention.  As described earlier in this review
(\S\ref{closurephase}), the closure phase is formed by summing the
interferometer phases on three baselines around a triangle of
telescopes, and this quantity is immune to atmospheric phase delays.
The lack of attention to precision closure phase methods is
understandable since few interferometers possess the requisite minimum
of three telescopes.  \citet{monnier2002} and \citet{segransan2002}
recently discussed how closure phases are immune to dominant
calibration problems of differential phase and that they can also be
used to solve for all the parameters of a binary system without
needing to measure any visibility amplitudes.  For reference, a
typical closure phase for a binary with brightness ratio of 10$^4$ is
$\sim$0.01~degrees as long as the component separation is resolved by
the interferometer -- the same magnitude effect as for differential
phase methods.

Current published measurement precision of closure phases is only 0.5
to 5 degrees \citep{tuthill2000, benson1997, young2000b}. Improving
the three orders of magnitudes needed to detect even the brightest
possible exoplanet is a daunting challenge.  While there are surely
unconsidered systematic effects (perhaps due to birefringence or
drifts in optical alignment) which will degrade the sensitivity of the
precision closure phase technique, the lack of any ``showstopper''
effects, like differential atmospheric dispersion for the differential
phase methods, strongly argues for the further development of the
closure phase technique.  

\subsubsection{Nulling}

Another approach being pursued for planet detection is nulling
\citep{bracewell1978}.  The initial nulling experiments with the MMT
\citep{hinz1998} have continued \citep{hinzthesis}, and ultimately
will be applied on the Large Binocular Telescope Interferometer
\citep{hinz2001b}.  This project is still many years away, but offer
an alternative approach to the ``precision'' phase methods above.

In the nearer term, the Keck Interferometer will be applying nulling
in the mid-infrared in order to measure and characterize the zodiacal
dust around nearby stars.  This source of infrared radiation is
expected to be the dominant background for an eventual space-based 
planet detection interferometer, the so-called ``Terrestrial Planet
Finder'' mission (more information in \S\ref{spaceinterferometers}).
\citet{serabyn2001} describe the ``fully symmetric'' nulling combiner
being implemented on the Keck Interferometer, and initial on-sky tests
are expected to begin in 2003.  A more complete description of the
observing strategy and expected sensitivity has been documented in
\citet{kuchner2003}.

Nulling can also be applied on large single-apertures, and then are
called nulling coronagraphs.  New clever designs in coronagraphy are
competing with nulling interferometry for space mission concepts to
detect    terrestial planets around other stars, and I recommend
interesting papers on optimally-shaped 
and apodized pupils \citep{spergel2002,papa2001},
bandpass-limited image-masks \citep{kuchner2002}, and phase-mask-based
approaches \citep[e.g.,][]{guyon1999,rouan2000}.

\subsubsection{Spectroscopy}
There have been only a few significant results combining spectroscopy
and interferometry; fortunately, this is about to change.  The
near-infrared AMBER instrument, slated to arrive at the VLTI
interferometer in 2003, will combine three telescope beams together
and disperse the light with 3 different spectral resolutions, the
maximum is R$\simge$10000.  This resolution will allow interferometry
on individual spectral lines in the 1-2.5$\mu$m regime, opening up
shock-excited emission lines, CO-absorption/emission features, and even
emission from YSO jets to be probed in novel and exciting ways for the
first time.  We can expect the value of interferometric observations
to be greatly enhanced by these new capabilities.

\subsubsection{Polarimetry}
Imaging stars in polarized light with interferometers also promise
fascinating new insights into many areas of astrophysics, although
this capability is difficult to implement with current interferometers.
\citet{vakili2002} discuss interesting applications of combining the
high spectral resolution of AMBER with polarimetry, and highlight the
new capabilities for imaging scattered light and potentially even
measuring stellar magnetic fields from the Zeeman effect.
Experimental efforts \citep{RP1997} in this area have been very
limited compared to the theoretical progress
\citep{RP2000}; this situation should be remedied soon.

\subsubsection{New Observables}
Along with greater spectral coverage and more telescopes come new
interferometric observables. While \S\ref{precision} discussed
possible applications of {\em differential phase} and {\em
  differential closure phase}, there are other interferometric
observables yet to be exploited for precision interferometry.

Measuring the diameter of a star by precisely locating the first null
of the visibility pattern is immune to amplitude calibration errors.
This could be done by using a well-calibrated spectrograph to search
for the null, either measuring fringe amplitudes or looking for the
signature phase-flip across the null (e.g., Mozurkewich, private
communication).  This technique is similar to the method of A.
Michelson in measuring the diameter of Betelgeuse
\citep{michelson1921}, where the baseline was adjusted in order to
find the visibility minimum as detected by his eyes.

The {\em closure amplitude} (requires sets of 4~telescopes) is an
important quantity in radio interferometry to compensate for unstable
amplifier gains and varying antenna efficiencies that can be linked to
individual telescopes \citep[e.g.,][]{readhead1980}.  Closure
amplitudes are not practical for current optical interferometers
partially because most fringe amplitude variations are not caused by
telescope-specific gain changes but rather by changing coherence
(e.g., due to changing atmosphere).  However, the introduction of
spatial filtering (e.g., single-mode fibers) should make the closure
amplitude a useful tool for optical interferometry soon \citep[see
discussion in][]{monnier_mss}.  

Necessarily, most new observables have yet to be used in practice or
described in print.  I mention here a few possibilities that this
author has considered to encourage future experimentation.  For
instance, it may be possible to use closure amplitudes in the
case when fringe jitter causes loss of visibility contrast in a
fringe-tracking interferometer, due to the way in which small random
phase errors degrade coherence. Also, the {\em Closure Differential
  Phase} is a recently defined quantity \citep{monnier2002},
introduced to overcome one limitation of current phase-referencing
techniques, namely, that differential phase (and differential closure
phase) methods requires assumptions about the source structure of the
phase calibrator.  

\subsubsection{Sensitivity (iKeck and VLTI)}
Another area where we expect immediate progress is in observing new
classes of faint objects for the first time.  The Keck and VLTI
interferometers will have the capability of observing sources as faint
as K$\sim$11 magnitude (down to K$\sim$20 with phase referencing),
opening up extragalactic sources for the first time.  By the time this
review is printed, I expect that the first optical interferometric
observations of the core of an Active Galactic Nuclei (AGN) will be
announced.  Size measurements of AGN should offer new constraints on
models of the infrared continuum and, when coupled with high spectral
resolution, could determine the physical origin of observed broad line
regions and possibly even measure dynamical black hole masses.

In terms of galactic sources, this increase in sensitivity will allow
a broad census of sources to be taken, including YSOs spanning a broad
range of ages, luminosities, and distances and binary systems of all
masses.  For instance, infrared observations of pre-main-sequence
binaries allow unique probes of the evolution of binary fraction
\citep[e.g.,][]{ghez1993} as well as important measurements of masses
of young stellar objects \citep{tamazian2002}.  I expect
interferometer observations to play an increasingly important role in
this area as the sensitivity increases.

Of course the additional sensitivity will permit new projects too,
such as tracking the motions of stars orbiting the black hole at the
center of the Milky Way with an order of magnitude greater precision
than possible today with single-aperture telescopes
\citep[e.g.,][]{genzel2002}.  Precision astrometry may allow even new
tests of General Relativity near super massive black holes at the
center of nearby galaxies.

In addition, the MIDI instrument for the VLTI will allow sensitive
measurements in the mid-infrared for the first time. While the ISI
interferometer pioneered interferometry in this wavelength range,
MIDI+VLTI will be first to probe a wide range of sources with
resolution of $\sim$0.''01 and down to N$\sim$4 mag ($\simge$
100$\times$ fainter than the ISI).  Mid-infrared observations are
sensitive to emission from relatively cool dust and can peer through
thicker layers of dust than possible in the visible or near-infrared.
There are great possibilities for advancing our understanding of young
and evolved stars both, and studying dust distributions in a variety
of environments.

\subsection{Space Interferometry}
\label{spaceinterferometers}

The greatest limitations to optical interferometers arise from
atmospheric turbulence.  It dramatically limits the sensitivity, the
ability to do imaging, and forces the engineering to be clumsy and
complicated.  Space is naturally an ideal place for interferometry,
with no atmosphere to corrupt the phase nor limit the coherent
integration time. And long baselines are obviously possible by
combining light intercepted by separate spacecraft flying in
formation.

\subsubsection{Critical Technologies Needed}
In order to successfully build space interferometers, many
technologies must first be developed.  To this day, there has not been
any dedicated space interferometer flown \citep[except for the Fine
Guidance Sensors on the Hubble Space Telescopes; e.g.][]{franz1991}.

For interferometers deployed on a single structure, one has to contend
with truss vibrations, thermal and gravitational gradients, and an
unusually large number of mechanisms (failures of which could end the
mission).  There are issues with propellant and power consumption for
maneuvering the array to point around the sky.  The Space
Interferometry Mission (SIM) is in advanced planning stages and is
being designed to measure accurate positions of stars with
micro-arcsecond resolution.  SIM is a ``simple'' 2-element
interferometer on a deployable truss ($\sim$10\,m maximum baseline),
and will be the first space mission to attempt space interferometry.

Ultimately, one would want to have baselines much longer than $\sim$10
meters, and this will require separate, free-flying spacecraft.  For a
space interferometer consisting of ``free-flyers,'' there are other
problems.  For instance, maintaining the physical distances between
space telescopes to sub-micron tolerances is indeed a challenge.
Probably this can not be done; however by monitoring the spacecraft
drifts in real-time using laser metrology, the changing distances can
be compensated for by onboard (short) delay lines.  Some engineering
missions have been proposed to test ideas, but have yet to really
get-off-the-ground (e.g., the NASA Starlight mission was recently
cancelled).  NASA and ESA should give such a test mission a high
priority since the science potential for a free-flyer interferometer
is so much greater than for one limited to a single structure.

\subsubsection{Review of Current NASA \& ESA Missions}
There are a number of mission concepts involving space interferometry
being considered by NASA and the European Space Agency (ESA). As
mentioned before, the only one in advanced design stages is the NASA
Space Interferometry Mission (SIM).  In
Table~\ref{table:space_interferometers}, I summarize some of the
missions that are being proposed, and their main science drivers.
Considering the unreliability of expected launch dates, I have omitted
these from the table -- it is unlikely any of these will fly before 2010 (2020?).

NASA and ESA have spent much energy on designing missions to detect
Earth-like planets around nearby stars, and to measure their crude
reflectance (or emission) spectra.  With luck, an extrasolar planet
spectrum could encode distinctive atmospheric spectral features
indicating the presence of life (biomarkers) on the distant planet
\citep[e.g.,][]{woolf2002}.  While originally envisioned as an
infrared interferometer mission, concepts involving a visible-light
coronagraph have been proposed lately. This mission is known at the
Terrestrial Planet Finder (TPF) in NASA, and as IRSI-Darwin at ESA.
The summary table also includes a few TPF follow-on missions, such as
``Life Finder.''  These missions are very futuristic, and testify to
NASA's ebullient imagination.

Another area of interest is imaging the far-infrared and
sub-millimeter sky at high angular resolution using space
interferometry.  These wavelengths are difficult to access from the
ground due to water absorption in the atmosphere.  Because of this,
the angular resolution of current observations are very limited
($\sim$30''); compared to all other wavelengths, the sky has been
surveyed with the lowest resolution in the far-IR.

The proposed NASA mission ``Submillimeter Probe of the Evolution of
Cosmic Structure'' (SPECS) would be a separate-telescope space
interferometer (possibly tethered together and not ``free-flying'')
designed to map the sky with great sensitivity at a resolution
comparable to that currently achievable at other wavelengths.
($\sim$0.010'').  This would avoid the confusion-limited regime
encountered by current low-angular-resolution galaxy count surveys,
and allow the evolution of cosmic structure to be investigated back to
high redshift.  The SPIRIT mission is meant as a precursor to SPECS to
test out various aspects on a single platform.

The X-ray community has also proposed a space interferometer, which
would boast micro-arcsecond resolution and be capable of studying the
hot material at the event horizon of nearby Black Holes.  Bolstered
by successful lab experiments \citep{cash2000}, plans for a
free-flying x-ray interferometer called the Micro-Arcsecond X-ray
Imaging Mission (MAXIM) have begun.  Controlling distances between
macroscopic mirrors to picometer-precision, as is needed for X-ray
interferometry, is indeed a daunting challenge. However, a MAXIM
precursor mission with only a few meter baseline would have orders of
magnitude greater resolution than the Chandra X-ray telescope and
stands some chance of being flown.

\begin{table}

\footnotesize
\caption {Proposed Space Interferometers}
\label{table:space_interferometers}
\begin{center}
\begin{tabular}{l|l}
\footnotesize

Acronym  & Full Name \&   Primary Science Drivers \\
\hline
SIM (NASA) & Space Interferometry Mission  \\
& Precision astrometry; exosolar planets \\
\hline
FKSI (NASA) & Fourier-Kelvin Space Interferometer \\
& Find Jovian planets (nuller); map circumstellar disks \\
\hline
SMART-3 (ESA) & SMART-3 \\ 
 & Test free-flying concept for ESA IRSI-Darwin mission \\
\hline
IRSI-Darwin (ESA) & Infra-Red Space Interferometer (one concept: Darwin) \\
& Image terrestial planets (IR nuller); measure spectra\\
\hline
TPF (NASA) & Terrestrial Planet Finder \\
& Image terrestial planets (IR nuller); measure spectra \\
\hline
SPIRIT (NASA) & Space Infrared Interferometry Trailblazer  \\
 & Far-IR, sub-mm galaxy counts; precurser to SPECS \\
\hline
SPECS (NASA) & Submillimeter Probe of the Evolution of Cosmic Structure\\
 & High-resolution map of High-Z universe (far-IR, sub-mm) \\
\hline
SI (NASA) & Stellar Imager \\
 & Image surfaces of stars (visible, ultraviolet) \\
\hline
MAXIM (NASA) & Micro-Arcsecond X-ray Imaging Mission \\
 & Map black hole accretion disks and event horizons (X-rays) \\
\hline
MAXIM Pathfinder (NASA) & MAXIM Pathfinder \\
 & Demonstrate feasibility of X-ray interferometry; achieve 100~$\mu$-arcsecond 
resolution\\
\hline
LF (NASA) & Life Finder  \\
 & Search for biomarkers in planet spectra; TPF extension \\
\hline
PI (NASA) & Planet Imager  \\
 &Image surfaces of terrestial planets, 25x25 pixels \\
& (requires 6000km baselines, futuristic!) \\
\hline
\end{tabular}
\end{center}

\end{table}

\subsection{Future Ground-based Interferometers}

While it is interesting to speculate about the future of space
interferometry, we recognize that it will be expensive, difficult, and
slow-paced.  In the next 10 or 20 years, we can expect more affordable
and rapid progress to be possible from the ground.  In this concluding
section, I review some of the necessary characteristics of an Optical
Very Large Array (OVLA).  \citet{ridgway2000} discusses many of these
considerations, and I refer the reader to his interesting report for 
further details.

\subsubsection{Design Goals}
The main design goal of a next-generation optical interferometer array
will be to allow the ordinary astronomer to observe a wide-range of
targets without requiring extensive expert knowledge in interferometer
observations.  An imaging interferometer with great sensitivity could
fulfill this promise by providing finished images, the most intuitive
data format currently in use. It will not be a specialty instrument
with narrow science drivers, but a general purpose facility to advance
our understanding in a wide range of astrophysical areas.

\subsubsection{Optical Very Large Array}
One way to achieve this design goal is to scale up the existing
arrays.  Simply put, this main goal will require an array with a large
number of telescopes ($\simge$20 to allow reliable aperture synthesis
imaging) and with large-aperture telescopes corrected by adaptive
optics (preferably using laser guide stars for full-sky coverage),
allowing a reasonably faint limiting magnitude (roughly speaking,
brighter than $\sim$15th magnitude in the infrared with no phase
referencing).

This array would likely be reconfigurable, like the radio VLA, to
allow different angular resolutions to be investigated. The longest
baselines should cover a few kilometers ($\sim$0.1 {\em
  milli}-arcsecond resolution in the near-IR).  The main limitation of
such a system will be a small field-of-view, typically limited to the
diffraction-limited beam of an individual telescope (for 10-m class
telescopes, the instantaneous field of view would be only about
$\sim$50 milliarcseconds) -- although mosaicing would be possible, as
in the radio.  There are schemes which can image a larger field
simultaneously, but are probably not very practical.

With an even larger (billion-dollar) budget, one can partially combine
the goals of interferometry with the community priority for a 30~m
diameter telescope.  This clever idea was recently proposed by R.
Angel and colleagues at the University of Arizona.  In their ``20/20''
scheme, light from two extremely large telescopes (diameter $>$20
meters) would be combined in a Fizeau combination scheme, patterned
after the Large Binocular Telescope, maintaining the entire
field-of-view ($\sim$30'', limited by atmospheric turbulence) with the
resolution of the two-element interferometer.  Further, this scheme
maximizes raw collecting area and would boast potentially incredible
sensitivity ($>$20 mag!).  One demanding feature of this design is
that the two 20+~m telescopes would have to smoothly move along a
track in real-time to maintain the large field-of-view; this may not
be impossible, but is surely an interesting complication.  Further,
the imaging advantages of this system only work when the two-telescope
baseline is 5-10$\times$ as large as the telescope diameter, and hence
the ``20/20'' interferometer would have maximum baselines of only a
few hundred meters at most, not much better than current
interferometer arrays.  While granting that this system could allow
much fainter objects to be observed, this option would cost many times
more than a dedicated OVLA system described above.

\subsubsection{Technological Obstacles Needed to be Overcome}
If optical interferometry is to continue its impressive growth over
the coming decades, important breakthroughs must be made in critical
areas.  Here, I briefly list a few obvious improvements
which would make an OVLA more affordable.

The main advance needed to make the OVLA affordable will be the
development of ``cheap'' large aperture telescopes with adaptive
optics.  Currently, it costs millions of dollars to build even a
4~m-class telescope -- without adaptive optics.  Advances in lightweight
mirrors with adaptive optics designed-in from the beginning may change
the economics of the situation.  

Another area which could revolutionize optical interferometry is
advances in photonic bandgap fiber materials
\citep[e.g.,][]{mueller2002}.  These materials offer possibility of
extremely wide-bandwidth, low dispersion and low-loss single-mode
fibers, which could open up the possibility of practical fiber delay
lines. Such an advance would greatly simplify the optical beam-train
and engineering of an optical interferometer, making projects such as
'OHANA straightforward.  This would put optical interferometry on more similar
footing as radio interferometry, where cable delay lines (either
coaxial or fiber) are routinely used.

Combining dozens of telescopes may not be practical using bulk optics,
and solutions involving integrated optics should be pursued. The main
limitation of this technology is restricted wavelength coverage,
currently only proven shortward of 2.2~$\mu$m.  Development of
materials (e.g., lithium niobate) and fabrication processes that can
extend the coverage into the thermal infrared (1-5$\mu$m) would mean
that a general purpose interferometer could be built around an
integrated optics combiner.  Work is currently underway in Europe
towards this end, in particular in pursuit of mid-infrared nulling
capabilities for the ESA IRSI-Darwin mission \citep{haguenauer2002}.

Lastly, improved infrared detectors are crucial to maximizing the
scientific output of a future interferometer. It has already been
discussed here (see \S\ref{detectors}) that near-infrared detectors
remain limited by avoidable detector ``read'' noise, and 
a future OVLA must have better detectors.

\section{Conclusion}
After decades of development, optical interferometry is now poised to
play a major role in mainstream astronomy. The emergence of
well-funded interferometer ``facilities,'' in particular the Very
Large Telescope Interferometer and the Keck Interferometer, promise to
revolutionize the impact of high-resolution observations in many areas
of astrophysics.  Clearly, the main beneficiaries will be 
stellar astrophysics and galactic astronomy, in particular the areas
of star and planet formation, fundamental stellar properties, and all
stages of stellar evolution.  In addition, we can look forward to the first
extragalactic results.

Although the Keck and VLT interferometers hold immense promise, the
field is currently driven forward by the activities of many other
smaller groups, and scientific results will be dominated by these
workers for the near future.  While many experimental (astro)physics
fields have matured to the point where future progress rests in ``big
science'' collaborations and national research centers (e.g., NASA),
optical interferometry represents one of the few healthy and active
``experimental astrophysics'' endeavors left in astronomy where
university-based groups continue to make important technical
innovations and astronomical discoveries.

It is widely acknowledged that astronomy as a whole is experiencing a
golden age of progress, spurred on by observational advances across
the electromagnetic spectrum.  Optical Interferometry has expanded in
response to its own promising initial results, and we in the field
look forward to exploiting the significant infrastructure buildup just
now being completed.  I hope that the next review of optical
interferometry will vindicate my optimism in the field, and that the
pioneering discoveries reported here presage even grander exploits.
It is safe to predict that 
the next decade will be critical to the field of high-resolution
optical astronomy, since the scientific impact of current facilities
will wholly determine whether the substantial funding required for an
``Optical'' Very Large Array can be justified to the international
astronomical community.  

\ack 

Firstly, I must apologize for omitting many important works due to
space constraints, especially in the areas of speckle interferometry.
I thank J.-P. Berger, R. Millan-Gabet, and P. Lena for a careful
reading of the manuscript and important suggestions. Also, I
acknowledge useful conversations with E. Pedretti, A. Boden, 
P. Tuthill, and F.  Malbet.  
Lastly, I recognize formative discussions with S. Ridgway, C.
Haniff, D.  Buscher, and D.  Mozurkewich, whose ideas have helped
shape my perspective of the field of optical interferometry,
especially on the future of an OVLA.

\footnotesize
\bibliography{monnier,monnier2}
\bibliographystyle{apalike}

\end{document}